\documentclass[
 reprint,
superscriptaddress,
 amsmath,amssymb,
 aps,
]{revtex4-2}

\usepackage[dvipsnames]{xcolor}
\usepackage{dsfont}  %
\usepackage[percent]{overpic} %
\usepackage{float}
\usepackage{braket}
\usepackage{graphicx}%
\usepackage{dcolumn}%
\usepackage{bm}%

\usepackage[colorlinks=true,
            linkcolor=blue,
            citecolor=blue,
            urlcolor=blue]{hyperref}
\usepackage[normalem]{ulem}

\newcommand{\be}{\begin{equation}}
\newcommand{\ee}{\end{equation}}
\newcommand{\bee}{\begin{equation*}}
\newcommand{\eee}{\end{equation*}}
\newcommand{\dd}{\mathrm{d}}

\newcommand\braa[1]{{({#1}|}}
\newcommand\kett[1]{{|{#1})}}

\newcommand\brakett[2]{\left(#1 \! \,  \middle|\, \! #2\right)}

\begin{document}
\flushbottom

\preprint{APS/123-QED}

\title{Mixed-State Topology in One-Dimensional Open Floquet Systems}

\author{Görkem D. Dinc}
\affiliation{School of Physics, University of Melbourne, Parkville, VIC 3010, Australia}

\author{Alexander Schnell}
\affiliation{Technische Universität Berlin, Institut für Physik und Astronomie, 10623 Berlin, Germany}

\author{Andy M. Martin}
\affiliation{School of Physics, University of Melbourne, Parkville, VIC 3010, Australia}

\noaffiliation

\date{\today}%

\begin{abstract}

We investigate the non-equilibrium topology of the steady state of a periodically driven, dissipative Su--Schrieffer--Heeger (SSH) chain using the ensemble geometric phase (EGP) $\phi_{\mathrm{EGP}}$---a generalisation of the Zak phase to mixed states. The coupling to thermal reservoirs is treated microscopically within Floquet--Born--Markov theory without the rotating-wave approximation. We classify the steady state rather than the generator of the dynamics, using its purity spectrum as the analogue of a band structure. We show that, within the EGP framework, the $\mathbb{Z}\times\mathbb{Z}$ structure of isolated chiral-symmetric Floquet SSH systems generally does not carry over to the steady state. Instead, the steady state carries a single invariant whose type is set by an underlying symmetry. When this symmetry holds, the invariant is $\mathbb{Z}_2$-valued, with quantised $\phi_{\mathrm{EGP}}=\pm\pi/2$ at all times and finite temperatures. The EGP can switch within a driving period, tracking the presence of edge modes in the open-boundary purity spectrum and thereby establishing a bulk-boundary correspondence. The fraction of each period spent in the nontrivial regime varies continuously with the driving frequency, and the switching disappears if the  dependence of the dissipator on the drive is neglected. When the symmetry is broken, $\phi_{\mathrm{EGP}}$ is no longer quantised, but its winding over a drive period $T$ defines an invariant $\nu_T \in \mathbb Z$, with no purity-spectrum bulk-boundary correspondence. We propose that its changes signal temperature- and frequency-driven topological phase transitions. 
Our formalism extends to general quadratic fermionic systems with linear bath couplings.

\end{abstract}

\maketitle

\section{\label{sec:level1}Introduction}

Topological phases of matter are characterised by integer-valued invariants that remain robust under continuous deformations and give rise to protected edge modes \cite{Ozawa2019,Bernevig2025,Altland2010}. In one-dimensional systems \cite{Batra2020}, a prominent example is the Zak phase \cite{Zak1989,Delplace2011,Aihara2020}, whose nontrivial value is associated with the appearance of robust states localised at the boundaries of a finite chain. Models with nontrivial one-dimensional topology have been realised experimentally, for example, in quantum simulators using photonic waveguides \cite{Xiao2014,Wang2016,SaeiGharehNaz2018,Ozawa2019,Malkova2009,Queralto2020}, electrical circuits \cite{Lee2018}, ultracold atoms \cite{Atala2013} and exciton-polaritons in semiconductor structures \cite{St-Jean2017,Solnyshkov2016,Ozawa2019,Yu2025}.

In recent years, increasing attention has been directed to non-equilibrium topological phases, where new phenomena can emerge beyond equilibrium classifications. In particular, time-periodically driven (Floquet) systems exhibit quasienergy band structures with additional spectral gaps and topologically nontrivial winding of Floquet modes, enabling novel topological phases without static counterparts and supporting exotic edge modes~\cite{EckardtColl,Eckardt_2015,Nathan_2015,Kitagawa_2010,Wintersperger2020a,Schnell2024a,Rudner_2013,Gavensky2025}.

At the same time, realistic quantum systems are inherently open and subject to dissipation \cite{breuer2002theory, Weiss2012}. As a result, they are generally described by mixed states at finite temperature, making the extension of topological classifications, conventionally formulated for pure states, nontrivial \cite{Wawer2021,Hannukainen2022,Tran2017,Schnell2024a,yang2026,Yang2025,Uhlmann_1986,Uhlmann_1991,Uhlmann_1989,Viyuela_2014,Hou_2023}. A central challenge is to identify quantities that remain quantised and physically meaningful in such settings. The ensemble geometric phase (EGP) $\phi_{\mathrm{EGP}}$, introduced by Bardyn \emph{et al.}~\cite{Bardyn_2018}, provides a promising framework for characterising the topology of mixed states \cite{Unanyan2020,Molignini_2023,Huang2022,Wawer2021,Mink2019,hannukainen2026,Huang2025,Wawer2021_2}.

Molignini and Cooper~\cite{Molignini_2023} showed that the EGP correctly captures finite-temperature topological phase transitions in a dissipative Su--Schrieffer--Heeger (SSH) chain. In particular, they showed that the topological classification of the isolated SSH model can be extended to open systems in the presence of thermal reservoirs.
However, the interplay between periodic driving, dissipation, and mixed-state topology remains largely unexplored. In isolated one-dimensional chiral-symmetric Floquet systems, topology is characterised by a pair of invariants associated with the $0$  and $\pi$ gaps of the quasienergy spectrum \cite{Asboth_2014}. Whether such a classification persists in open, driven systems at finite temperature, and is reliably visible in the steady state, is an open question.

In this work, we investigate the non-equilibrium topology of the steady state of a periodically driven, dissipative SSH chain. 
Extending earlier work for the SSH chain in the absence of periodic driving \cite{Molignini_2023}, we implement Floquet--Born--Markov theory allowing for a microscopic description of the driven system in the presence of thermal reservoirs.
The steady state can be characterised by its covariance matrix, whose Hermitian purity spectrum provides a direct analogue of band topology for mixed states. Building on this framework, we identify two regimes determined by an underlying steady-state symmetry, which is based on the chiral structure of the model. In the symmetric regime, $\phi_{\mathrm{EGP}}=\pm \pi/2$ is quantised at all times, describing the topology of the steady state and predicting the appearance of edge states in the purity spectrum, thereby establishing a bulk-boundary correspondence between the EGP and the open-boundary purity spectrum. In the symmetry-broken regime, a winding number $\nu_T$ can be extracted and captures temperature- and frequency-driven topological phase transitions. Here, no bulk-boundary correspondence between $\nu_T$ and the open-boundary purity spectrum is found.
\begin{figure}
    \centering
    \includegraphics[width=0.45\textwidth]{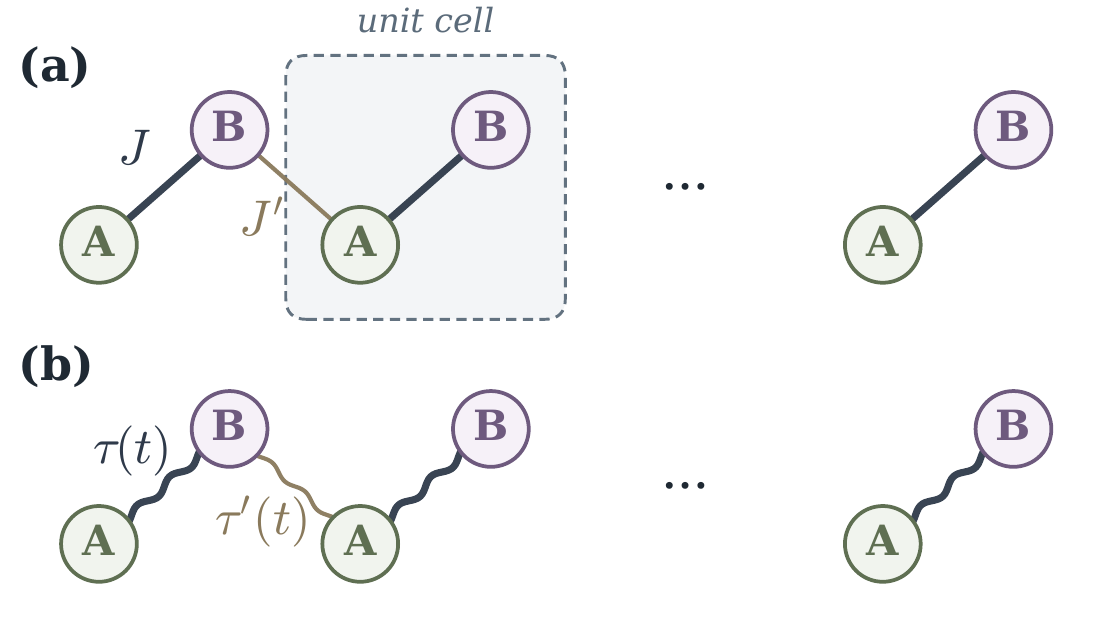}
    \caption{Sketch of the SSH model. Highlighted is  a unit cell consisting of two sites, A and B, connected by bonds indicating the intracell and intercell nearest-neighbour hoppings. (a) Fixed intracell and intercell hoppings $J$ and $J'$, respectively, and (b) time-periodically modulated hoppings $\tau(t) = J + 2V \cos(\omega t)$ (intracell) and $\tau'(t) = J' - 2V \cos(\omega t)$ (intercell). Here, $V$ and $\omega$ denote the driving strength and frequency, respectively.}
    \label{fig:sshmodel}
\end{figure}

This paper is organised as follows. In Sec.~\ref{sec:model}, we introduce the driven-dissipative SSH model. In Sec.~\ref{sec:mixedstatetopology}, we outline the description of mixed-state topology for quadratic fermionic systems and introduce the EGP. In Sec.~\ref{sec:theoreticalframework}, we present the Floquet--Born--Markov framework used to obtain the steady state. In Sec.~\ref{sec:results}, we discuss the relation between generator and steady-state topology, analyse topological phases in the different symmetry regimes, and comment on practical considerations for measurements of the EGP. We conclude in Sec.~\ref{sec:conclusion}.

\section{Model} \label{sec:model}

\subsection{SSH Chain}

The SSH model, as depicted in Fig.~\ref{fig:sshmodel}(a), is a one-dimensional lattice model of non-interacting fermions with two bands. The unit cell consists of two sites, A and B, with alternating intracell and intercell hopping amplitudes $J$ and $J'$, respectively. Throughout this work, the chain length $L$ denotes the number of unit cells, corresponding to $2L$ lattice sites.  For the SSH model, the real-space Hamiltonian (in dimensionless units and with periodic boundary conditions (PBC)) reads
\begin{equation}
H_\mathrm{SSH} = -\sum_{j=1}^{L}\bigl[J\,f_{j,A}^\dagger f_{j,B}
+ J'\,f_{j+1,A}^\dagger f_{j,B} + \text{h.c.}\bigr],
\label{eq:SSH_rs}
\end{equation} 
where $f_{j,A}$ ($f_{j,B}$) and $f_{j,A}^\dagger$ ($f_{j,B}^\dagger$) are fermionic annihilation and creation operators for unit cell $j$ on sublattices A (B), respectively. We identify $j=L+1$ with $j = 1$ and $\text{h.c.}$ denotes the Hermitian conjugate. The fermionic operators fulfil the fundamental fermionic anticommutation relations $\{f_i,f_j^\dagger\}=\delta_{ij}$. To consider the system under open boundary conditions (OBC), the summation of the second term is taken to $L-1$.

It is well established \cite{su_1979} that the SSH model hosts a topological phase with protected edge states when the intercell hopping $J'$ exceeds the intracell hopping $J$, and a trivial phase otherwise. The topological phase transition thus occurs at $J = J'$, where the bulk band gap closes. 
The quantisation of the associated topological invariant (the Zak phase $\phi_\mathrm{Z}$) is protected by the chiral (sublattice) symmetry of the model. Specifically, the SSH Hamiltonian anticommutes with the unitary, Hermitian, and (within a unit cell) local chiral symmetry operator $\Gamma$ (acting as $+1$ on A and $-1$ on B), $\Gamma H_\mathrm{SSH}\Gamma = -H_\mathrm{SSH}$, which quantises $\phi_\mathrm{Z}$. Physically, this reflects the bipartite structure of the lattice, where the Hamiltonian couples only A-sites to B-sites, never two sites of the same sublattice.

\subsection{Periodically Driven Isolated SSH Chain} \label{sec:isol_driv_SSH}

In time-periodically driven (Floquet) systems, the time dimension can be interpreted as an additional synthetic degree of freedom alongside the real-space lattice. As a result, Floquet systems exhibit richer topology than their static counterparts and can host exotic edge states associated with their quasienergy spectrum. First, consider an arbitrary isolated quantum system described by a Hamiltonian $H(t)$. The time evolution of the state $\ket{\psi(t)}$ is governed by the Schrödinger equation 
\be
i \hbar \frac{\partial}{\partial t} \ket{\psi(t)}= H(t) \ket{\psi(t)},  \label{eq:SE}  
\ee
where we restrict ourselves to the case of a time-periodic system Hamiltonian, i.e.\ $H(t+T)=H(t)$. To formally solve Eq.~(\ref{eq:SE}), one can introduce the time-evolution (or Floquet) operator 
\be
U(t,0)=\mathcal{T} \exp \biggl(-\frac{i}{\hbar}\int_0^t \mathrm{d} t_1  H(t_1) \biggr), \label{eq:Floop}
\ee
with time-ordering operator $\mathcal{T}$. In particular, one can extract the one-cycle time-evolution operator $U(T)$, whose repeated application describes the stroboscopic dynamics of Eq.~(\ref{eq:SE}).  
Since $U(T)$ is unitary, it can be expressed in terms of a so-called \textit{Floquet Hamiltonian} $H_\mathrm{F}$ \cite{EckardtColl,Eckardt_2015}, which is time-independent and can be extracted via 
\be
U(T)=\exp\Bigl(-\frac{i}{\hbar}H_\mathrm{F}T\Bigr)   \hspace{0.1cm} \Leftrightarrow \hspace{0.1cm} H_\mathrm{F}= \frac{i\hbar}{T} \log U(T). \label{eq:HU}
\ee
At stroboscopic times $nT, n\in \mathbb{N}$, the evolution is therefore generated by the time-independent Hamiltonian $H_\mathrm{F}$. In practice, it is often convenient to represent these operators in a special eigenbasis obtained from the fundamental solutions of Eq.~(\ref{eq:SE}), namely the Floquet states, which take the form \cite{Eckardt_2015}
\be
\ket{\psi_\mu (t)} = e^{-i\varepsilon_\mu t/\hbar}  \ket{\Phi_\mu (t)}, \label{eq:Floquetstates}
\ee
where $\varepsilon_\mu$ denote the \textit{quasienergies} and $\ket{\Phi_\mu(t)}$ are the time-periodic \textit{Floquet modes}, i.e.~$\ket{\Phi_\mu(t+T)}=\ket{\Phi_\mu(t)}$. Note that the Floquet modes are not uniquely defined since setting $\varepsilon_\mu \rightarrow \varepsilon_\mu + m \hbar \omega$ and $\ket{\Phi_\mu (t)} \rightarrow e^{im\omega t}\ket{\Phi_\mu (t)}$, with $m \in \mathbb{Z}$ and $\omega=2\pi /T$, leaves the Floquet states $\ket{\psi_\mu (t)}$ invariant. This is why the system `energies' $\varepsilon_\mu$ are only defined up to integer multiples of $\hbar \omega$ and are referred to as quasienergies. 
We can spectrally decompose $U(T)$ in terms of the Floquet modes $\ket{\Phi_\mu (t)}$, 
\be
U(T) = \sum_{\mu=1}^{2L} e^{-i\varepsilon_\mu T/\hbar} \ket{\Phi_\mu (T)} \bra{\Phi_\mu (0)},
\ee
and, 
due to the multibranch structure of the complex logarithm, we can obtain a family of Floquet generators 
\be
\begin{aligned}
H_{\{m_1,\ldots,m_{2L}\}}
&= \frac{i \hbar}{T} \log U(T) \\
&= \sum_{\mu=1}^{2L} (\varepsilon_\mu + m_\mu \hbar \omega) \,
   \ket{\Phi_\mu(T)} \bra{\Phi_\mu(0)} ,
\end{aligned}
\label{eq:HB}
\ee
with driving frequency $\omega$ and $m_\mu \in \mathbb{Z}$. Redefining the quasienergies as $\varepsilon_\mu \rightarrow \varepsilon_\mu + m_\mu \hbar \omega$ therefore corresponds to switching the branches of the complex logarithm in Eq.~(\ref{eq:HB}). Inspired by Bloch's theorem for spatially periodic systems, the principal branch of the quasienergy spectrum, i.e.~the convention
\be
    \varepsilon_\mu \in \Bigl(-\frac{\hbar \omega}{2}, \frac{\hbar \omega}{2} \Bigr] , 
\ee
is referred to as the \textit{first Floquet Brillouin zone (FBZ)}, 
while other branches are often called \emph{Floquet replicas (or copies)}. %
Henceforth, we set $\hbar = 1$.   

\begin{figure}
  \centering
  \includegraphics[width=0.4\textwidth]{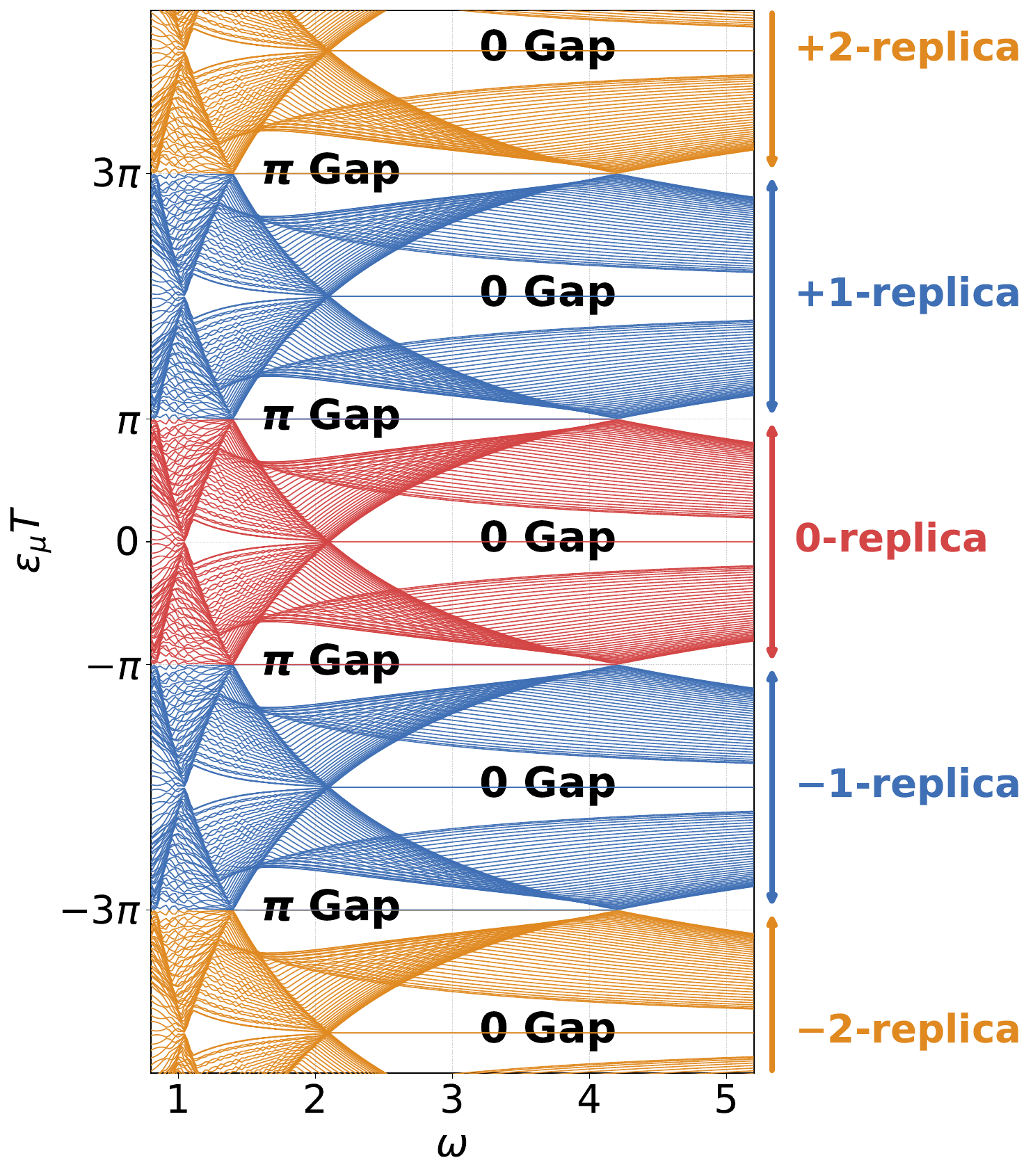}
  \caption[Quasienergy spectra]{Extended OBC quasienergy spectra $\varepsilon_\mu T$
    of the periodically driven SSH Hamiltonian $H_\mathrm{PDSSH}(t)$ in Eq.~\eqref{eq:SSHFloquet},
    showing the zeroth, first ($\pm 1$) and second ($\pm 2$) Floquet replicas as a function of driving frequency $\omega$
    for a chain length $L=40$, a driving strength of $V=-0.2$ and tunnelling amplitudes $J=0.8$, $J'=1.3$.
    Due to the periodic modulation, each Floquet replica exhibits a gap at
    $\varepsilon_\mu T = 0 \pmod{2\pi}$ ($0$ gap) and another at
    $\varepsilon_\mu T = \pi \pmod{2\pi}$ ($\pi$ gap), allowing topological edge states in either gap.}
  \label{img:quasienegspectrum_theory_branches}
\end{figure}

The fact that quasienergies are defined only up to multiples of $\omega$ has important consequences for the topology of Floquet systems. We now return to the SSH model in Eq.~(\ref{eq:SSH_rs}) and, motivated by Ref.~\cite{Lago_2015}, we add a time-periodic modulation to the tunnelling amplitudes, as sketched in Fig.~\ref{fig:sshmodel}(b). This is described by the Hamiltonian
\begin{equation}
\begin{aligned}
H_\mathrm{PDSSH}(t)
&= -\sum_{j=1}^L \Bigl[
    (J + 2V \cos(\omega t)) \,f_{j,A}^\dagger f_{j,B} \\
&\quad + (J' - 2V \cos(\omega t)) \,f_{j+1,A}^\dagger f_{j,B} \\
&\quad + \text{h.c.} \Bigr] ,
\end{aligned}
\label{eq:SSHFloquet}
\end{equation}
with driving strength $V$ and frequency $\omega$. This choice of driving preserves the chiral symmetry of the undriven SSH model.
In Fig.~\ref{img:quasienegspectrum_theory_branches}, we plot the extended quasienergy spectrum $\varepsilon_\mu T$ for OBC with tunnelling amplitudes $J=0.8$ and $J'=1.3$, for chain length $L=40$ and driving strength $V=-0.2$, as a function of the driving frequency $\omega$. 
We observe that the Floquet bands repeat, forming a quasienergy band structure with multiple spectral gaps. We highlight different Floquet replicas with different colours.
The time-periodic system exhibits two kinds of gaps, namely the $0$ gap at $\varepsilon_\mu T = 0 \pmod{2\pi}$ and the $\pi$ gap at $\varepsilon_\mu T = \pi \pmod{2\pi}$, which can host states in distinct frequency regimes. For $\omega>2$, edge states emerge inside the $0$ gap, while the $\pi$ gap hosts edge states for $\omega \in [1.5,4.2]$. Consequently, the periodic driving allows the system to support topologically protected edge states in both the $0$ gap and the $\pi$ gap, making Floquet systems particularly rich from a topological viewpoint.
Moreover, the micromotion, i.e.\ the time dependence of the Floquet modes $\ket{\Phi_\mu(t)}$ (see Eq.~\eqref{eq:Floquetstates}), can introduce additional topologically nontrivial winding.
However, since most invariants, including the Zak phase $\phi_\mathrm{Z}$ \cite{Zak1989}, are defined in terms of the (many-body) ground-state wavefunction, the periodicity of the quasienergy spectrum prevents a clear definition of a ground state, complicating the topological classification of Floquet systems. 

Several approaches exist to classify the topology of Floquet systems. 
A simple strategy is to apply the Altland--Zirnbauer classification to an effective Floquet Hamiltonian 
$H_\mathrm{F}$, which governs the stroboscopic evolution \cite{Ryu_2010}. 
However, it has been shown that this approach can miss topological contributions associated with the micromotion within one driving period, and therefore a more complete Floquet invariant is required
\cite{Rudner_2013,Kitagawa_2010,Nathan_2015,Jiang_2011,Roy_2017,Asboth_2014,Lago_2015}.
In isolated one-dimensional systems with chiral symmetry, the classification simplifies considerably. 
In this case the topology associated with the quasienergy gaps at $0$ and $\pi$ is characterised by a pair of \emph{winding numbers} $(\nu_0,\nu_\pi)$, 
which are obtained from chiral-symmetric decompositions of the Floquet operator evaluated at two special times within the driving period 
\cite{Jiang_2011,Roy_2017,Asboth_2014,Lago_2015}.
Following Ref.~\cite{Asboth_2014}, these invariants can be defined as follows. The effective Floquet Hamiltonian does not automatically inherit the chiral symmetry of the instantaneous Hamiltonian. Chiral symmetry is ensured if the period can be split at an intermediate time $t_1$ into two parts related by the chiral symmetry operator $\Gamma$~\cite{Asboth_2014}.
Let $F = U(\tau+t_1,\tau)$ [Eq.~\eqref{eq:Floop}] denote the evolution over the first part of the period, with $0 < t_1 < T$. If $t_1$ is chosen such that $\Gamma F^\dagger \Gamma = U(\tau+T,\tau+t_1)$, the one-period Floquet operators starting at the two times $\tau$ and $\tau+t_1$ take the form
\be
U = \Gamma F^\dagger \Gamma F, \qquad U' = F \Gamma F^\dagger \Gamma,
\ee
and both $U$ and $U'$, together with their effective Floquet Hamiltonians $H_\mathrm{F}, H_\mathrm{F}'$, are chiral-symmetric~\cite{Asboth_2014}. Consequently, they can each carry topological invariants.

Imposing PBC, in quasimomentum space and the canonical basis, the effective Floquet Hamiltonians can be decomposed into a block off-diagonal form
\be
H_{\mathrm F}^{(\prime)}(k)=
\begin{pmatrix}
0 & h^{(\prime)}(k) \\
\bigl(h^{(\prime)}(k)\bigr)^\dagger & 0
\end{pmatrix},
\ee
which allows for the definition of the winding numbers
\be
\nu^{(\prime)}  = \frac{1}{2\pi i}\int_{-\pi}^{\pi} \mathrm{d}k \,
\frac{\mathrm{d}}{\mathrm{d}k}\ln \det h^{(\prime)}(k),
\ee
where the integration is taken over the first Brillouin zone.
The invariants corresponding to the edge states in the $0$ and $\pi$ gaps then read as
\be
\nu_0 = \frac{\nu+\nu^{\prime}}{2}, \quad \nu_\pi = \frac{\nu-\nu^{\prime}}{2} .
\ee
More details and a simplification of $\nu_0,\nu_\pi$ are given in Ref.~\cite{Asboth_2014}.
For chiral-symmetric periodically driven SSH models, this leads to a $\mathbb{Z} \times \mathbb{Z}$ topological classification associated with the $0$  and $\pi$-quasienergy gaps \cite{Jiang_2011,Lago_2015,Asboth_2014}. If chiral symmetry is broken but particle--hole symmetry remains, the classification reduces to $\mathbb{Z}_2 \times \mathbb{Z}_2$, so that only the parity of the number of end states at quasienergy $0$ and $\pi$ is protected~\cite{Jiang_2011,Roy_2017,Asboth_2014}.

\subsection{Periodically Driven-Dissipative SSH Chain} \label{sec:drivdissmodel}

\begin{figure}
    \centering
    \includegraphics[height=0.48\linewidth]{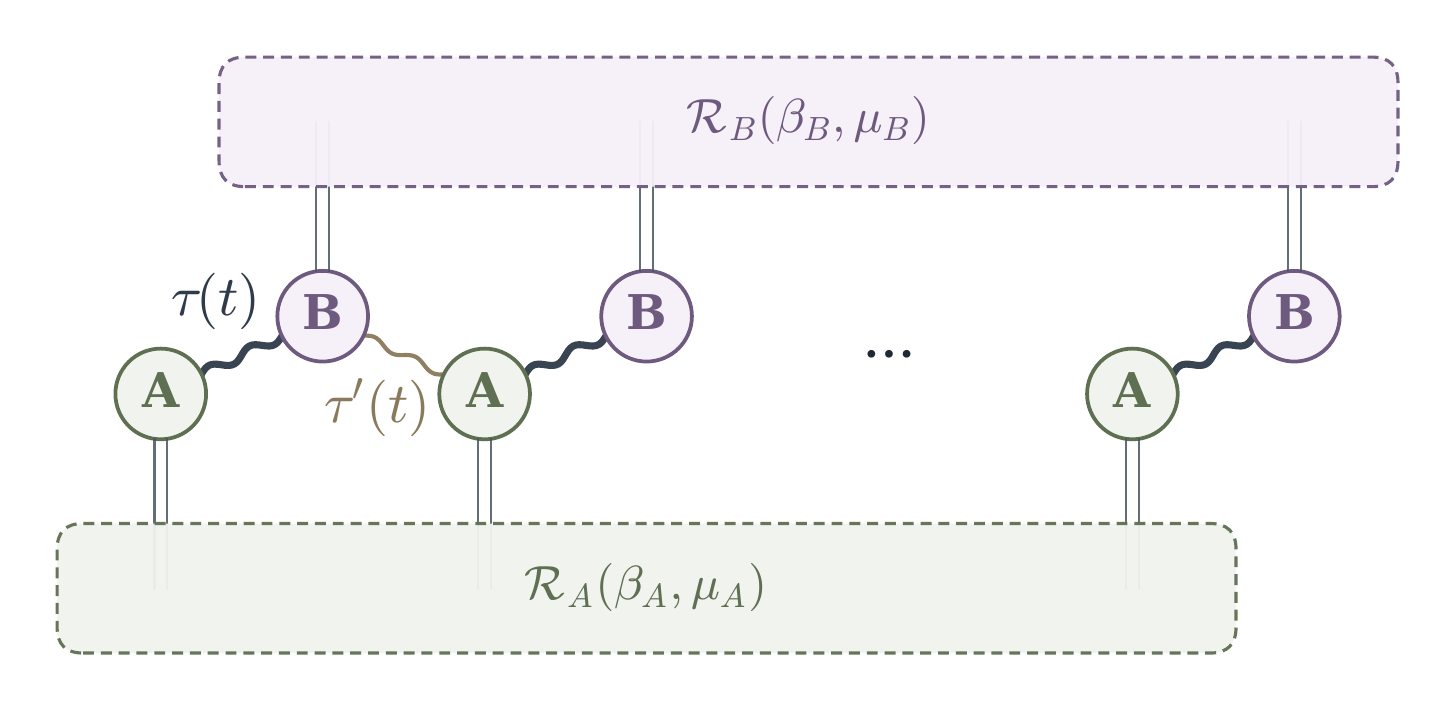}  
    \caption{Sketch of the system--bath configuration of interest in this work. Sublattice A (B) of the SSH chain is coupled dissipatively to a Markovian fermionic thermal bath $\mathcal{R}_A$ ($\mathcal{R}_B$) at inverse temperature $\beta_A$ ($\beta_B$) and chemical potential $\mu_A$ ($\mu_B$). Here, the tunnelling amplitudes $\tau(t)$ and $\tau'(t)$ are time-periodically modulated with driving frequency $\omega$ and driving amplitude $V$, as in Fig.~\ref{fig:sshmodel}(b). The drive is chosen to preserve the chiral symmetry of the system.}
    \label{img:drivenSSHdiss}
\end{figure}

The aim of this paper is to investigate topological phase transitions in periodically driven-dissipative quantum systems. We consider a total time-dependent system--bath decomposition according to
\be
H(t) = H_{\mathcal S}(t) \otimes \mathds{1}_{\mathcal{R}} + \mathds{1}_{\mathcal{S}} \otimes H_{\mathcal{R}} + 
\kappa \sum_\ell S_\ell \otimes R_\ell , \label{eq:H_sysbath} 
\ee
where the considered system $\mathcal{S}$ is the isolated periodically driven SSH chain, with system Hamiltonian $H_{\mathcal S}(t) = H_\mathrm{PDSSH}(t)$ given by Eq.~(\ref{eq:SSHFloquet}). Motivated by the findings in Ref.~\cite{Molignini_2023}, we couple the chain to two fermionic Markovian thermal reservoirs $\mathcal{R}_A$ and $\mathcal{R}_B$, at inverse temperatures $\beta_A$ and $\beta_B$ and chemical potentials $\mu_A$ and $\mu_B$, respectively, cf.~Fig.~\ref{img:drivenSSHdiss}. The bath coupling operators describing the dissipative dynamics are given by 
\begin{equation}\label{eq:Xell}
S_\ell = S_{j,I,\alpha}
= \begin{cases}
f_{j,I} + f_{j,I}^\dagger, & \alpha=1,\\[1ex]
i\bigl(f_{j,I} - f_{j,I}^\dagger\bigr), & \alpha=2 ,
\end{cases}
\end{equation}
with index $\ell = (j,I, \alpha)$ corresponding to the injection and removal of particles at site $j$ from (to) a bath $I = A,B$ with inverse temperature $\beta_I$ and chemical potential $\mu_I$, as illustrated in Fig.~\ref{img:drivenSSHdiss}. Note that we couple each site to two of the quadratures via the index $\alpha$. With the given choice, the $S_\ell$ are all Hermitian and the dissipation is local, i.e.~we assume that the coherence length of the bath is much shorter than the distance between the individual sites of the chain. Here, the $R_\ell$ are the bath operators acting on the reservoir $\mathcal{R}_I$ and are encoded in the bath correlation functions (explicitly introduced later) within the Born--Markov approximation.

The necessary theoretical framework to investigate this open system within the  Floquet--Born--Markov approach will be introduced in Sec.~\ref{sec:theoreticalframework}.
We emphasise that we illustrate our results using the SSH chain and a particular choice of coupling. Our framework, however, extends to arbitrary quadratic fermionic Hamiltonians $H_{\mathcal S}(t)$ with linear coupling operators $S_\ell(t)$.

\section{Mixed-state topology} \label{sec:mixedstatetopology}

In isolated one-dimensional quantum systems, like the SSH model, an invariant which predicts the topological behaviour of the system is the Zak phase $\phi_\mathrm{Z}$, given by
\be
\phi_\mathrm{Z} =  2\pi P = \Im\log\langle\psi_0|T_\mathrm{tw}|\psi_0\rangle
\label{eq:zakphaseresta} ,
\ee
where $P$ is the electric polarisation, $\ket{\psi_0}$ is the many-body ground state and $T_\mathrm{tw}$ the twist operator, which can be extracted from the many-body centre-of-mass operator $X$ via 
\begin{equation}
T_\mathrm{tw} = \exp\Bigl(\frac{2\pi i}{L}X\Bigr),
\qquad
X=\sum_{j=1}^{L} j\,n_j,
\label{eq:T_def} 
\end{equation}
where $n_j=f_{j}^\dagger f_{j}$ is the number operator on site $j$. For the SSH model, $X$ takes the form
$
X = \sum_j\bigl[j\,f_{j,A}^\dagger f_{j,A} + (j+\tfrac12)\,f_{j,B}^\dagger f_{j,B}\bigr] 
$. 
Up to a factor of the electron charge, $\phi_\mathrm{Z}$ gives the electronic contribution to the bulk polarisation and coincides with the single-particle Zak phase for band insulators under PBC \cite{Resta_1998}. Note that this construction relies heavily on the assumption that the system is described in terms of single pure-state wavefunctions and, further, on a Hermitian and gap-protected band structure. For isolated quantum systems, one is thus generally concerned with zero-temperature states.
However, realistic systems are generally mixed (e.g.\ thermal states at finite temperature) and thus described by density operators $\varrho$. 
This raises the question of how topological invariants, such as the Zak phase $\phi_\mathrm{Z}$, can be generalised to mixed states and open quantum systems.

\subsection{Purity Spectrum}

\begin{figure}
    \centering
    \includegraphics[width=0.38\textwidth]{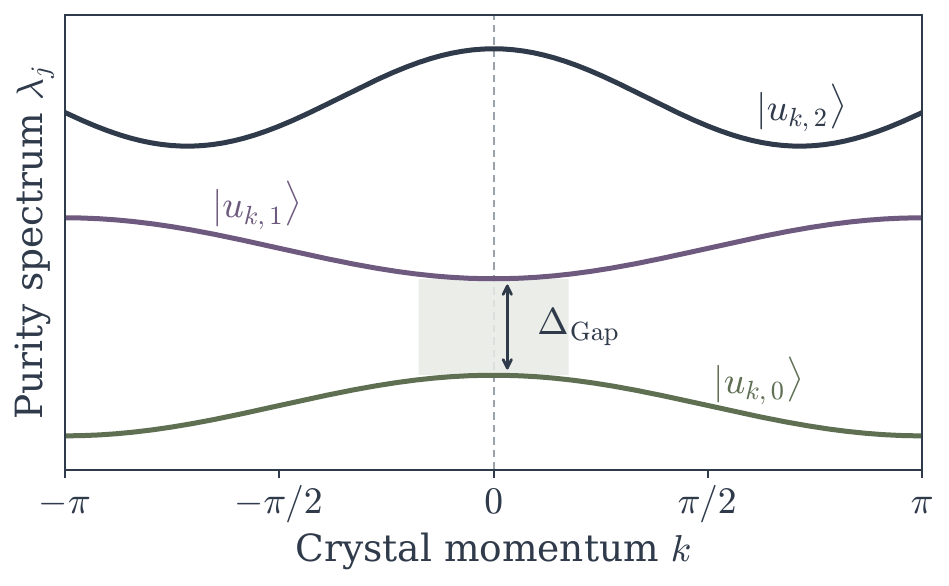}
    \caption{Schematic purity spectrum $\lambda_j$ of a Gaussian mixed state as a function of crystal momentum $k$, exhibiting purity bands $\ket{u_{k,j}}$ and a finite purity gap $\Delta_\mathrm{Gap}$ between the lowest bands, required for the well-definedness of topological invariants like the EGP $\phi_\mathrm{EGP}$.}
    \label{img:Puritygapsketch}
\end{figure}

To address this, we adopt a density-operator ($\varrho$)-based approach to classify the mixed (non-equilibrium) steady-state topology under specific symmetries of the system, following Refs.~\cite{Bardyn_2013,Bardyn_2018}.
In these works, the authors show that the problem of classifying mixed-state topology simplifies when we restrict ourselves to fermionic systems with quadratic Hamiltonians $H_{\mathcal S}$ and linear bath-coupling operators $S_\ell$. Then, in the long-time limit $t \rightarrow \infty$, the density matrix $\varrho(t)$ of the system takes a Gaussian quadratic form
\be
\varrho(t) \sim \exp{ \left( -\tfrac{i}{4}\,\mathbf{w}^T \mathbb G(t)\, \mathbf{w} \right)} ,
\label{eq:Gauss-gen}
\ee
where $\mathbf{w}=(w_1,w_2,\dots)^T$ is a vector of Majorana operators (explicitly defined later), and $\mathbb G$ is a real antisymmetric matrix such that $i \mathbb G$ is Hermitian \cite{Bardyn_2013}. It is convenient to characterise the Gaussian state in the single-particle basis using the covariance matrix,
\be
\begin{split} 
    C_{jk}(t)&=\frac{i}{2}\mathrm{Tr}\left([w_j,w_k]\varrho(t)\right) \\
    &= i \Bigl(\mathrm{Tr}(w_jw_k\varrho(t))-\delta_{jk} \Bigr), %
\end{split}
    \label{eq:covariance}
\ee
where  $C$ fully captures the correlations of the Majorana operators $w_n,w_m$. 
In this representation, since the covariance matrix is real and antisymmetric ($C^T=-C$), the mixed-state topology can be fully studied using an effective `single-particle' Hamiltonian, the so-called fictitious Hamiltonian $H_{C}$ \cite{Bardyn_2013}, which takes the form
\be
H_{C} = i \sum_{mn} C_{mn} w_m w_n .
\label{eq:H_fict}
\ee
In Ref.~\cite{Bardyn_2013}, the spectrum of the positive semidefinite matrix $[iC]^2$ is defined as the \emph{purity spectrum} and coincides (up to a square root) with the positive part of the spectrum of $H_{C}$. In the following, however, we will refer to the eigenvalues $\lambda_j$ of $H_{C}$ as the purity spectrum. The purity eigenvalues lie in the interval $\lambda_j \in [-1,+1]$, where states with $\lambda_j = \pm 1,\ \forall j$ correspond to pure and states with $\lambda_j = 0,\ \forall j$ to fully mixed Gaussian states. This purity spectrum is the (Gaussian) open-system analogue of the isolated eigenenergy (or quasienergy) spectrum and can show a band structure with purity (Bloch) bands $\ket{u_{k,0}},\ket{u_{k,1}}, \dots$, as illustrated in Fig.~\ref{img:Puritygapsketch}. Just as in the isolated case, where the energy spectrum can have a spectral gap (and important electronic and topological properties depend on this gap and its closings), the purity spectrum can also have a finite so-called \emph{purity gap} $\Delta_\mathrm{Gap}$, which can, from a homotopy viewpoint, be used to topologically classify the fictitious Hamiltonian $H_{C}$ (or, analogously, the covariance matrix $iC$) \cite{Bardyn_2013}. In this context, whenever we have a finite purity gap $\Delta_\mathrm{Gap}$ in the purity spectrum (see Fig.~\ref{img:Puritygapsketch}) and detect a gap-closing point, we expect the system to undergo a topological phase transition \cite{Bardyn_2013,Bardyn_2018}. Note that the existence of a finite purity gap is thus crucial for the definition (more specifically, the well-definedness) of invariants.

While the purity spectrum encodes information about the structure of the mixed state and may indicate possible topological regimes, to reliably detect topological features in ensembles described by mixed Gaussian states, one needs a geometric quantity that generalises the Berry or Zak phase to density matrices. This motivates the introduction of the EGP.

\subsection{Ensemble Geometric Phase}

The EGP $\phi_{\mathrm{EGP}}$, introduced by Bardyn \emph{et al.}~\cite{Bardyn_2018}, is well-defined whenever the purity gap remains finite and for a general density matrix (not necessarily a Gaussian state) is defined as
\begin{equation}\label{eq:phiE}
\phi_{\mathrm{EGP}} = \Im\log\text{Tr}\bigl[\varrho\,T_\mathrm{tw}\bigr],
\end{equation}
where $T_\mathrm{tw}$ is again the twist operator as defined in Eq.~(\ref{eq:T_def}). Note the similarities between the EGP $\phi_{\mathrm{EGP}}$ and the Zak phase $\phi_\mathrm{Z}$, as defined in Eq.~(\ref{eq:zakphaseresta}). Instead of computing the expectation value of $T_\mathrm{tw}$ in the ground state, here we take the ensemble average of $T_\mathrm{tw}$ with respect to the density operator $\varrho$. Although the EGP is defined for general density matrices, it is generally hard to compute. 
For Gaussian systems, however, the trace simplifies drastically into terms involving Pfaffians of the covariance matrix, and one can make use of Grassmann calculus \cite{Bravyi_2005}. Further, although a purity gap is not strictly required to define $\phi_{\mathrm{EGP}}$, the existence of a finite purity gap ensures that $\phi_{\mathrm{EGP}}$ is well-defined and can be quantised \cite{Bardyn_2018}. The EGP is constructed such that in the zero-temperature limit $\beta \rightarrow \infty$, $\varrho$ projects onto the ground state and $\phi_{\mathrm{EGP}}$ reduces exactly to the Zak phase $\phi_\mathrm{Z}$. For driven-dissipative systems, the EGP $\phi_{\mathrm{EGP}}$ is typically evaluated for the non-equilibrium steady state (NESS) $\varrho_{\mathrm{NESS}}$. 
If the system is described by a quantum master equation,  this state can in principle be obtained by exact diagonalisation of the Liouvillian, although the computation becomes challenging for larger systems. 

\section{Theoretical Framework} \label{sec:theoreticalframework}

Having introduced the key concepts of mixed-state topology and the EGP, we now turn to the formalism of open quantum systems and show how the problem of computing the NESS $\varrho_\text{NESS}$ simplifies in the case of quadratic fermionic Hamiltonians $H_{\mathcal S}(t)$ and linear bath couplings $S_\ell$  by mapping onto Majorana operators. After focusing on the autonomous case, we show how, in the case of periodically driven open quantum systems, the NESS can be obtained in this formalism within the correct Floquet--Born--Markov approach \cite{Kohler_1997,wustmann2010statistical}.

\subsection{\texorpdfstring{Open Quantum Systems and Gaussian Quadratic Steady States}{Open Quantum Systems and Gaussian Quadratic Steady States}} \label{sec:open_qs}

We consider an open quantum system, which we decompose according to Eq.~(\ref{eq:H_sysbath}). Under sufficiently weak coupling between system and environment, the time evolution of the reduced density operator of the system $\varrho_\mathcal{S}(t)$ is governed by the Born--Markov master equation \cite{breuer2002theory,Weiss2012}
\begin{equation}
\partial_t \varrho_{\mathcal{S}}(t) = \hat{\mathcal{L}}(t)\bigl[\varrho_{\mathcal{S}}(t)\bigr] = \hat{\mathcal{C}}(t)\bigl[\varrho_{\mathcal{S}}(t)\bigr] + \hat{\mathcal{D}}(t)\bigl[\varrho_{\mathcal{S}}(t)\bigr] , \label{eq:RME}
\end{equation}
where 
$
\hat{\mathcal{C}}(t)\bigl[\cdot\bigr]  =  -i[H_{\mathcal S}(t), \cdot] %
$
describes the coherent evolution of the system and the Redfield dissipator $\hat{\mathcal{D}}(t)\bigl[\cdot\bigr]$ takes the form \cite{Kohler_1997}
\begin{equation}
\begin{split} 
\hat{\mathcal{D}}(t)\bigl[\cdot\bigr] ={}& \sum_{\ell, \ell'} \int_{0}^{\infty} \mathrm{d}\tau \, \Gamma_{\ell'\ell}^{I}(t,\tau) \bigl[\tilde{S}_{\ell}(t-\tau,t)\,\cdot\,, S_{\ell'}\bigr] \\
&+ \text{h.c.} ,
\end{split}
\end{equation}
where
$
\Gamma^I_{\ell'\ell}(t,\tau) \equiv \mathrm{Tr}_{\mathcal{R}_I} [\varrho_{\mathcal{R}_I} \tilde{R}_{\ell'}(t)\tilde{R}_{\ell}(\tau) ]
$
are the bath correlation functions. The tilde denotes operators in the interaction picture, i.e.~$\tilde A(t_1,t_0) = U_0(t_1,t_0)^\dagger A U_0(t_1,t_0)$ with $ U_0(t_1,t_0) = \mathcal{T}\exp[-i\int_{t_0}^{t_1} \mathrm{d}\tau (H_{\mathcal S}(\tau)  +  H_{\mathcal{R}})]$. Here, the index $I$ in the spectral functions $\Gamma_{\ell'\ell}^{I}(t,\tau)$ indicates that different sites can couple to different reservoirs $I$ at inverse temperatures $\beta_I$ and chemical potentials $\mu_I$.
Note that Eq.~\eqref{eq:RME} is valid both in the case of an autonomous- as well as in the case of a time-periodically driven system.
For quadratic Hamiltonians $H_{\mathcal S}(t)$ and linear bath-coupling operators $S_\ell$, we can perform a transformation from $2L$ fermionic creation/annihilation operators $\{f_j^\dagger,f_j\}$ onto $4L$ Hermitian Majorana operators $\{w_\alpha\}$, with the mapping given by 
\begin{align}
    w_{2j - 1} &= f_j + f_j^\dagger, \hspace{0.8cm} 
    w_{2j} = i (f_j - f_j^\dagger) .
\end{align}
 Note that the new operator basis fulfils the anticommutation relations $\{w_j,w_k\}=2\delta_{jk}$ (with which they generate a Clifford algebra) and that Majorana particles are their own antiparticles, i.e.\ $w_j^\dagger=w_j$. 
The Majorana-basis Hamiltonian $\mathbb H_\mathrm{Majo}(t)$ and coupling operators $s_\ell$ can be obtained from the mapping
\begin{align}
H_{\mathcal S}(t)
&= \sum_{j,k=1}^{4L} w_j [\mathbb H_\mathrm{Majo}]_{jk}(t) w_k
 = \mathbf{w}^T \mathbb H_\mathrm{Majo}(t)\, \mathbf{w} \\
&\equiv \brakett{w}{\mathbb H_\mathrm{Majo}(t)\, w} ,
\label{eq:H_traf} \\[1ex]
S_\ell(t)
&= \sum_{j=1}^{4L} s_{\ell,j}(t) w_j
 = \mathbf{s}_\ell^T(t)\mathbf{w}
 \equiv  \brakett{s_\ell(t)}{w} ,
\label{eq:X_traf}
\end{align}
where $\kett{s_\ell(t)}$ is a vector of scalar-valued or operator-valued symbols.
For systems where such a mapping is possible, the steady-state solution is given by a general Gaussian quadratic state, Eq.~\eqref{eq:Gauss-gen}, and, thus, all relevant information about the state is fully contained in the covariance matrix $C(t)$, Eq.~\eqref{eq:covariance}.
One can show \cite{Zunkovic_2010,Prosen_2011,Prosen_2010,Prosen_2010_spectral} that in the Majorana representation, Eq.~(\ref{eq:RME}) can be mapped onto a differential equation for the matrix ${\tilde{C}}(t)\equiv -i C(t)$, given by 
\begin{equation}
\begin{aligned}
\frac{\mathrm{d}}{\mathrm{d}t} \tilde{C}(t)
&= -4i \bigl[ \mathbb H_\mathrm{Majo}(t), \tilde{C}(t) \bigr] \\ 
&\quad + 4i\bigl( \mathbb M_i(t) - \mathbb M_i^{T}(t) \bigr)  \\
&\quad -4\bigl( \mathbb M_r(t)\tilde{C}(t)  + \tilde{C}(t) \mathbb M_r^{T}(t) \bigr). \label{eq:ddtC}
\end{aligned}
\end{equation}
We observe that the coherent part of $\tilde{C}(t)$ is, up to a factor of four, governed by a von Neumann-type time evolution.
Here ${\mathbb M_r}=(\mathbb M+\mathbb M^*)/2$ denotes the real and ${\mathbb M_i}=(\mathbb M-\mathbb M^*)/(2i)$ the imaginary part of the bath matrix $\mathbb M(t)$, defined as
\begin{equation}
\mathbb M(t) = \sum_{\ell'} \kett{s_{\ell'}} \braa{z_{\ell'}(t)}. \label{eq:bathma}
\end{equation}
Herein,  $\braa{z_{\ell'}(t)}$ encodes the bath correlations via 
\begin{equation}
\begin{aligned}
\braa{z_{\ell'}(t)}
&= \sum_\ell \int_0^\infty \dd \tau \,
   \Gamma_{\ell'\ell}^I(\tau)\,
   \braa{f_\ell(t-\tau,t)} ,
\end{aligned}
\label{eq:zdef}
\end{equation}
where $\braa{f_\ell(t-\tau,t)}$ is a propagator of the Heisenberg dynamics in the closed system, i.e.~$\tilde{S}_{\ell}(t-\tau,t)=\brakett{f_\ell(t-\tau,t)}{w}$. In the case of a quadratic system Hamiltonian $H_{\mathcal S}(t)$, it takes the form 
\begin{align}
\braa{f_\ell(t-\tau,t)}
&= \braa{s_\ell(t)} \mathcal{T}
\exp \Bigl[-4i \int_{t}^{t-\tau} \mathbb H_\mathrm{Majo}(t')\,\mathrm{d}t' \Bigr]
\nonumber\\
&\equiv \braa{s_\ell(t)} \ \mathbb U_\mathrm{Majo}(t-\tau,t), \label{eq:MajFloOp}
\end{align}
where we have defined the Majorana time-evolution operator $\mathbb U_\mathrm{Majo}(t-\tau,t)$ (cf.~Eq.~(\ref{eq:Floop})).
In this decomposition, the prefactor of \emph{four} arises from the normalisation convention for Majorana operators and the structure due to the canonical anticommutation relations, which together determine the energy scaling.
Note that we can map Eq.~(\ref{eq:ddtC}) onto the form of a differential Lyapunov equation
\be
\frac{\mathrm{d}}{\mathrm{d}t} \tilde{C}(t) = - \mathbb X(t) \tilde{C}(t) - \tilde{C}(t) \mathbb X^T(t) + i \mathbb Y(t) , \label{eq:SylvFloqapp}
\ee
by defining the operators
\be
\begin{aligned}
\mathbb X(t)
&= 4\bigl( i  \mathbb H_\mathrm{Majo}(t) + \mathbb M_r(t) \bigr) , \\
\mathbb Y(t)
&= 4\bigl( \mathbb M_i(t) - \mathbb M_i^{T}(t) \bigr) . \label{eq:X_Y_def}
\end{aligned} 
\ee
This form allows for a direct numerical solution using standard Sylvester equation solvers.
We first restrict ourselves to the autonomous case where $H_{\mathcal S}(t)=H_{\mathcal S}$ and $S_\ell(t)=S_\ell$ are time-independent. In that case, we have $\braa{f_\ell(t)}=\braa{s_\ell} \exp [-4i\mathbb H_\mathrm{Majo}t]$ and can spectrally decompose $\braa{z_{\ell'}}$ as \cite{Prosen_2010}
\begin{equation}
\begin{aligned}
\braa{z_{\ell'}}
= \pi \sum_\ell \sum_{m}& \braa{s_\ell} \!\left[
   \tilde{\Gamma}_{{\ell'}\ell}^I(+4E_m) \!\,
   \kett{u_m^*}\,\!\braa{u_m} \right. \! \\
&\quad \left. {}+ \tilde{\Gamma}_{{\ell'}\ell}^I(-4E_m) \!\, 
   \kett{u_m}\,\!\braa{u_m^*} \right], \label{eq:z_spec_stat}
\end{aligned}
\end{equation}
where $\kett{u_m}$ and $\kett{u_m^*}$ are the eigenvectors of the static Hamiltonian $\mathbb H_\mathrm{Majo}$ in Majorana representation, with corresponding eigenvalues $E_m,-E_m$ (the eigenvalues always come in $\pm E_m$ pairs and the corresponding eigenvectors can and should be chosen orthogonal \cite{Prosen_2010}). We have also used the Fourier transform 
\be
\tilde\Gamma_{\ell'\ell}^{I}(\Omega)
= \frac{1}{2\pi} \int_{-\infty}^\infty \mathrm{d}\tau\, 
\Gamma_{\ell'\ell}^{I}(\tau)\,e^{-i\Omega\tau} ,
\ee
of the bath correlation functions $\Gamma^I_{\ell'\ell}$, instead of the half-sided Fourier transform  that is typically used for the Redfield master equation \cite{breuer2002theory}. This neglects the Cauchy principal-value term in the bath spectral functions, which physically amounts to neglecting the Lamb-shift Hamiltonian. Such an approximation is generally justified in the weak-coupling limit, especially when one is interested in the global topological properties of the system (since the Lamb-shift Hamiltonian only introduces a small shift of the system energies \cite{breuer2002theory}). In Ref.~\cite{Prosen_2010}, it is also argued that this exactly gives us the Gibbs state as the steady state of the (equilibrium) thermal Redfield model.

In the static case, this form can be implemented numerically to solve for the NESS ${\tilde{C}}_\mathrm{NESS}$, which is the steady-state solution of Eq.~(\ref{eq:ddtC}) or Eq.~(\ref{eq:SylvFloqapp}) (continuous Lyapunov equation \cite{Prosen_2011}). All physically relevant higher-order observables can then be computed by utilising Wick's theorem.

\subsection{\texorpdfstring{Floquet--Born--Markov Description in Majorana Basis}{Floquet--Born--Markov Description in Majorana Basis}}

In this work, however, we are interested in investigating time-periodically driven systems. Consider an open quantum system with time-periodic system Hamiltonian $H_{\mathcal S}(t+T)=H_{\mathcal S}(t)$ and time-independent coupling operators $S_\ell$ to one of the baths. 
Since the additional time dependence is only in the coherent part of the dynamics, namely in the Hamiltonian $H_{\mathcal S}(t)$, and the system--bath coupling operators $S_\ell$ are left unchanged, one might think that we can simply plug the Majorana representation of the new time-dependent Hamiltonian $H_{\mathcal S}(t)$ into Eq.~(\ref{eq:ddtC}), while leaving the bath matrix $\mathbb M$ static and unchanged, and solve for the steady-state solution of Eq.~(\ref{eq:ddtC}). 

However, the time-periodic drive in the Hamiltonian reorganises the energy levels of the system, such that the coupling operators (describing discrete jumps between the system and reservoir at specific allowed energy levels, cf.~Eq.~\eqref{eq:z_spec_stat}) need to be adjusted as well. Because of that, the bath operator $\braa{z_{\ell'}(t)}$ cannot simply be decomposed as in the static case leading to Eq.~(\ref{eq:z_spec_stat}), but the system needs to be analysed within the proper Floquet--Born--Markov approach \cite{Kohler_1997,wustmann2010statistical}. Here, since for Floquet systems we are dealing with a quasienergy spectrum, we have, in theory, infinitely many (in practice, only a few dominant)  Floquet replicas (or FBZs) contributing to possible quantum jumps, as illustrated in Fig.~\ref{img:quasienegspectrum_theory_branches}.
There are limiting cases (such as weak driving or high driving frequency) in which neglecting the effect of the drive on the dissipator is a qualitatively valid approximation \cite{kolisnyk_floquet_2024}, and in this context topological phase transitions have been studied for driven-dissipative systems \cite{molignini_2019,Prosen_2011}. However, in general, the effect is non-negligible, which is why, in the following, we derive a new differential equation for $C(t)$ by means of the Floquet--Born--Markov approach to take into account the impact of the drive. 

In order to do that, we follow a similar approach as in Refs.~\cite{Kohler_1997,wustmann2010statistical}, but rather than the  Redfield Eq.~(\ref{eq:RME}), we consider directly the relevant Eq.~(\ref{eq:ddtC}),
with $\mathbb M(t)$ defined as in Eq.~(\ref{eq:bathma}), in Majorana representation of the relevant operators. We further do not transform into the Floquet frame to solve for the steady-state solution. Recall that the Floquet modes $\kett{\Phi_{\eta}(t)}$ (see Eq.~(\ref{eq:Floquetstates})), which we obtain from the time-periodic Majorana Hamiltonian $\mathbb H_\mathrm{Majo}(t+T)=\mathbb H_\mathrm{Majo}(t)$ with $T = 2\pi / \omega$, constitute an orthonormal set of basis states at all times $t$. We will henceforth refer to the corresponding Floquet Hamiltonian $\mathbb H_\mathrm{Majo}^{\mathrm{F}}$ as the Floquet-Majorana Hamiltonian.

To investigate the relevant degrees of freedom of the Floquet system, we expand $\braa{z_{\ell'}(t)}$, Eq.~(\ref{eq:zdef}), in terms of the time-periodic Floquet modes $\kett{\Phi_{\eta}(t)}$.
Note that since $\mathbb H_\mathrm{Majo}(t)$ is antisymmetric and purely imaginary, the corresponding Majorana Floquet operator satisfies $\mathbb U_{\mathrm{Majo}}(T,0) \in \operatorname{SO}(4L)$. Consequently, the quasienergies come in pairs $\pm\varepsilon_\eta$, and the associated Floquet modes $\kett{\Phi_{\eta}(t)}$ and $\kett{\Phi_{\eta}^{*}(t)}$ can be chosen orthogonal such that
\begin{equation}
\brakett{\Phi_{\eta}(t)}{\Phi_{\nu}(t)} = 0,
\qquad
\brakett{\Phi_{\eta}(t)}{\Phi_{\nu}^*(t)} = \delta_{\eta\nu}
\end{equation}
for all $\eta,\nu$. We express the Floquet modes in terms of their Fourier components, 
\begin{equation*}
\begin{aligned}
\kett{\Phi_{\eta}(t)} &= \sum_{K \in \mathbb{Z}} e^{-iK \omega t}\kett{\Phi_{\eta}(K)} , \\
\kett{\Phi^{*}_{\eta}(t)} &= \sum_{K \in \mathbb{Z}} e^{+iK \omega t}\kett{\Phi^{*}_{\eta}(K)} ,
\end{aligned}
\end{equation*}
with $\kett{\Phi^{*}_{\eta}(K)}=\kett{\Phi_{\eta}(K)}^{*}$.
Then, introducing the projectors
\be
\begin{aligned}
\Pi_\eta(K,Q) &= \kett{\Phi_\eta(K)} \braa{\Phi_\eta^*(Q)}, \\
\Pi_\eta^*(Q,K) &= \kett{\Phi_\eta^*(Q)} \braa{\Phi_\eta(K)},
\end{aligned}
\ee
leads to the decomposition (see Appendix~\ref{sec:app_FBM} for details) 
\begin{equation} 
\begin{split}
& \braa{z_{\ell'}(t)}
= \pi \sum_\ell^{4L} \sum_\gamma^{2L} \sum_{Q,K \in \mathbb{Z}}
   e^{-i(K-Q) \omega t} \braa{s_\ell}\\ 
   & \times \Bigl[\tilde{\Gamma}_{{\ell'}\ell}^I(-\Delta^{Q}_{\gamma})
   \Pi_\gamma(K,Q) + 
   \tilde{\Gamma}_{{\ell'}\ell}^I(+\Delta^{K}_{\gamma})
   \Pi_\gamma^*(Q,K) \Bigr], 
\end{split}
\label{eq:zdash} 
\end{equation}
where the time dependence is now explicit. Here, we introduced the shorthand notation $\Delta^{N}_{\gamma }=\varepsilon_\gamma + N\omega$.
The bath is sampled at the quasienergies $\pm\varepsilon_\gamma$, shifted by an integer number of drive quanta. The case $-\Delta^{Q}_{\gamma}$ corresponds to quasienergy $-\varepsilon_\gamma$, with
harmonic index $n=-Q$, and the case $+\Delta^{K}_{\gamma}$ to quasienergy
$+\varepsilon_\gamma$, with $n=+K$. For $\varepsilon_\gamma$ in the first FBZ, the integer $n$ counts the drive quanta exchanged in the corresponding transition.
Note that, in contrast to the time-autonomous case,
cf.~Eq.~(\ref{eq:z_spec_stat}), the factor of four in front of the
quasienergies $\varepsilon_\gamma$ is already incorporated in the definition of $\mathbb U_{\mathrm{Majo}}(T,0)$, Eq.~(\ref{eq:MajFloOp}). 

For a general time-dependent $\mathbb{H}_\mathrm{Majo}(t)$ and $\mathbb{M}(t)$, the solution of Eq.~(\ref{eq:SylvFloqapp}) for the covariance matrix can be obtained using the ansatz~\cite{Prosen_2011}
\begin{equation}
\tilde{C}(t)
=
\mathbb{Q}(t)\tilde{C}(0)\mathbb{Q}^{T}(t)
-i\mathbb{P}(t)\mathbb{Q}^{T}(t),  \label{eq:gen_sol}
\end{equation}
with explicit solutions 
\begin{equation}
\begin{aligned}
\mathbb{Q}(t)
&=
\mathcal{T}
\exp\!\left(
-\int_{0}^{t}\mathrm{d}t'\,\mathbb{X}(t')
\right), \\
\mathbb{P}(t)
&=
-\mathbb{Q}(t)
\int_{0}^{t}
\mathrm{d}t'\,
\mathbb{Q}^{-1}(t')
\mathbb{Y}(t')
\mathbb{Q}^{-T}(t'),
\end{aligned} \label{eq:QandT}
\end{equation}
where $\mathbb{X}(t),\mathbb{Y}(t)$ are defined in Eq.~(\ref{eq:X_Y_def}) and $\mathbb{Q}^{-T}=(\mathbb{Q}^{T})^{-1}=(\mathbb{Q}^{-1})^{T}$.
The covariance matrix $\tilde{C}_{\mathrm{NESS}}$ of the steady-state solution is obtained from the discrete Lyapunov equation \cite{Prosen_2011}
\begin{equation}
\mathbb{Q}(T)\tilde{C}_{\mathrm{NESS}}
-
\tilde{C}_{\mathrm{NESS}}\mathbb{Q}^{-T}(T)
=
i\mathbb{P}(T),
\end{equation}
which follows from the ansatz after imposing the periodicity condition
$
\tilde{C}_{\mathrm{NESS}}(0)=\tilde{C}_{\mathrm{NESS}}(T)\equiv \tilde{C}_{\mathrm{NESS}}.
$
The steady-state covariance matrix $\tilde{C}_{\mathrm{NESS}}(t)$ at intermediate times $0\le t\le T$ can then be computed by propagating $\tilde{C}_{\mathrm{NESS}}$ according to Eq.~\eqref{eq:gen_sol}.

With that, we have all the necessary tools to calculate the NESS covariance matrix $C_{\text{NESS}}(t)$ within the correct Floquet--Born--Markov approach and can use it to compute observables like the EGP $\phi_{\mathrm{EGP}}(t)$ by utilising Wick's theorem.
 
\section{Results} \label{sec:results}

To address the challenge of topologically classifying driven-dissipative steady states, we consider the periodically driven, dissipative SSH system introduced in Sec.~\ref{sec:drivdissmodel}. Within the Floquet--Born--Markov approximation, for our model, the system--bath correlation functions are  conveniently expressed in frequency space, in which they take the block-diagonal form  \cite{Molignini_2023}
\begin{equation}\label{eq:GammaDiag}
\tilde\Gamma(\Omega)
= \mathrm{diag}\bigl(\tilde\Gamma_A(\Omega), \tilde\Gamma_B(\Omega),
\tilde\Gamma_A(\Omega), \tilde\Gamma_B(\Omega), \dots\bigr),
\end{equation}
with two-by-two matrices
\begin{equation}\label{eq:GammaI}
\tilde\Gamma_I(\Omega)
= \kappa^2 g_I(\Omega)\bigl[f_+(\Omega)
\mathds{1} - f_-(\Omega)\sigma_y\bigr],
\end{equation} 
indexing over the quadrature index $\alpha$.
Here, we use
$
f_{\pm}(\Omega) \equiv n_{I}(\Omega) \pm \big(1 - n_{I}(-\Omega)\big), 
$
and the Fermi--Dirac distribution of reservoir $\mathcal{R}_I$ is $n_{I}(\Omega) = (e^{\beta_{I}(\Omega-\mu_{I})}+1)^{-1}$, with $g_I(\Omega)=\mathrm{const.}$ denoting its spectral density. We assume a uniform coupling between the bath of free fermions and all unit cells.
Using the mappings in Eqs.~(\ref{eq:H_traf}) and (\ref{eq:X_traf}), we calculate the relevant operators $\mathbb H_{\mathrm{Majo}} (t)$ and $s_{\ell,j}$ in Majorana representation for our system--bath configuration and present them in Appendix~\ref{sec:appb}. Note that $\mathbb H_{\mathrm{Majo}} (t)$ is periodic in time, and, 
importantly, the configuration and choice of drive preserve the chiral symmetry of the SSH model. 
To illustrate our results, in Secs.~\ref{sec:quasirapidity} and \ref{sec:res_sym} we start by considering a driven-dissipative SSH chain of length
$L=35$, with OBC, intracell and intercell hopping amplitudes of $J=1$ and  $J'=1.4J$, respectively, a driving strength of $V=0.8J$, equal inverse bath temperatures $ \beta_A=\beta_B=10/J$ and vanishing chemical potentials $\mu_A=\mu_B=0$ (corresponding to half-filled fermionic reservoirs). To satisfy the Born--Markov assumption, the dissipation is chosen to be of strength $\kappa=0.1J$ (entering the Born--Markov rates through $\kappa^2$) and spectral densities of $g_A=g_B\equiv g=1$, such that we are in the weak-coupling regime.  
In theory, a countably infinite number of FBZ branches may contribute to the dynamics. Thus, we have to carefully choose a cutoff of the branches (i.e.~Floquet Brillouin zones) $K \in \{0, \pm 1, \pm 2, \dots\}$, when constructing the bath operators $\braa{z_{\ell'}(t)}$ in Eq.~(\ref{eq:zdash}).
To calculate the Floquet--Born--Markov rates, in the following we cut off the Floquet spectrum  at $|K_\mathrm{max}|=5$. 

The NESS is Gaussian and can therefore be fully described by the covariance matrix $C_{\text{NESS}}(t)$. Physical insights into the steady state can be gained by analysing the correlations $W_{jk} \equiv \langle w_j w_k \rangle  $ between Majorana quasi-particles on different sites $w_j, w_k$. In Appendix~\ref{sec:app_correlations} we show $W$ for different driving regimes and times within a micromotion period $T$ and show that the NESS remains structured and correlated for the chosen parameters.

\subsection{\texorpdfstring{Quasirapidity Spectrum and the Limits of Generator-Based Diagnostics}{Quasirapidity Spectrum and the Limits of Generator-Based Diagnostics}} \label{sec:quasirapidity}

To gain physical information about the system dynamics and the stability of modes, it is insightful to inspect the spectrum of the full Liouvillian $\hat{\mathcal{L}}$, before we turn to the EGP to analyse the topology of our time-periodic steady state. Similar to the isolated case (cf. Eq.~(\ref{eq:Floop})), one can introduce a dynamical map (superoperator)
\be
\hat{\mathcal U}(t,0)=\mathcal{T} \exp \biggl(\int_0^t \mathrm{d} t_1  \hat{\mathcal{L}}(t_1) \biggr), 
\ee
which, in the dissipative case, is non-unitary. Similarly to the unitary case, the spectrum of the Floquet operator $\hat{\mathcal U}(T,0)$ gives rise to a `quasi-eigenvalue spectrum'. In Ref.~\cite{Prosen_2011} it is shown that the full Liouvillian spectrum of $\hat{\mathcal U}(T,0)$ is fully determined by the spectrum of the propagator $\mathbb Q(T)$, Eq.~(\ref{eq:QandT}). The eigenvalues of $\mathbb Q(T)$ are Floquet multipliers $\exp(-{\beta_jT})$, which, unlike in the unitary case where they live on the unit circle, live on the unit disc. The $\beta_j$ are the complex `rapidities' \cite{Prosen_2010}. Here, $\Im (\beta_j )$ dictates the coherent and $\Re (\beta_j )$ the dissipative dynamics (decay rates) of the system. As in the static case \cite{Prosen_2010}, the NESS solution $C_{\mathrm{NESS}}(t)$ is unique if the whole rapidity spectrum is strictly away from the imaginary axis, i.e.\ $\Re(\beta_j)>0,\forall j$. The imaginary part $\Im (\beta_j )$ is defined modulo integer multiples of $\omega$, since the mapping $\Im (\beta_j ) \rightarrow \Im (\beta_j )+N\omega$ leaves the multipliers $\exp(-{\beta_jT})$ invariant, resembling a `quasirapidity' structure.
 
\begin{figure}
    \centering
    \includegraphics[width=0.47\textwidth]{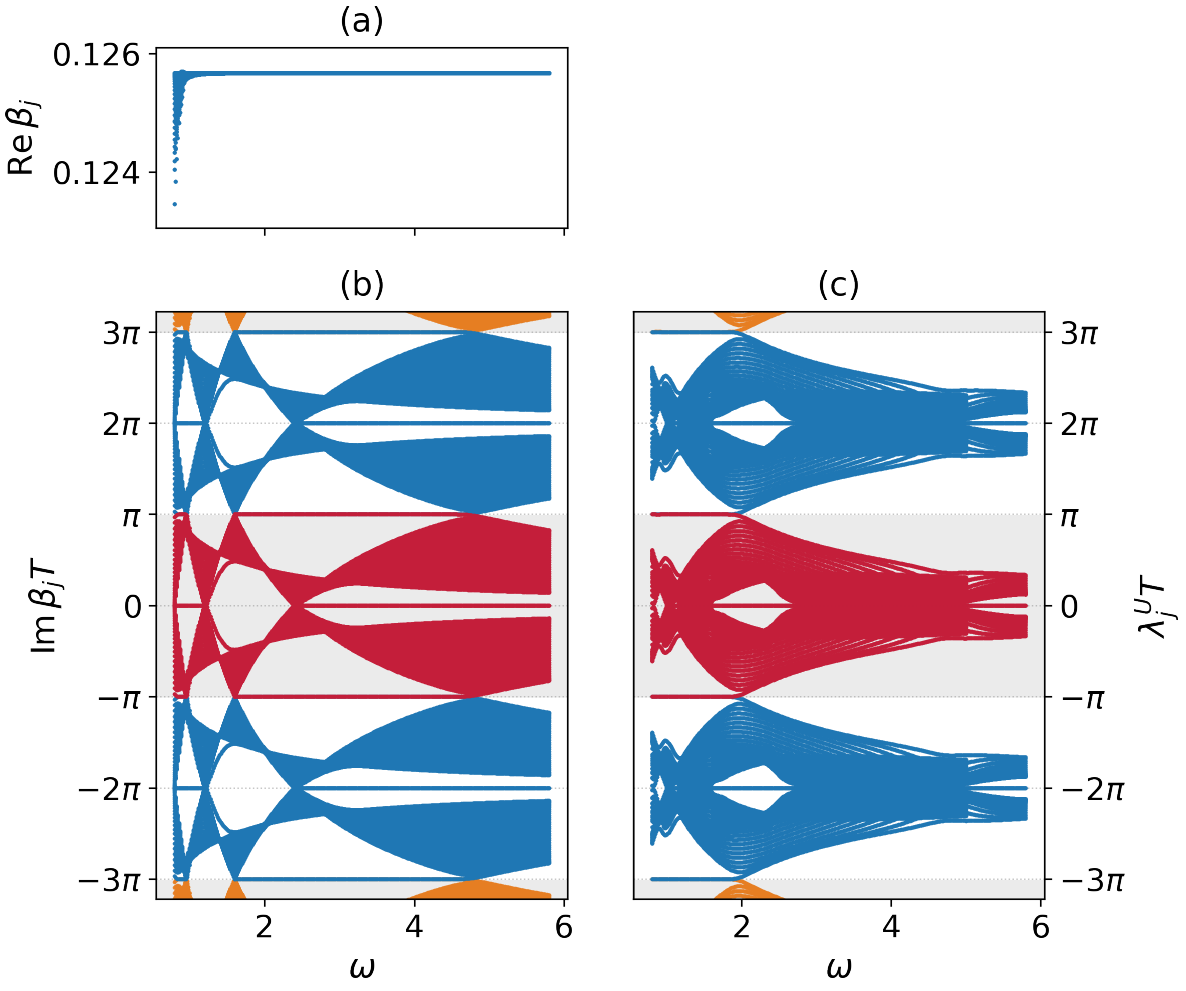}
    \caption{(a),(b) Liouvillian rapidity spectrum $\beta_j$ and (c)  quasipurity spectrum $\lambda_j^UT$ as a function of the driving frequency $\omega$ for a chain length of $L=35$, OBC, hopping amplitudes $J=1, J'=1.4J$, a driving strength of $V=0.8J$, inverse bath temperatures $ \beta_A=\beta_B=10/J$, chemical potentials $\mu_A=\mu_B=0$, dissipation strength $\kappa=0.1J$ and spectral densities of $g_A=g_B=1$. To calculate the Floquet--Born--Markov rates, we cut off the Floquet spectrum  at $|K_\mathrm{max}|=5$.~(a) shows the uniqueness of $C_{\mathrm{NESS}}(t)$, as $\Re(\beta_j)>0,\forall j$. (b) The imaginary part coincides with the quasienergy spectrum, since Lamb-shift corrections are neglected.~(c) While the quasipurity spectrum shows separated states at $0$ and $\pm \pi$, its physical meaning is questionable.}
    \label{img:test1}
\end{figure}

In Fig.~\ref{img:test1} we depict  (a) their real part $\Re(\beta_j)$ and (b) their imaginary part $\Im(\beta_j)$ as functions of $\omega$.
The real part of the rapidity spectrum, Fig.~\ref{img:test1}(a), confirms that the NESS solution $C_{\mathrm{NESS}}(t)$ is unique, as  $\Re (\beta_j) > 0, \forall j$. Note that the mandatory zero eigenvalue of any generator of the dynamical map $\hat{\mathcal{U}}(T,0)$ does not appear as a vanishing rapidity, since the $\beta_j$ determine the spectrum of $\hat{\mathcal{U}}(T,0)$ without coinciding with it~\cite{Prosen_2011}, and $\Re(\beta_j)>0\ \forall j$ is precisely the condition for a unique NESS. Further, for $\omega \gtrsim 0.9$ we observe that, independent of $\omega$, all modes decay with the same rates. In Appendix~\ref{sec:app_independence}, we derive an expression for $\mathbb X(t)$ for the specific system investigated in this work and find the decay rates $\Re (\beta_j)=4\pi\kappa^2 g \approx 0.1257$ that are constant and equal for all modes $j$. This is verified in Fig.~\ref{img:test1}(a), except for low frequencies, where the small deviations arise from Floquet truncation errors.
The imaginary part, Fig.~\ref{img:test1}(b), shows the structure of a quasienergy band diagram (similar to the one of the isolated system, Fig.~\ref{img:quasienegspectrum_theory_branches}), which is expected since in the limit of vanishing coupling, the spectrum of the (Majorana) Floquet Lindbladian \cite{Schnell_2020,Schnell_2021,Dinc_2025} is fully determined by the (Majorana) Floquet Hamiltonian $\mathbb H_{\mathrm{Majo}}^\mathrm{F}$. 
At weak coupling, here, this also holds because we neglect Lamb-shift terms~\cite{breuer2002theory,wustmann2010statistical}. We observe protected states both in the $0$ and the $\pi$ gaps, with the latter arising solely from the periodic drive. Note that for high frequencies $\omega \gtrsim 4.8$, the $\pi$ gap stays open and is trivial, which is expected since in the high-frequency regime, the system approaches a static time-averaged Floquet Liouvillian $\mathcal{L}_{\text{Majo}}^\mathrm{F}$ \cite{Schnell_2020,Schnell_2021}. 
Given the `quasiband' structure generated by $\mathbb X(t)$, the operator $\mathbb X(t)$ would be a natural starting point for a topological classification of the driven-dissipative model. This route has been followed by van Caspel \emph{et al.}~\cite{Caspel_2019} for a time-periodically quenched, dissipative Kitaev chain within a phenomenological quadratic Lindblad setting, neglecting the impact of the drive on the dissipator. In the following, we discuss the limitations of a generator-based approach when one is concerned with the topology surviving in the steady state.

In Ref.~\cite{Lieu2020}, it was shown that in the case of static quadratic Lindbladians $\mathcal{L}$, topological features of the spectrum need not be visible in the steady state of the system. That is, they prove that it is always possible to interpolate between the physical steady state of the system and a fully mixed (i.e.\ trivial) one, without changing topological features in the spectrum used to topologically classify the system. We do not know whether an equivalent interpolation exists for our time-dependent Floquet--Born--Markov master equation, and do not extend the result of Ref.~\cite{Lieu2020} to this setting. Nonetheless, this establishes that generator topology and steady-state topology are logically distinct objects in general, which we will independently confirm in Secs.~\ref{sec:res_sym} and \ref{sec:windingnotop} in our case.

This does not mean, however, that they never coincide in practice. Van Caspel \emph{et al.}~\cite{Caspel_2019} find a $\mathbb{Z}\times\mathbb{Z}$ classification by winding numbers $(\nu_0,\nu_\pi)$, closely analogous to the chiral-symmetric construction in the isolated case (cf. Sec.~\ref{sec:isol_driv_SSH}), and report that signatures of their generator-based winding numbers appear in the steady-state current under weak dissipation. Their construction (based on Liouvillian eigenmodes) is insensitive to their operator $\mathbb{Y}$ and fully determined by $\mathbb{X}$, while $\mathbb{Y}$ determines the steady-state covariance matrix $C_{\text{NESS}}(t)$. This makes their approach an indirect diagnostic when one is primarily concerned with the topology of the steady state rather than that of the generator, since any link to the steady state must be established separately.

For our chosen system--bath correlation function $\tilde\Gamma(\Omega)$, this creates an even larger structural problem. In Appendix~\ref{sec:app_independence}, leveraging the fact that for our system $g_I(\Omega)$ is even, i.e.\ $g_I(\Omega)=g_I(-\Omega)=g$, and that Lamb shifts are neglected, we derive the expression
\be
\mathbb X(t)
= 4\Bigl[ i  \mathbb H_\mathrm{Majo}(t) + \pi  \kappa^2 g \sum_{\ell} \kett{s_{\ell}}\braa{s_\ell} \Bigr] ,
\ee
such that $\mathbb X(t)$ is `blind' to the bath parameters $\beta_{A,B}$ and $\mu_{A,B}$.
While this need not hold for arbitrary system--bath correlation functions, in our case this means that any topological classification based on $\mathbb X(t)$ would be fully independent of $\beta_{A,B}$ and $\mu_{A,B}$. As we will see in the following, however, the NESS and its symmetries, as well as the EGP and its quantisation, depend fundamentally on both $\beta_{A,B}$ and $\mu_{A,B}$.
At the expense of the richer $(\nu_0,\nu_\pi)$-type classification, we therefore do not adopt a generator-level construction as the primary approach. In exchange, we take the time-periodic steady state $C_{\text{NESS}}(t)$ itself as the starting point for our topological analysis, in order to characterise the topology of the object that is physically present in the long-time limit.

On the level of the steady state, motivated by isolated Floquet topology, another natural approach would be to define and analyse a `quasipurity' spectrum, potentially giving rise to two independent topological invariants of the steady state, which we discuss in the following.
Note that the time-periodicity of the steady-state covariance matrix $C_{\text{NESS}}(t)$
is inherited by the fictitious Hamiltonian $H_C(t)$, Eq.~(\ref{eq:H_fict}). Since $H_C(t)$ is Hermitian, analogously to the isolated periodically driven case, we formally define a unitary `fictitious Floquet operator'
\be
U_C(T)=\mathcal{T} \exp \biggl(-i\int_0^T \mathrm{d} t \ H_C(t) \biggr), \label{eq:fic_Floop}
\ee
whose eigenvalues $\exp({-i\lambda_j^UT})$ are on the unit circle, leading to two special points $\lambda_j^U=0$ and $\lambda_j^U=\pi/T$ and thus to a `quasipurity' spectrum $\lambda_j^U$, which we plot in Fig.~\ref{img:test1}(c) for our system. 
As we can see, we obtain regimes with finite spectral gaps and separated states in the $0$- and $\pi$-`quasipurity gaps'.
However, firstly, unlike the system Hamiltonian $H_\mathrm{PDSSH}(t)$, $H_C(t)$ does \emph{not} generate any physical dynamics. Secondly, the purity spectrum is naturally bounded to the interval $ \lambda_j \in [-1,+1]$, where states at $\lambda_j \in \{ -1,+1 \}$ correspond to pure Gaussian states. The physical meaning of a quasipurity spectrum bounded to $\lambda_j^U \in[-\pi/T,+\pi/T]$ (or any other branch of the logarithm) is thus questionable. Unlike genuine purity eigenvalues, whose approach to $\pm 1$ signals a mode becoming pure, there is no analogous interpretation for  $\lambda_j^U$ approaching $\pm \pi /T$. Therefore, while this construction of a quasipurity spectrum and a fictitious Floquet unitary is mathematically consistent, it lacks a clear physical justification. The $0$--$\pi$ structure visible in the quasirapidity
spectrum of Fig.~\ref{img:test1}(b) therefore does not descend to the steady-state purity spectrum.

For these reasons, the observable we use to classify the topology of the time-periodic NESS of our system will be the EGP $\phi_{\mathrm{EGP}}$ \cite{Bardyn_2018}, defined as $\phi_{\mathrm{EGP}}(t) = \Im\log\text{Tr}\bigl[\varrho(t)\,T_\mathrm{tw}\bigr]\equiv\Im\log Z(t)$. It is directly tied to a genuine physical expectation value at all times, rather than to a formally constructed effective unitary. In Sec.~\ref{sec:res_sym} we consider a transformation $\Lambda$, based on the chiral structure of the system, and show that in the symmetric regime, where $\Lambda C_{\text{NESS}}(t)\Lambda= C_{\text{NESS}}(t)$, the EGP itself is quantised at all times $t$, $\phi_\mathrm{EGP}(t)\in \{\pm \pi/2 \}$, and determines the topology of the steady state. In  Sec.~\ref{sec:windingnotop}, we consider the general, symmetry-broken case, $\Lambda C_{\text{NESS}}(t)\Lambda \neq C_{\text{NESS}}(t)$, and show that while $\phi_{\mathrm{EGP}}(t)$ loses its quantisation, a winding number $\nu_T \in \mathbb Z$, extracted from the closed trajectory traced out by $Z(t) \in \mathbb{C}$ in the complex plane within a Floquet period $T$, can be used to classify the topology of the driven-dissipative steady state and gives rise to frequency- and temperature-driven topological phase transitions. When $C_{\text{NESS}}(t)$ is symmetric under $\Lambda$, $Z(t)$ is pinned to the imaginary axis, confining the
trajectory to $i\mathbb{R}\setminus\{0\}$. Since it never leaves the imaginary
axis, it cannot encircle the origin and $\nu_T$ vanishes. The
symmetry that pins $\phi_{\mathrm{EGP}}$ is exactly what forbids $Z(t)$ from
winding. The topology of the steady state is then carried by the sign of
$\Im Z(t)$, or, equivalently, $\phi_{\mathrm{EGP}}(t) = \pm\pi/2$.

\subsection{\texorpdfstring{Quantised Steady-State Topology in the Symmetric Regime}{Quantised Steady-State Topology in the Symmetric Regime}} \label{sec:res_sym}

We start by discussing under what conditions $\phi_{\mathrm{EGP}}(t)$ is quantised and how that quantisation arises from a symmetry of the driven-dissipative system and its steady state.
Since the NESS $C_{\mathrm{NESS}}(t)$ is time-dependent, $Z(t)$ will be time-dependent. For the model considered in this work, at time $t$, one finds 
\begin{equation}
Z(t) = c\,\Omega\,\mathrm{Pf}\bigl(C_{\mathrm{NESS}}(t)\bigr)\Bigl[ \mathrm{Pf}\bigl(\mathcal{N}_{1}(t)\bigr)
- \mathrm{Pf}\bigl(\mathcal{N}_{2}(t)\bigr) 
\Bigr], \label{eq:def_Z}
\end{equation} 
where we have defined 
$
\mathcal{N}_{1,2}(t) = \mathcal{K}_{1,2} - C^{-1}_{\mathrm{NESS}}(t)  
$. 
Here, the prefactors are  
\begin{equation*}
c = \exp \bigl( i \pi(2L+3)/2 \bigr) , \quad
\Omega = \prod_{\substack{k=1 \\ k \neq L-1}}^{2L} \cos\bigl( \pi(k+1)/(2L) \bigr) ,
\end{equation*}
and the constant matrices $\mathcal K_1$ and $\mathcal K_2$ are given by
\begin{equation}
\begin{aligned}
\mathcal{K}_1 &= 
\Bigl[\bigoplus_{\substack{k=1 \\ k \neq L-1}}^{2L} 
\tan\Bigl( \frac{\pi(k+1)}{2L} \Bigr) \sigma_y^k \Bigr] 
\;\oplus\; \sigma_y^{L-1}, \\
\mathcal{K}_2 &= 
\Bigl[\bigoplus_{\substack{k=1 \\ k \neq L-1}}^{2L} 
\tan\Bigl( \frac{\pi(k+1)}{2L} \Bigr) \sigma_y^k\Bigr] \;\oplus\; \mathbf{0}^{L-1}_{2 \times 2} .
\end{aligned} \label{eq:Kmats}
\end{equation}
This analytic expression of $Z(t)$ can be derived by making use of Grassmann calculus~\cite{Bravyi_2005} and is done explicitly for the time-independent case in Ref.~\cite{Molignini_2023}.

In the static case, the quantisation of $\phi_{\mathrm{EGP}}$ was shown in Ref.~\cite{Molignini_2023}. We briefly show that in the special case of $\mu_{A,B}=0$, $\phi_{\mathrm{EGP}}(t)$ is quantised at all times $t$ and refer to Appendix~\ref{sec:app_quantisation} for details. Suppose the inverse temperatures $\beta$ are nonzero and consider the following transformation \cite{Molignini_2023}
\be
f_{n,A} = -\tilde{f}_{n,A}^{\dagger}, \qquad f_{n,B} = \tilde{f}_{n,B}^{\dagger} , 
\ee
which relies on the chiral structure of the model, as it acts differently on the A and B sublattices. This defines a transformation $\Lambda$ across the whole chain  in the Majorana basis (see Appendix~\ref{sec:app_quantisation}), which is real, symmetric and involutory, so $\Lambda = \Lambda^\dagger= \Lambda^{-1}$.
Regardless of $\mu$ and $\beta$, at all times $t$ we find 
$
\Lambda \mathbb{X}(t) \Lambda = \mathbb{X}(t)$ and $\Lambda\mathcal{K}_{1,2} \Lambda = - \mathcal{K}_{1,2} 
$.
Further, if $\mu=0$ we also have $\Lambda \mathbb{Y}(t) \Lambda = \mathbb{Y}(t)$ for all $t$, which, in turn, results in the symmetry being inherited by the steady-state covariance matrix $C_{\text{NESS}}(t)$ \cite{Bardyn_2013}, 
\be
\Lambda C_{\text{NESS}}(t) \Lambda = C_{\text{NESS}}(t) ,
\ee
and its inverse $C^{-1}_{\mathrm{NESS}}(t)$. Consider the regime $\mu=0$, such that $C_{\text{NESS}}(t)$ is symmetric under $\Lambda$ for the following. 
Consider the factors in the definition $\phi_{\mathrm{EGP}}(t)=\arg(Z(t))$, Eq.~(\ref{eq:def_Z}).
Importantly, since $\mathcal{K}_{1,2}$ are fully imaginary, we find that
\be
\mathcal{N}_{1,2}^*(t) = - \mathcal{K}_{1,2} - C^{-1}_{\mathrm{NESS}}(t) = \Lambda \mathcal{N}_{1,2}(t) \Lambda ,  \label{eq:ident}
\ee
i.e.\ complex conjugation has the same effect on $\mathcal{N}_{1,2}$ as the chosen transformation $\Lambda$. Taking the Pfaffian of both sides and using $\det\Lambda=+1$ shows that $\mathrm{Pf}\bigl(\mathcal{N}_{1,2}(t)\bigr)$ is real (see Appendix~\ref{sec:app_quantisation}). Since $\mathrm{Pf}\bigl(C_{\mathrm{NESS}}(t)\bigr)$ and $\Omega$ are real as well, while $c^*=-c$, Eq.~(\ref{eq:def_Z}) gives
\be
Z^*(t) = - Z(t) \Longrightarrow  Z(t) = \left|Z(t)\right| e^{\pm i \pi/2} ,
\ee
which implies the quantisation of the EGP
\be
\phi_{\mathrm{EGP}}(t) = \frac{\pi}{2}
\ \mathrm{mod}\,\pi ,
\ee
provided $Z(t) \neq 0$.
\begin{figure}
    \centering
    \includegraphics[width=0.48\textwidth]{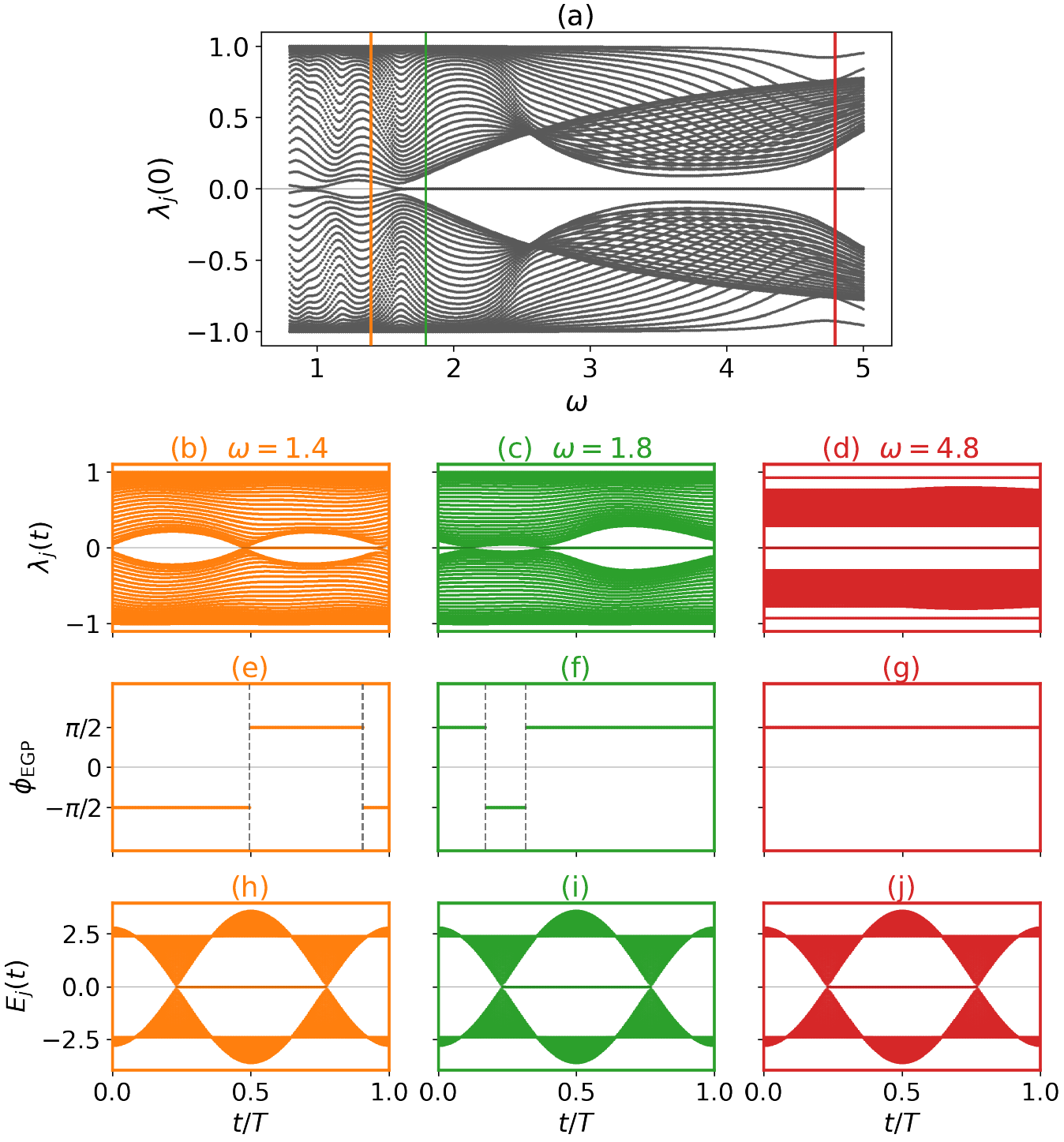}
    \caption{Analysis of the OBC purity spectra, Hamiltonian energy spectra as well as the EGP $\phi_\mathrm{EGP}(t)$, for the same system parameters as in Fig.~\ref{img:test1}. (a)~OBC purity spectrum, obtained from the stroboscopic covariance matrix $C_{\text{NESS}}(0)$ as a function of $\omega$. Note the finite purity gap and $0$-purity states outside of the regime $\omega \in [1.0,1.5]$. (b)--(d)~Instantaneous OBC purity spectra, obtained from $C_{\text{NESS}}(t)$ for $\omega=1.4$, $\omega=1.8$ and $\omega=4.8$ (highlighted by coloured vertical lines in (a)), indicating intra-period gap closings and vanishing (emerging) $0$-purity states. (e)--(g)~Instantaneous EGP $\phi_\mathrm{EGP}(t)$, calculated from PBC for the different frequencies. Topologically nontrivial values ($\phi_\mathrm{EGP}(t)=+\pi/2$) correctly predict the emergence of edge states in the purity spectra, when OBC are imposed (cf. (b)--(d)). (h)--(j)~Instantaneous energy eigenspectrum of $\mathbb H_\mathrm{Majo}(t)$ for the different frequencies. Unlike the purity spectrum, it is independent of $\omega$. Note the two reflection-symmetric points in the spectrum at $t=0$ and $t=T/2$.}
    \label{img:test2}
\end{figure}
Having shown the quantisation of $\phi_{\mathrm{EGP}}(t)$, in this section we focus on the symmetric case, setting $\mu=0$, and investigate the relation between the existence of protected states in the OBC purity spectrum and nontrivial values of $\phi_{\mathrm{EGP}}$. We emphasise that, while $\phi_{\mathrm{EGP}}(t)$ is a bulk quantity and in the subsequent calculations we compute it from PBC, the plots showing $\phi_{\mathrm{EGP}}$ are visually unaltered by the choice of boundary conditions, similar to what was observed in Ref.~\cite{Bardyn_2018}.

In Fig.~\ref{img:test2}(a), we first plot the purity spectrum $\lambda_j(0)$, obtained from the stroboscopic steady-state covariance matrix $C_{\text{NESS}}(0)$, as a function of the driving frequency $\omega$ for OBC. We observe purity bands and a well-defined purity gap, which closes twice, at $\omega \approx1$ and $\omega \approx 1.5$. For low to intermediate frequencies, the bulk purity eigenvalues cluster near $\lambda=\pm1$. Apart from the regime $\omega \in [1.0,1.5]$, the system hosts $0$-purity states in its spectrum, which are separated from the bulk. We recall from isolated Floquet systems that considering only the stroboscopic topology can miss contributions coming from within the Floquet period $T$ \cite{Asboth_2014,Rudner_2013}. 
We thus investigate whether there are any topological intra-period effects which cannot be understood from the stroboscopic steady state alone. 
To this end, in Fig.~\ref{img:test2}(b)--(d) we plot the purity spectrum $\lambda_j(t)$ as a function of $t/T$ within a micromotion period $t/T\in [0,1]$ for OBC. 
Interestingly, we can see that for lower frequencies (b),(c), there are intra-period gap closings present and the exact time $t^*/T$ of the gap closings depends on $\omega$. 
For high frequencies, on the other hand, no gap closings occur and the spectrum within a period looks almost constant. %

Since, as discussed, for our system the EGP is quantised at all times $t$,  in Fig.~\ref{img:test2}(e)--(g) we plot the EGP $\phi_\mathrm{EGP}(t)$ within a period $T$, calculated from PBC. We observe that for low frequencies, at the gap-closing points $t^*/T$ in the OBC purity spectrum plots in (b)--(d), the EGP jumps by $\pi$ and switches from topologically nontrivial ($\phi_\mathrm{EGP}=+\pi/2$) to topologically trivial ($\phi_\mathrm{EGP}=-\pi/2$)  for some finite time within the period, and then back to topologically nontrivial again. Interestingly, this fraction within a period that is spent in the trivial regime is dependent on the driving frequency $\omega$. Comparing to (b)--(d) we verify that topologically nontrivial values correctly predict the appearance of $0$-purity states in the purity spectra, when OBC are imposed. We have verified that the $0$-purity states correspond to purity eigenvectors that are exponentially localised at the boundaries of the chain and separated from the bulk. Hence, we will refer to them as edge states and we find bulk-boundary correspondence between the EGP and the OBC purity spectrum. As a comparison to the purity spectrum, in Fig.~\ref{img:test2}(h)--(j) we show the instantaneous Hamiltonian spectrum of $\mathbb H_\mathrm{Majo}(t)$ within a driving period $T$. Note that here, the spectrum is independent of $\omega$ and, importantly, the Hamiltonian spectrum has a symmetry along the time axis which is not present for the purity spectra. 
These results can be understood as follows.

Recall that the chosen drive preserves the chiral symmetry of the system and is even in time around $s=0$ and $s=T/2$. Thus, $H_\mathrm{PDSSH}(t)$ has two time-reflection symmetric points at $s=0$ and $s=T/2$ (cf. Fig.~\ref{img:test2}(h)--(j)), where
\be
H_\mathrm{PDSSH}(s-t)  = H_\mathrm{PDSSH}(s+t).
\ee
Together with chiral symmetry which holds at every instant, it
allows for the chiral-symmetric decomposition of the Floquet operators $U(\cdot,\cdot)$ and the definition of the winding numbers $(\nu_0,\nu_\pi)$, as outlined in Sec.~\ref{sec:isol_driv_SSH}. This symmetry transfers to the operators $\mathbb{X}(t)$ and $\mathbb Y(t)$ in Eq.~(\ref{eq:SylvFloqapp}), which determine the time evolution of $C(t)$. However, this reflection symmetry is generically not inherited by the steady-state covariance matrix. That is, there is no $s \in [0,T)$ such that
\be 
    C_\mathrm{NESS}(s-t) = C_\mathrm{NESS}(s+t) \ \ \forall\, t ,
\ee
as reflected in the asymmetry of the purity spectrum about $s = 0, T/2$ in
Fig.~\ref{img:test2}(b)--(d). We have verified this numerically for  $C_\mathrm{NESS}(t)$. Consequently, no two-time decomposition, as available at the level of the generator, exists for the state, and any
topological invariant of the full drive cycle must be computed from the entire period rather than from two snapshots.
However, interestingly, this broken symmetry in the steady state enables new possibilities of control, which we outline in the following. 

\begin{figure}
    \centering
    \includegraphics[width=0.48\textwidth]{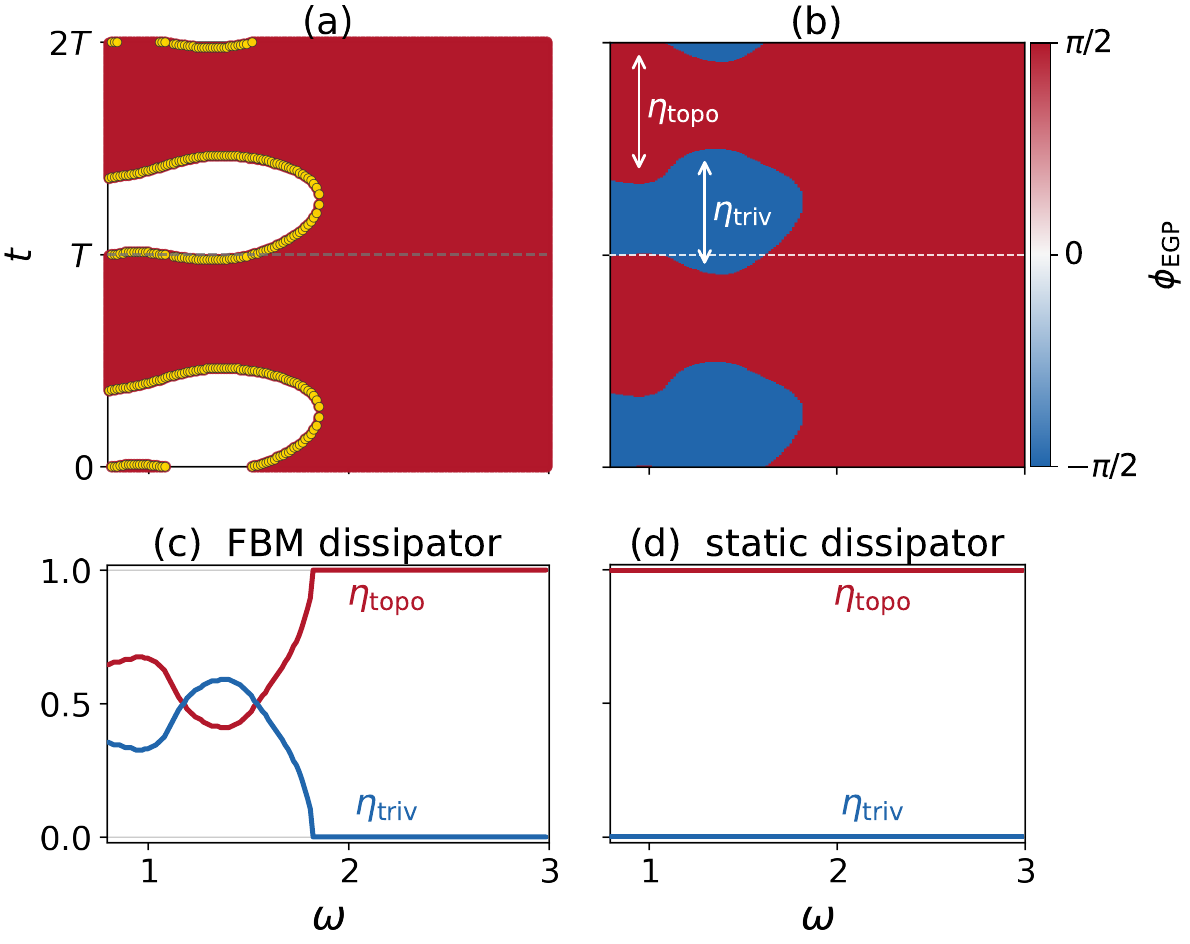}
    \caption{Intra-period topological switching of the NESS for the same parameters as in Fig.~\ref{img:test1}. (a)~OBC purity-gap closing times $t/T$ against $\omega$, indicated by yellow markers. Red regions show $0$-purity states in the $(\omega,t)$-parameter space. Closings are identified by $\lambda_{\mathrm{lowest}}(t)<\Delta_{\mathrm{close}}$ with $\Delta_{\mathrm{close}}=0.05$. (b)~PBC EGP $\phi_{\mathrm{EGP}}(\omega,t)=\arg Z(t)$ over the same window. White arrows indicate the topological and trivial fraction $\eta_{\mathrm{topo}}$ and $\eta_{\mathrm{triv}}$, respectively. Nontrivial values $\phi_{\mathrm{EGP}}=+\pi/2$ track the OBC edge states in~(a). (c)~Topologically nontrivial and trivial fractions $\eta_{\mathrm{topo}}(\omega)$ and $\eta_{\mathrm{triv}}(\omega)$ for the Floquet--Born--Markov (FBM) dissipator. The switching window ends at $\omega^*\approx 1.8$, beyond which $\eta_{\mathrm{topo}}=1$. (d)~Same quantities for a static dissipator (drive neglected in the bath rates). The intra-period switching is erased and $(\eta_{\mathrm{triv}},\eta_{\mathrm{topo}}) =(0,1) \ \forall\omega$. }
\label{img:test3}
\end{figure}

Recall that the Hamiltonian gap-closing time fractions $t^*_H/T$ %
are independent of $\omega$. The gap-closing condition follows directly from the drive. Recalling the definitions of
the hoppings $\tau(t) = J + 2V\cos(\omega t)$ and
$\tau'(t) = J' - 2V\cos(\omega t)$, the instantaneous gap closes when $\tau(t^*_H) = \tau'(t^*_H)$, i.e.
\begin{equation}
J + 2V\cos(\omega t_H^*) = J' - 2V\cos(\omega t_H^*) ,
\end{equation}
which is solved if
$
\cos(\omega t_H^*) = (J'-J)/(4V)
$.
Since $\omega$ and $t$ enter only through $\omega t = 2\pi\,t/T$, this
condition is independent of $\omega$, cf. Fig.~\ref{img:test2}(h)--(j), giving two gap-closing times per period,
\begin{equation}
    \frac{t^{*}_{H,1}}{T} = \frac{1}{2\pi}\arccos\!\left(\frac{J'-J}{4V}\right),
    \qquad
    \frac{t^{*}_{H,2}}{T} = 1 - \frac{t^{*}_{H,1}}{T},
    \label{eq:tstar_two_solutions}
\end{equation}
symmetric around $t/T = 0$ and $t/T = 1/2$. The gap-closing points exist provided $|J'-J| \le 4V$.
Looking at the purity spectrum in Fig.~\ref{img:test2}(b)--(d), however, we clearly observe a frequency-dependent `lag/shift' from the purity-gap closing times 
\be
t^*_p=t^*_p(\omega) ,
\ee
to the instantaneous Hamiltonian gap-closing times 
$
t^*_H
$.
 Plotting the purity-gap closing times $t^*_{pj}(\omega)/T$ in a $2$D parameter space $(\omega,t)$ for OBC in Fig.~\ref{img:test3}(a) (yellow dots), we can see that we either have zero or two  closings of the purity gap,
where the two gap closings $t^*_{p1}(\omega), t^*_{p2}(\omega)$ \emph{continuously} merge into each other and annihilate each other at the critical frequency $\omega^*\approx1.8$. The gap-closing points are numerically determined by setting a condition $\lambda_\mathrm{lowest}(t_{pj}^*) < \Delta_\mathrm{close}$, where we set $\Delta_\mathrm{close}=0.05$. We have verified, using the PBC bulk purity spectrum, that the gap-closing points coincide with those obtained from the OBC spectrum (up to finite-size effects). 
In red is shown where in the parameter space $(\omega,t)$ we find edge states between the gap-closing points $t^*_{p1}(\omega), t^*_{p2}(\omega)$ in the OBC purity spectrum. To analyse the steady-state topology, in (b), we further depict the EGP $\phi_\mathrm{EGP}(\omega,t)$ calculated from PBC. We see that nontrivial values ($\phi_\mathrm{EGP}(\omega,t)=+\pi/2$) correctly predict the existence of edge states in the OBC purity spectrum in (a) (up to numerical inaccuracies and finite-size effects).  We have confirmed that, in the thermodynamic limit $L\rightarrow \infty$, $\phi_\mathrm{EGP}(t)$ remains constant throughout the
period whenever the PBC purity gap stays open, and changes only at gap
closings. For finite $L$ the closings appear as soft minima of depth $\propto 1/L$. Accordingly, the purity gap at the switching times vanishes as $L \rightarrow \infty$ and, thus, any fixed detection threshold $\Delta_{\mathrm{close}}>0$ eventually counts these events as closings once $L$ is large enough.

We can see that the fraction of time spent in the nontrivial (trivial) regime, which we term the \emph{topologically nontrivial (trivial) fraction}, denoted by $\eta_\mathrm{topo}$ ($\eta_\mathrm{triv}$), changes smoothly as a function of $\omega$. To quantify the nontrivial fraction, we define
\be
\eta_{\mathrm{topo}} = \frac{1}{T} \int_0^T \theta_{\mathrm{topo}} \bigl[\phi_{\mathrm{EGP}}(t)\bigr]\,\mathrm{d}t  \in [0,1], 
\ee
where we introduced the step function
\begin{equation}
\theta_{\mathrm{topo}}\bigl[\phi_{\mathrm{EGP}}(t)\bigr] =
\begin{cases}
1, & \phi_{\mathrm{EGP}}(t) = +\pi/2,\\[4pt]
0, & \phi_{\mathrm{EGP}}(t) = -\pi/2.
\end{cases}
\end{equation}
This $\eta_{\mathrm{topo}}$ then immediately lets us categorise the driven-dissipative steady state into the cases
\begin{equation}
\eta_{\mathrm{topo}}  \
\begin{cases}
=0, & \text{fully trivial cycle},\\[4pt]
\in (0,1), & \text{topological switching regime},\\[4pt]
=1, & \text{fully topological cycle}.
\end{cases}
\end{equation}
Trivially, for the trivial fraction we find
$
\eta_{\mathrm{triv}} = 1-\eta_{\mathrm{topo}}  
$.
We remark that $\eta_{\mathrm{topo}}$ ($\eta_{\mathrm{triv}}$) is not a topological invariant but quantifies the intra-period topological switching of the steady state and can be tuned externally via the driving frequency. In Fig.~\ref{img:test3}(c) we plot $\eta_{\mathrm{topo}}$ and $\eta_{\mathrm{triv}}$ as functions of $\omega$ and can clearly see its continuous, frequency-dependent behaviour, approaching $(\eta_{\mathrm{triv}},\eta_{\mathrm{topo}}) = (0,1)$ for frequencies $\omega>\omega^*$, giving the frequencies at which the dissipative steady state no longer switches topological regimes within a period. For $\omega>\omega^*$ the steady state is either fully trivial or fully topological throughout the entire period.
We have checked the behaviour for different system sizes $L$ and find that both $\eta_{\mathrm{topo}}$ and $\omega^*$ converge for $L \approx 12$ and are stable for $L\ge12$.  We note that this behaviour is a direct consequence of the dissipation and the non-adiabatic dynamics generated by the Floquet drive. The dissipation further ensures convergence to a unique time-periodic steady state whenever the real parts of the rapidities are nonzero, in contrast to an isolated setting, where the state generally remains dependent on the initial condition. This interplay of drive and dissipation therefore provides a mechanism for controlling the duration of the intra-period topological regimes through the driving frequency, a feature that is not directly captured by the Hamiltonian or Liouvillian generator alone. 
\indent As discussed above, since the steady-state covariance matrix depends on both $\mathbb{X}(t)$ and $\mathbb{Y}(t)$, the topology of the steady state need not be inherited from the topology of the generator alone. This is consistent with our numerical observation that the critical frequency $\omega^*$, at which $t^*_{p1}(\omega)$ and $t^*_{p2}(\omega)$ annihilate in Fig.~\ref{img:test3}(a), does not coincide with the critical points of the quasirapidity spectrum in Fig.~\ref{img:test1}(b).

Interestingly, the concrete construction of the dissipator seems to be crucial for the obtained results. In Fig.~\ref{img:test3}(d), we plot $\eta_{\mathrm{topo}}$ and $\eta_{\mathrm{triv}}$ as functions of $\omega$, but naively neglect the impact of the drive on the dissipator and instead use the bath-operator construction given by Eq.~(\ref{eq:z_spec_stat}). In Appendix~\ref{sec:stat_diss}, we show full diagnostics as in Fig.~\ref{img:test3}(a)--(c) for the static-dissipator case. 
We observe that the intra-period switching physics is erased, and the topological and trivial fractions are fixed to values $(\eta_{\mathrm{triv}},\eta_{\mathrm{topo}}) = (0,1) \ \forall \omega$. This demonstrates that, even though the Hamiltonian part $H_\mathrm{PDSSH}(t)$ is unchanged in both cases, the correct treatment of the drive dependence of the dissipator within the Floquet--Born--Markov approach is essential for capturing the intra-period topological switching observed in the steady state. This raises the question of whether a rotating-wave approximation \cite{wustmann2010statistical} would produce a comparable suppression of the intra-period dynamics. 

Having established that the pointwise quantisation of 
$\phi_\mathrm{EGP}(t)$ relies on the $\Lambda$ symmetry, we now ask what topological structure survives once that symmetry is broken.

\subsection{\texorpdfstring{Winding-Number Topology in the Symmetry-Broken Regime}{Winding-Number Topology in the Symmetry-Broken Regime}}
\label{sec:windingnotop}

When the symmetry is broken, i.e.~$\Lambda C_{\text{NESS}}(t)\Lambda \neq C_{\text{NESS}}(t)$, $Z(t)$ is no longer pinned to the imaginary axis and traces a closed curve in the complex plane. In adiabatic protocols, the accumulated change $\Delta\phi_{\mathrm{EGP}}$ along a slow parameter path is a winding number, both for Lindblad-type~\cite{Bardyn_2018} and genuine finite-temperature Redfield-type~\cite{Molignini_2023} dissipative extensions of Rice--Mele models, with protection by the purity gap discussed in Ref.~\cite{Bardyn_2018}. Here, in contrast, the Floquet NESS is itself intrinsically time-periodic and the cycle is traversed at finite driving frequency, with no adiabatic assumption.
\begin{figure}
    \centering
    \includegraphics[width=0.48\textwidth]{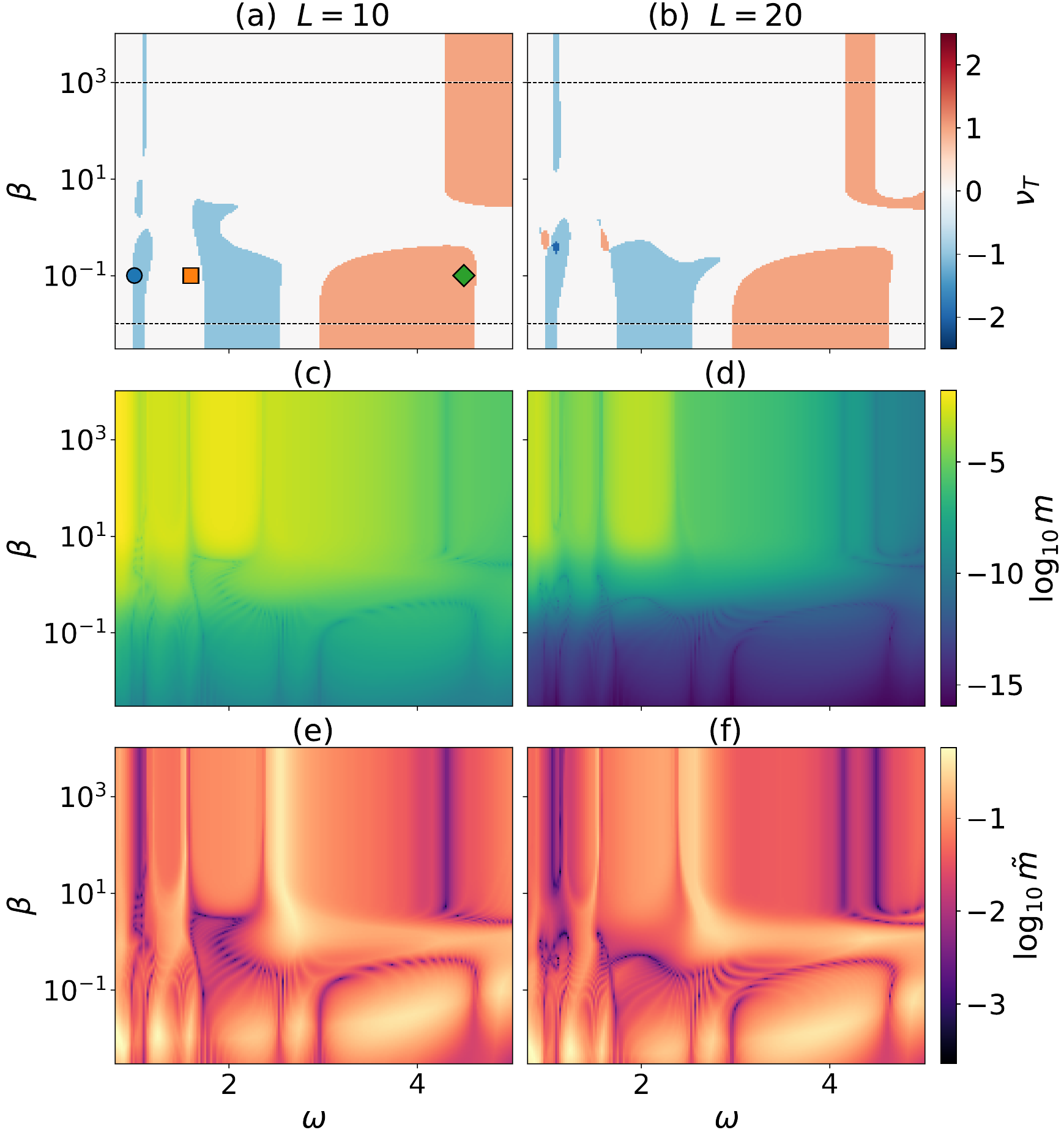}
    \caption{Winding number $\nu_T(\omega,\beta)$ and minimum trajectory distance $m(\omega,\beta)$ for the symmetry-broken NESS with $\mu_A=-\mu_B=2J$ and PBC. Remaining parameters as in Fig.~\ref{img:test1}, except for the chemical-potential bias. (a),(b)~$\nu_T$ for $L=10$ and $L=20$. Extended plateaus with $\nu_T=\pm 1$ appear for both sizes. Dashed horizontal lines mark the cuts $\beta=10^{3}$ and $\beta=10^{-2}$ analysed in Fig.~\ref{img:test52}, and the symbols in~(a) indicate the $(\omega,\beta)$ points whose trajectories are shown in Fig.~\ref{img:test53}. (c),(d)~Corresponding $\log_{10}m$, with $m=\min_{t\in[0,T]}|Z(t)|$. Boundaries between regions with different $\nu_T$ coincide with $m\to 0$, while $m\to 0$ without a change of $\nu_T$ is also possible. At small $\beta$ the $\nu_T=\pm 1$ plateaus agree for $L=10$ and $L=20$. At large $\beta$ and high $\omega$ they do not (cf. Fig.~\ref{img:test52}). (e),(f)~Relative margin $\tilde{m}=m/\max_{t\in[0,T]}|Z(t)|$ as an indicator for the robustness (with respect to changing the system size $L$) of $\nu_T(\omega,\beta)$: $\nu_T(\omega,\beta)$ is most robust when $\tilde{m}$ is large.}
\label{img:test51}
\end{figure}

Since $Z(t)\in \mathbb{C}$ is periodic, $Z(t+T)=Z(t)$, its evolution over a drive period defines a closed loop in the complex plane. Provided that $Z(t)\neq 0$ throughout the cycle, this loop lies in the punctured complex plane $\mathbb{C}\setminus\{0\}$. Since $\pi_1(\mathbb{C}\setminus\{0\})\simeq\mathbb{Z}$, the homotopy class of the loop is characterised by the integer winding number~\cite{Bardyn_2018, Molignini_2023}
\begin{equation}
\nu_T \equiv  \frac{\Delta \phi_{\mathrm{EGP}}}{2\pi} = \frac{1}{2\pi}\int_0^{T}\partial_t \arg(Z(t))\,\mathrm{d}t \in \mathbb{Z}, \label{eq:nu_T}
\end{equation} 
which counts signed encirclements of $Z=0$ during a Floquet period $T$. The question we ask is whether this integer forms extended, drive- and bath-dependent  phases of the driven-dissipative NESS, protected by a measure of the minimum distance (within a period $T$) of the time-parametrised path (trajectory) $t\mapsto Z(t)$ from the origin
\begin{equation}
 m =   \min_{t\in[0,T]} |Z(t)| , \label{eq:def_m}
\end{equation} 
and whether those phases survive the limit $L \rightarrow \infty$. We emphasise that $\nu_T$ classifies the NESS, not any Floquet operator. In the following, we break the symmetry by setting a chemical potential bias $\mu_A=-\mu_B=2J$ between the A and B baths. We investigate various system sizes $L$. 

We plot $\nu_T(\omega,\beta)$ in Fig.~\ref{img:test51}(a) and (b) for chain lengths of $L=10$ and $L=20$, respectively. As we can observe, for most of the parameter space, we find $\nu_T=0$. However, we find extended regions of parameter space exhibiting quantised winding numbers $\nu_T=\pm1$, for both $L$. For low frequencies, we observe a thin window, where $\nu_T=-1$ for almost all $\beta$, except for small regimes at intermediate temperatures. For low to intermediate frequencies and low $\beta$, we find an extended region where $\nu_T=-1$, where the shape of the region stays consistent for both $L$. In the intermediate- to high-frequency regime, we find separate dome-like regions where $\nu_T=+1$ for low \emph{or} high temperatures. At small $\beta$ the $\nu_T=\pm 1$ regions agree for $L=10$ and $L=20$.
At large $\beta$ and high $\omega$ they do not, which we will discuss further below.
The plot demonstrates that topological phase transitions can occur both driven by frequency and temperature. Importantly, it shows that the transition frequencies are not directly determined by the generator $\mathbb X(t)$, as we can see when comparing the findings with the rapidity spectrum in Fig.~\ref{img:test1}. Specifically, for fixed $\omega$, we find temperature-driven topological phase transitions 
\be 
\nu_T(\beta): \quad  0 \leftrightarrow -1 , \quad \mathrm{or} \quad \nu_T(\beta): \quad  0 \leftrightarrow +1 .
\ee
Further, for fixed $\beta$, as we tune the driving frequency, we can change not only the number of times the origin is encircled but also the direction of encircling. For example, at approximately $\beta \le 10^{-1}$, we find topological phase transitions
\be 
\nu_T(\omega): \quad  \ldots \leftrightarrow -1 \leftrightarrow 0 \leftrightarrow +1 \leftrightarrow \ldots . 
\ee
Notably, and in contrast to the common intuition that temperature is detrimental to topology (cf.~Ref.~\cite{Molignini_2023}), both the extended regions of nontrivial $\nu_T$ and most of the temperature- and frequency-driven transitions between them are found at high temperatures, $\beta \lesssim 1$, rather than at low temperatures. Since $\nu_T$ depends only on the phase of $Z(t)$, temperature reshapes the trajectory of $Z(t)$ rather than washing it out.
Moreover, in Fig.~\ref{img:test51}(b), we find small neighbourhoods at low $\omega$ and intermediate $\beta$ where $\nu_T \neq0$, as well as a small region in the direct vicinity of $(\omega,\beta) \approx (1,5\times10^{-1})$, where $\nu_T=-2$. While in Fig.~\ref{img:test51}(b) these are genuine observations and not numerical artifacts, they are only a few cells across and not stable under changing $L$. Before we analyse the different regimes and trajectories $t\mapsto Z(t)$ more closely, 
as a numerical consistency check, we verify that regions of different winding numbers are separated robustly. 

For this, we consider $m$, defined in Eq.~(\ref{eq:def_m}), to investigate whether boundaries between regions of different $\nu_T$ correspond to singularities, where for some $t^*$, $Z(t^*)\rightarrow0$ (and, in turn, $m \rightarrow0$), such that the invariant becomes ill-defined and can change. 
We plot $ m(\omega,\beta)$ in Fig.~\ref{img:test51}(c) and (d) for chain lengths of $L=10$ and $L=20$, respectively, and see that it produces the expected behaviour with respect to the phase boundaries in Fig.~\ref{img:test51}(a). 
We further see regions where $m \rightarrow0$ without a change of $\nu_T$, which can happen since $m \rightarrow0$ is a necessary, though not sufficient, condition for $\nu_T$ to change. Moreover, note that for $L=20$ and hot baths, in Fig.~\ref{img:test51}(d) we observe that $m$ approaches very small values, even away from boundaries. We discuss this behaviour and its practical consequences in Sec.~\ref{sec:measurability}.  The panel demonstrates the numerical robustness of the topological invariant as it only changes at singularities. We remark that the protecting gap for $\nu_T$ in this symmetry-broken regime is given by $m$ and not the purity gap $\Delta_{\mathrm{Gap}}$.
\begin{figure}
    \centering
    \includegraphics[width=0.48\textwidth]{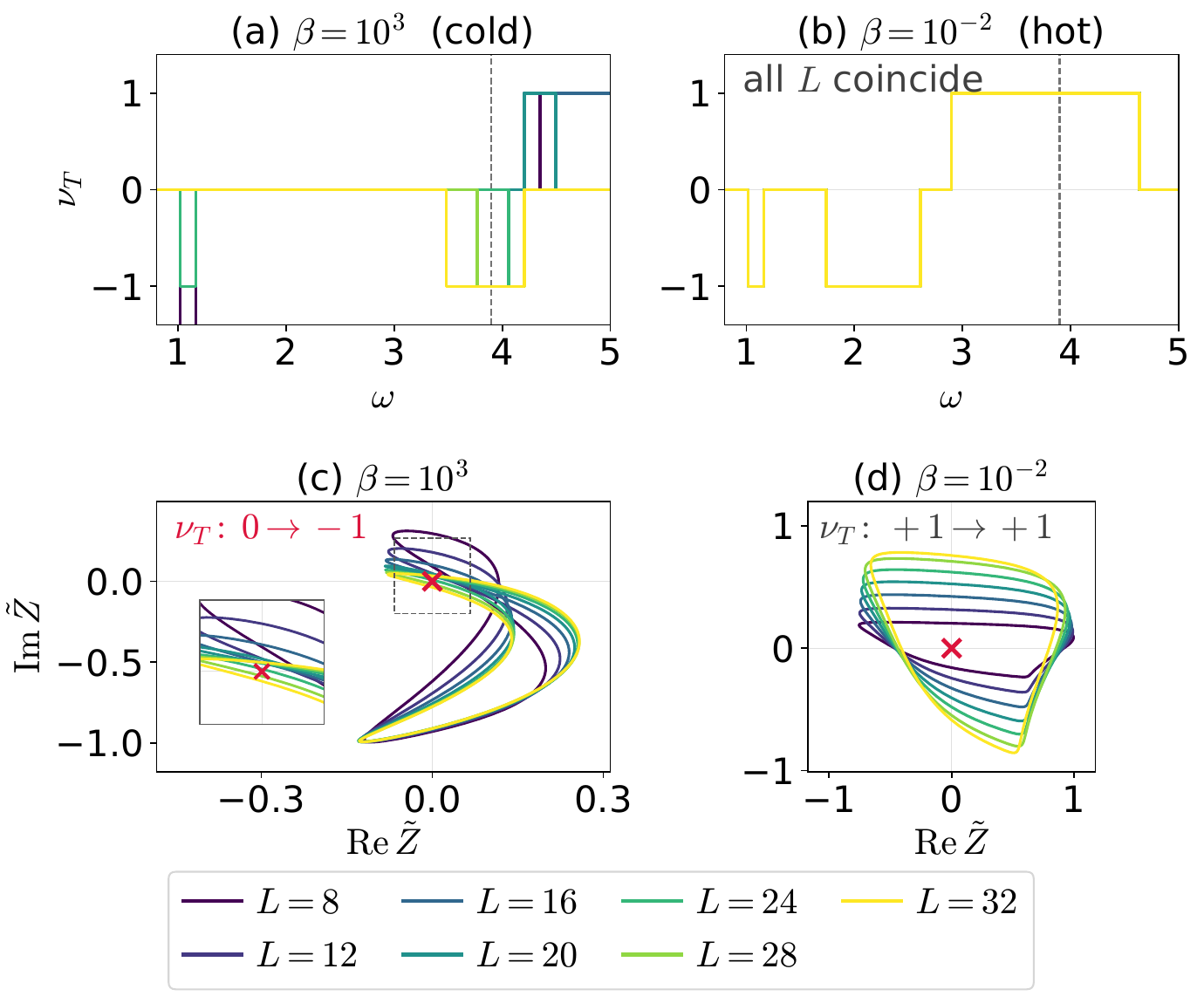}
    \caption{System-size dependence of $\nu_T$ along the cuts marked in Fig.~\ref{img:test51}, for the same parameters as in Fig.~\ref{img:test51}. (a)~$\nu_T(\omega)$ at $\beta=10^{3}$ (cold bath) for chain lengths $L\le 32$. Different $L$ produce different plateaus. A $\nu_T=+1$ plateau leaves the window while a $\nu_T=-1$ plateau enters from below near $\omega\approx 4$. (b)~$\nu_T(\omega)$ at $\beta=10^{-2}$ (hot bath). The curves for all $L$ coincide, so the transitions are already converged by $L=8$. (c),(d)~Normalised trajectories $\tilde{Z}(t)=Z(t)/\max_{t\in[0,T]}|Z(t)|$ at the dashed line $\omega=3.9$ in~(a) and~(b), for $\beta=10^{3}$ and $\beta=10^{-2}$, respectively. The red cross marks the origin. In~(c) varying $L$ drags the loop through $Z=0$, so $\nu_T$ changes. In~(d) the loop stays away from the origin and $\nu_T$ is $L$-independent.}
\label{img:test52}
\end{figure}

Next, we analyse the stability of transitions of $\nu_T$ for different bath temperatures as we increase the system size $L$. In Fig.~\ref{img:test52}(a) and (b), we plot $\nu_T(\omega)$ for varying system sizes $L\le 32$ and fixed inverse temperatures $\beta=10^3$ (cold bath) and $\beta=10^{-2}$ (hot bath), respectively, corresponding to the dashed lines in Fig.~\ref{img:test51}(a) and (b). In Fig.~\ref{img:test52}(a) different $L$ create different plateaus, while for Fig.~\ref{img:test52}(b) the curves for all $L$ coincide. Note that in the high-frequency regime ($\omega >3.5$) in Fig.~\ref{img:test52}(a), increasing $L$ leads to a $\nu_T=+1$ plateau vanishing, and a different $\nu_T=-1$ plateau emerging. Thus, while for hot baths the winding number $\nu_T$ is already `converged' for $L=8$ and topological transitions are robust under scaling, for cold baths this is not the case and different $L$ lead to different winding numbers $\nu_T$ and different transition frequencies. 
To understand the mechanism behind the $L$-sensitivity in the cold-bath regime, for $\omega=3.9$ (dashed lines in (a),(b)) we plot the normalised (relative) trajectories $\tilde{Z}(t)=Z(t)/(\max_{t\in [0,T]}|Z(t)|)$ for fixed $\beta=10^3$ and $\beta=10^{-2}$ in Fig.~\ref{img:test52}(c) and (d), respectively. As we can see in (d), the loop stays well away from the origin (red cross). 
Changing $L$ deforms the loop but not sufficiently to hit the origin, such that $\nu_T$ is independent of $L$, leading to the coinciding transitions in Fig.~\ref{img:test52}(b) for all $L$ that we consider. In the cold-bath case, Fig.~\ref{img:test52}(c), however, the trajectory moves closer to the origin, such that a modest deformation caused by varying $L$ drags the trajectory through the origin, as illustrated in the inset in Fig.~\ref{img:test52}(c), leading to the migrating plateaus observed in Fig.~\ref{img:test52}(a).  Thus, for the hot bath, where $\tilde{Z}(t)$ stays away from the origin under
increasing $L$, $\nu_T(\omega)$ has already converged by $L=8$,
and we expect the transitions in Fig.~\ref{img:test52}(b) to survive the thermodynamic limit $L \rightarrow\infty$.
For the cold bath at high $\omega$, $\tilde{Z}(t)$ sits close to the origin,
so the same $L$-deformation crosses $Z=0$. In this case
the plateaus in Fig.~\ref{img:test52}(a) are nonzero integers at each finite $L$,
but they are not stable when changing $L$.   

Motivated by these observations, we introduce the relative margin 
\be
\tilde{m}= \min_{t\in[0,T]} |\tilde{Z}(t)| = \frac{m}{\max_{t\in[0,T]} |Z(t)|} ,
\ee
which provides a measure of the relative distance of the loop from the origin, thereby serving as an indicator of the robustness of the winding number $\nu_T$, under changing the system size $L$, deep within plateaus and away from transitions. Here, $\tilde{m}\in[0,1]$, with $\tilde{m}=1$ corresponding to a trajectory with constant modulus and $\tilde{m} \rightarrow 0$ meaning that (the minimum of) the trajectory approaches the origin $\tilde{Z}=0$. In Fig.~\ref{img:test51}(e) and (f), we plot $\tilde{m}$ over $(\omega,\beta)$, for $L=10$ and $L=20$, respectively. We observe that $\tilde{m}$ is smallest at and around phase transitions, since there $m \rightarrow 0$. Away from the phase boundaries $\tilde{m}$ becomes meaningful and regions where $\tilde{m}$ is largest ($\log_{10}\tilde{m}\gtrsim-0.5$) correspond to regions where $\nu_T$ is most robust under changing $L$, thereby verifying the observations made in Fig.~\ref{img:test51}. We remark that the margin $\tilde{m}$ serves as a simple diagnostic for the robustness of $\nu_T$. It is incomplete since it characterises the shape of the trajectory relative to its own size, and therefore does not by itself determine whether a given perturbation, caused by changing $L$, is large enough to drag the loop through (or away from) the origin. Further, the NESS trajectories are nontrivial loops, possibly containing self-intersections. Being extracted from only two points, $\tilde{m}$ discards the nontrivial geometry of these trajectories. 

\begin{figure}
    \centering
    \includegraphics[width=0.47\textwidth]{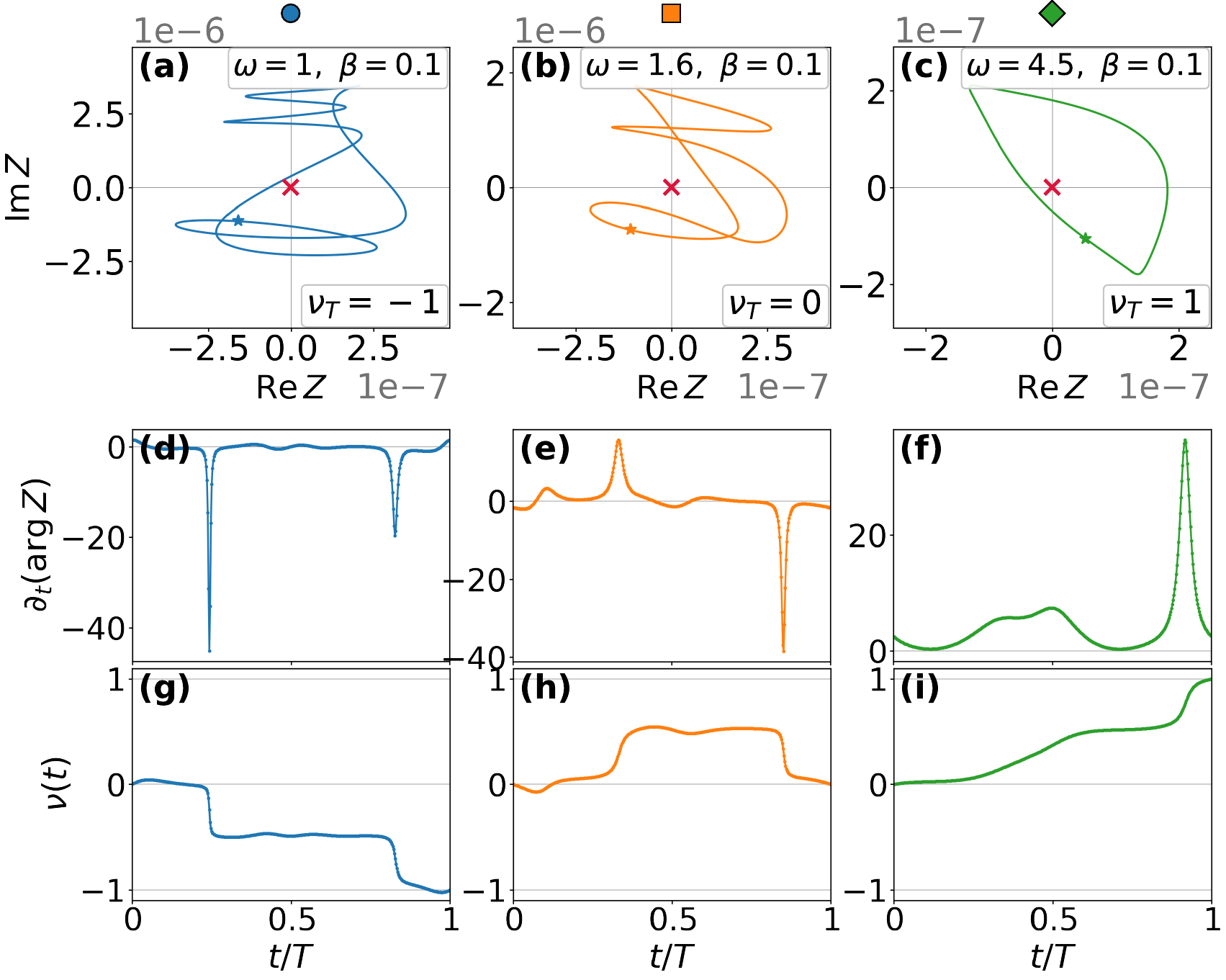}
    \caption{Trajectory micromotion $t\mapsto Z(t)$ for three points marked in Fig.~\ref{img:test51}(a), at $L=10$, $\beta=10^{-1}$ and the same remaining parameters as in Fig.~\ref{img:test51}. (a)--(c)~Closed trajectories $Z(t)$ in the complex plane for $(\omega,\beta)=(1.0,10^{-1})$, $(1.6,10^{-1})$ and $(4.5,10^{-1})$, giving $\nu_T=-1$, $0$ and $+1$, respectively. Stars mark $t=0=T$ and the red cross marks the origin. Note the substantial deformation of the trajectories, including several self-intersections. (d)--(f)~Phase-accumulation rate $\partial_t\arg(Z(t))$ within one Floquet period. (g)--(i)~Cumulative winding $\nu(t)$, with $\nu(T)=\nu_T$. The net winding is dominated by a few localised peaks of $\partial_t\arg(Z(t))$. At lower $\omega$ the accumulation is more temporally localised, while at higher $\omega$ it is more distributed over the period.}
\label{img:test53}
\end{figure}

Having discussed the winding number $\nu_T$ and its stability under scaling as functions of $\omega$ and $\beta$, we next analyse the geometries and dynamics of the trajectories $t\mapsto Z(t)$, for which in Fig.~\ref{img:test53}(a)--(c), we plot the loops $Z(t)$ in the complex plane at the three points $(\omega,\beta)$, marked by different symbols in Fig.~\ref{img:test51}(a).
The stars denote the start and end points $t=0=T$ and the red crosses denote the origins. 
As we can see, all three points give us different winding numbers $\nu_T$. 
While the trajectory in Fig.~\ref{img:test53}(b) does not encircle the origin ($\nu_T=0$), the trajectories in Fig.~\ref{img:test53}(a) and Fig.~\ref{img:test53}(c) give $\nu_T=-1$ and $\nu_T=+1$, respectively. We note that despite substantial deformation of the trajectories, including several self-intersections, the invariant is robust and quantised at the same value over extended regions, rather than being noisy across the $2$D parameter space $(\omega,\beta)$, as we can see in Fig.~\ref{img:test51}(a) and (b). The winding number $\nu_T$ thus robustly extracts a winding out of a highly nontrivial trajectory geometry. 
In order to gain insights into the accumulation of winding within a cycle we investigate the phase-accumulation rate $\partial_t \arg(Z(t))$ in the integrand in Eq.~(\ref{eq:nu_T}) and the cumulative winding $\nu(t)$ at time $t$, given by
\begin{equation}
\nu(t) = \frac{1}{2\pi}\int_0^{t}\partial_{t'} \arg(Z(t'))\,\mathrm{d}t',
\end{equation} 
such that $\nu(T)=\nu_T$. For the loops considered in Fig.~\ref{img:test53}(a)--(c), we plot both $\partial_t \arg(Z(t))$ and $\nu(t)$ as functions of $t$ within a Floquet period $T$ in Fig.~\ref{img:test53}(d)--(f) and Fig.~\ref{img:test53}(g)--(i), respectively. 
Firstly, we observe that, rather than a slow and uniform accumulation of the winding number, a few single points within the period seem to dominantly contribute. 
Specifically for lower frequencies $\omega$, we find highly temporally localised accumulation while for higher $\omega$ the accumulation is more distributed. 
For Fig.~\ref{img:test53}(d) and (f), we see two spikes or hills in the negative and positive direction, respectively, which lead to a negative and positive accumulation of winding. We confirm this also in Fig.~\ref{img:test53}(g) and (i). In Fig.~\ref{img:test53}(e) we observe a few spikes, which, however, have different signs, such that the net accumulation of winding cancels out when integrated over a period, as we show in Fig.~\ref{img:test53}(h), giving $\nu_T=0$.  The panels show that the integral leading to $\nu_T$ is dominated by a few localised points where $\partial_t \arg(Z(t))$ exhibits peaks within the period $T$.

Finally, we address what $\nu_T$ physically represents. Recall that
$\phi_{\mathrm{EGP}}/2\pi$ is the mixed-state generalisation of the
polarisation. In isolated systems, topological phases are tied to quantised charges transported per cycle in Thouless pumps. As found in the adiabatic finite-temperature setting of Ref.~\cite{Molignini_2023}, however, this
identification with a transported current breaks down in our case. We attribute this to the dissipation continuously injecting and removing charges at each site. Tracking $C_{\mathrm{NESS}}(t)$ throughout the cycle, we find no signature of net transport across the chain in the phases with $\nu_T \neq 0$. Similar to what was found in the adiabatic finite-temperature
setting of Ref.~\cite{Molignini_2023}, we stress that this does not
diminish $\nu_T$ as a topological quantity. It remains a quantised property
of the EGP itself and a genuine many-body observable \cite{Bardyn_2018}. 

As we show in Appendix~\ref{sec:nu_purity}, $\nu_T$ and the steady-state purity spectrum are not directly correlated. We find no bulk-boundary correspondence between $\nu_T$ and the instantaneous OBC purity spectrum in the symmetry-broken regime. This is expected since
the purity spectrum characterises $C_{\mathrm{NESS}}(t)$ at a single
instant, whereas $\nu_T$ depends on the full closed trajectory of $Z(t)$
over the period. In this context, the gap
protecting $\nu_T$ is the minimum trajectory distance $m$ of
Eq.~\eqref{eq:def_m}, not the purity gap $\Delta_{\mathrm{Gap}}$. In this regard, we further find no connection between $\nu_T$ and the quasipurity spectrum. Our observations, however, do not rule out that
$\nu_T$ could be  related to some other measurable consequence of the steady
state beyond the EGP itself.

\subsection{\texorpdfstring{Scaling of the Modulus $|Z(t)|$ and Experimental Accessibility}{Scaling of the Modulus |Z(t)| and Experimental Accessibility}} \label{sec:measurability}

\begin{figure}
    \centering
    \includegraphics[width=0.47\textwidth]{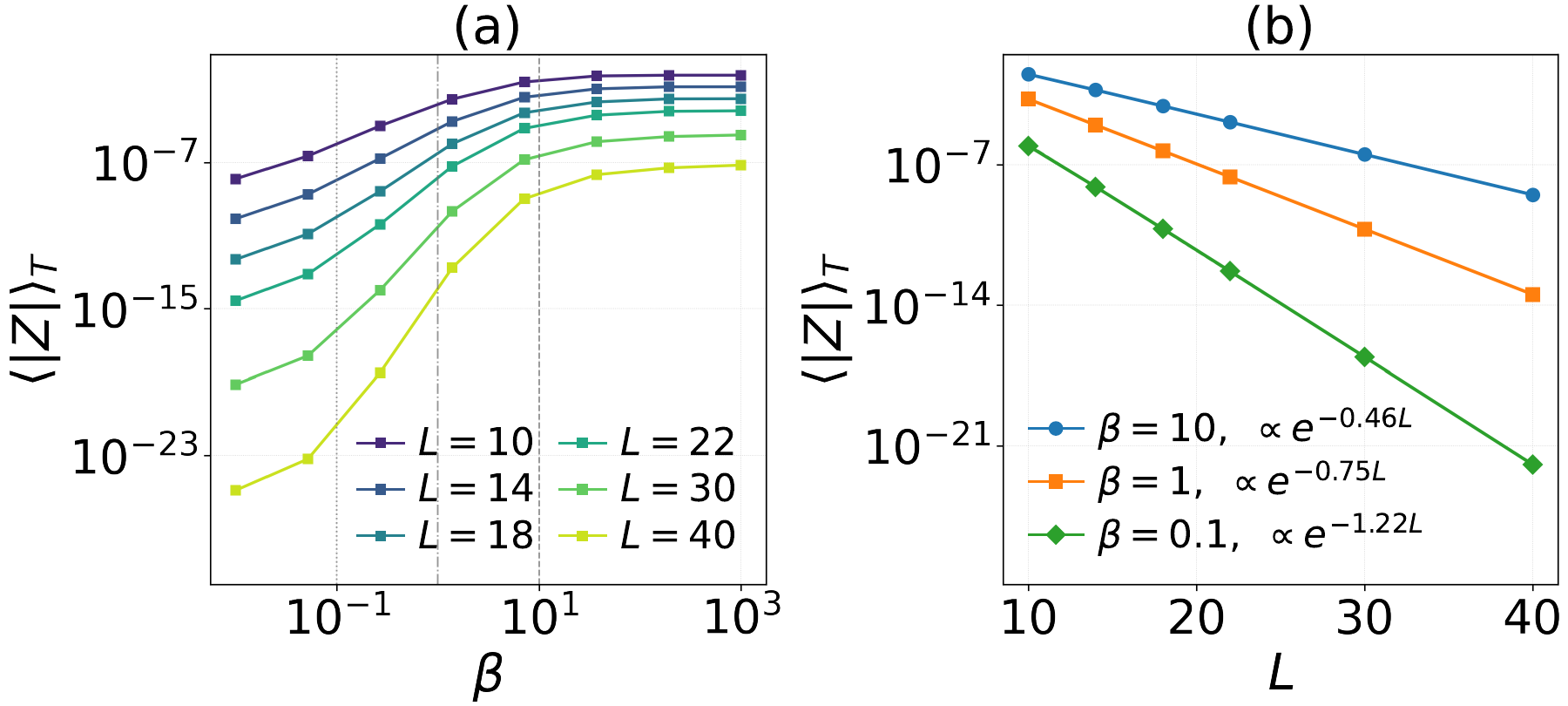}
    \caption{Period-averaged modulus $\langle|Z(t)|\rangle_T$ for the driven-dissipative NESS under PBC, $\omega=1.6$ and the same remaining system parameters as in Fig.~\ref{img:test51}.
  (a)~$\langle|Z(t)|\rangle_T$ as a function of the inverse temperature $\beta$ for system sizes $L=10,14,18,22,30,40$.
  (b)~$\langle|Z(t)|\rangle_T$ plotted against $L$ at $\beta=10$, $1$, and $0.1$ (vertical lines in (a)), with solid lines showing exponential fits $\propto e^{rL}$.
  The overall scale of $Z$ is strongly suppressed with increasing $L$, obeying $\langle|Z(t)|\rangle_T \propto e^{-a(\beta)L}$, with $a(\beta)>0$.}
  \label{fig:Zamp_avg_beta_L}
\label{img:Measurability}
\end{figure}

Before we conclude, we will quantify how $|Z|$
scales with system size and temperature, and discuss consequences for experimental measurements of the EGP $\phi_\mathrm{EGP}=\arg Z$ and the winding number $\nu_T$.
In Fig.~\ref{img:test51}(c),(d) we have observed that $m$, which is exactly the intra-period minimum of the modulus $|Z(t)|=|\text{Tr}\bigl[\varrho(t)\,T_\mathrm{tw}\bigr]|$, depends on the system size $L$ and the inverse temperature $\beta$. 
Specifically, it can reach very small values in certain regimes. We explore the behaviour of the modulus of the expectation value $|Z(t)|$ in Fig.~\ref{img:Measurability}. In Fig.~\ref{img:Measurability}(a) we show the period-averaged modulus $\langle|Z(t)|\rangle_T$ as a function of the inverse temperature $\beta$, for different $L$. We observe that increasing $\beta$ leads to larger moduli $\langle|Z(t)|\rangle_T$, with larger $L$ resulting in curves shifting towards lower moduli $\langle|Z(t)|\rangle_T$. In Fig.~\ref{img:Measurability}(b) we plot $\langle|Z(t)|\rangle_T$ against $L$ for different $\beta$ (vertical lines in Fig.~\ref{img:Measurability}(a)). Importantly, we find the scaling $\langle|Z(t)|\rangle_T\propto e^{-a(\beta)L}$, with $a(\beta)>0$, which is an expected, generic feature of the EGP \cite{Bardyn_2018,Molignini_2023}, and thoroughly discussed for a static thermal (extended) SSH model in Ref.~\cite{hannukainen2026}. Thus, the modulus $|Z(t)|$ becomes exponentially suppressed as we increase the system size and vanishes in the thermodynamic limit $L \rightarrow\infty$. This suppression, however, concerns the practical sensitivity in measurements, that is, how well the EGP can be resolved experimentally, rather than whether the EGP is defined. The EGP $\phi_\mathrm{EGP}(t)$ and the winding number $\nu_T$ depend only on the argument of $Z(t)$ and remain well-defined as long as the trajectory avoids the origin $Z=0$.
In Sec.~\ref{sec:res_sym}, we have observed that $\phi_\mathrm{EGP}(t)$ remains well-defined and pinned, even for large $L$. As we discussed in Sec.~\ref{sec:windingnotop}, the relative margin $\tilde{m}$, which does not suffer from this exponential suppression (unlike the gap $m$ itself), indicates the robustness of $\nu_T$. For high bath temperatures, $\nu_T$ remains pinned even for large $L$, where $\vert Z\vert$ is strongly suppressed. We refer to Ref.~\cite{Bardyn_2018} for details on experimental settings, platforms and procedures to measure the EGP, noting that they discuss slow protocols. How measurements can be carried out for our fast, non-adiabatic protocol remains an open question which we leave for future work.

Finally, Ref.~\cite{hannukainen2026} discusses and analytically derives the exponential suppression of the modulus $|Z|$ for a thermal (extended) SSH chain and the practical complications for experiments. They propose two other diagnostics which agree with the EGP on the topological phase, but do not suffer from an exponentially suppressed signal for large system sizes. This raises the question of whether there are diagnostics for our driven-dissipative model that are experimentally more practical while agreeing with $\phi_\mathrm{EGP}(t)$ and $\nu_T$ on the identified topological classification. 

\section{Conclusion} \label{sec:conclusion}

We studied the non-equilibrium topology of the steady state of a periodically driven, dissipative Su--Schrieffer--Heeger (SSH) chain using the ensemble geometric phase (EGP) $\phi_{\mathrm{EGP}} = \Im\log\text{Tr}\bigl[\varrho\,T_\mathrm{tw}\bigr]\equiv\Im\log Z$. We employed a Floquet--Born--Markov treatment of thermal reservoirs, without invoking the rotating-wave approximation, to find the time-periodic steady state of the system, which can be fully characterised by its covariance matrix, whose Hermitian purity spectrum enables a direct analogue of band topology for mixed states.

Within the EGP framework, the topology of the driven-dissipative steady state is carried by a single invariant whose type is fixed by a symmetry of the steady state. It is $\mathbb{Z}_2$-valued and protected by the purity gap when the symmetry holds, and $\mathbb{Z}$-valued and protected by the minimum trajectory distance when it is broken. 

The $\mathbb{Z}\times\mathbb{Z}$ classification of isolated chiral-symmetric Floquet systems can be extended to the generator of the driven-dissipative dynamics~\cite{Caspel_2019}. In our setting, however, such a generator-based construction is blind to the bath temperatures and chemical potentials, which fundamentally determine the steady state and its symmetries. Moreover, the classification generally does not descend to the steady state within the EGP framework. The time-reflection symmetry underlying the two-time construction of Refs.~\cite{Asboth_2014,Caspel_2019}, present in the operators of the dynamics, is generically not inherited by the steady state, and the purity spectrum has no quasiperiodic structure. We therefore based the topological analysis on the steady state itself.

We have demonstrated that for a specific choice of bath parameters, the steady state is symmetric under a transformation based on the chiral structure of the system, leading to a quantisation of the EGP, $\phi_{\mathrm{EGP}}(t)=\pm\pi/2$, %
at all times $t$ and finite temperatures. This is because the symmetry pins $Z(t)$ to the imaginary axis, forbidding it from winding around the origin $Z=0$, and the topology is carried by the sign of $\Im Z(t)$.
In this regime, we found bulk-boundary correspondence between the EGP and the OBC purity spectrum. %
In addition, we found that %
there is intra-period switching of the EGP, indicating that considering only the topology of the stroboscopic steady state is insufficient, since the steady state can pass through purity-gap closings within the period. We introduced a topologically nontrivial fraction $\eta_\mathrm{topo}\in[0,1]$, quantifying the fraction of the Floquet period during which the steady state is in the topologically nontrivial regime. %
We showed that it varies continuously with the drive frequency.
Importantly, we showed that the correct treatment of the impact of the drive on the dissipator within the Floquet--Born--Markov theory is crucial to capture topological effects, by demonstrating that a naive treatment which neglects this impact erases the intra-period switching. This raises the question of whether a (moderate) rotating-wave approximation \cite{wustmann2010statistical} would lead to a similar loss of intra-period structure, and whether secular or non-secular terms are necessary to observe the switching.

When we break the symmetry, $Z(t)$ traces out a closed trajectory in the complex plane, and a winding number $\nu_T \in \mathbb{Z}$, counting the signed encirclements of the origin $Z=0$ within a drive period, can be used to classify the topology of the steady state. %
We found frequency- and temperature-driven topological phase transitions of the steady state, and extensive regions with nontrivial $\nu_T$. %
For high temperatures, the phases have already converged for small system sizes $L$, while for cold baths the plateaus are genuine integers at each finite $L$ but shift under changing $L$. We found no bulk-boundary correspondence between $\nu_T$ and the instantaneous OBC purity spectrum, as expected for a property of the full loop, nor any connection between $\nu_T$ and the quasipurity spectrum. %

Moreover, isolated Floquet systems are particularly interesting because their independent $0$- and $\pi$-gap structure gives rise to an anomalous bulk-boundary correspondence with no static analogue \cite{Asboth_2014,Lago_2015,Rudner_2013,Nathan_2015}. It remains open whether markers beyond the EGP and constructions beyond one dimension can recover the anomalous Floquet signatures in a driven-dissipative steady state while still accounting for all bath parameters. Another open question is whether the closed path traced by the steady state over a period carries further topological structure accessible to other mixed-state markers, in particular ones that, unlike $|Z(t)|$, are not exponentially suppressed with system size.

Lastly, we emphasise that our formalism extends to arbitrary quadratic fermionic systems with linear bath couplings. It would therefore be interesting to investigate topological phase transitions in other systems, especially those with different underlying symmetries.

\begin{acknowledgments}

G.D.D. acknowledges support from a fellowship of the German Academic Exchange Service (DAAD).

\end{acknowledgments} 

\appendix

\section{\texorpdfstring{Operators in Majorana Basis}{Operators in Majorana Basis}} \label{sec:appb}

In the Majorana basis, the relevant operators take the form 
\begin{equation}
\setlength{\arraycolsep}{2pt}
\mathbb H_{\mathrm{Majo}}(t) = \frac{i}{4}
\begin{pmatrix}
0 & 0 & 0 & -\tau & 0 & 0 & 0 & \cdots \\
0 & 0 & \tau & 0 & 0 & 0 & 0 & \cdots \\
0 & -\tau & 0 & 0 & 0 & -\tau' & 0 & \cdots \\
\tau & 0 & 0 & 0 & \tau' & 0 & 0 & \cdots \\
0 & 0 & 0 & -\tau' & 0 & 0 & 0 & \cdots \\
0 & 0 & \tau' & 0 & 0 & 0 & \tau & \cdots \\
0 & 0 & 0 & 0 & 0 & -\tau & 0 & \cdots \\
\vdots & \vdots & \vdots & \vdots & \vdots & \vdots & \vdots & \ddots
\end{pmatrix} ,
\end{equation}
with $\tau \equiv \tau(t) = J + 2V\cos{(\omega t)}$, $\tau' \equiv \tau'(t) = J' - 2V\cos{(\omega t)}$, and 
$
s_{\ell,j} = \delta_{\ell,j}
$, 
obtained from the mapping in Eqs.~(\ref{eq:H_traf}) and (\ref{eq:X_traf}). Note that $\mathbb H_{\mathrm{Majo}}(t + T) = \mathbb H_{\mathrm{Majo}}(t)$ with $T = 2\pi / \omega$.

\section{\texorpdfstring{Expression of $\braa{z_{\ell'}(t)}$ in Terms of Floquet Modes $\kett{\Phi_{\eta}(t)}$}{Expression of (z(t)| in Terms of Floquet Modes |Phi(t))}} \label{sec:app_FBM}
Here we derive Eq.~(\ref{eq:zdash}) of the main text.
Recall that 
\begin{equation}
\begin{aligned}
\braa{z_{\ell'}(t)}
&= \sum_\ell \int_0^\infty \dd \tau \,
   \Gamma_{\ell'\ell}^I(\tau)\,
   \braa{f_\ell(t-\tau,t)} ,
\end{aligned}
\label{eq:app_zdef}
\end{equation}
where $\braa{f_\ell(t-\tau,t)}$ can be decomposed in terms of the orthogonal Floquet modes
\be
\begin{aligned}
\braa{f_\ell(t-\tau,t)}
&=
\sum_\gamma
e^{+i\varepsilon_\gamma\tau}\,
\brakett{s_\ell}{\Phi_{\gamma}(t-\tau)}\,\!
\braa{\Phi_{\gamma}^{*}(t)}
\\
&\quad
{}+
e^{-i\varepsilon_\gamma\tau}\,
\brakett{s_\ell}{\Phi_{\gamma}^{*}(t-\tau)}\,\!
\braa{\Phi_{\gamma}(t)}.
\end{aligned}
\ee
We express the Floquet modes in terms of their Fourier components,
\begin{equation*}
\begin{aligned}
\kett{\Phi_{\eta}(t)} &= \sum_{K \in \mathbb{Z}} e^{-iK \omega t}\kett{\Phi_{\eta}(K)} , \\
\kett{\Phi^{*}_{\eta}(t)} &= \sum_{K \in \mathbb{Z}} e^{+iK \omega t}\kett{\Phi^{*}_{\eta}(K)} ,
\end{aligned}
\end{equation*}
with $\kett{\Phi^{*}_{\eta}(K)}=\kett{\Phi_{\eta}(K)}^{*}$.
Introducing the projectors
\be
\begin{aligned}
\Pi_\eta(K,Q) &= \kett{\Phi_\eta(K)} \braa{\Phi_\eta^*(Q)}, \\
\Pi_\eta^*(Q,K) &= \kett{\Phi_\eta^*(Q)} \braa{\Phi_\eta(K)},
\end{aligned}
\ee
then leads to the decomposition
\be
\begin{aligned}
& \braa{f_\ell(t-\tau,t)}=  \sum_\gamma \sum_{Q,K \in \mathbb{Z}} e^{-i(K-Q)\omega t} \\
& \times \Bigl[ e^{-i(-[\varepsilon_\gamma+Q\omega]) \tau}
\braa{s_\ell}\Pi_\gamma(K,Q) \\
&\quad {}+ e^{-i(\varepsilon_\gamma+K\omega) \tau} \braa{s_\ell}\Pi_\gamma^*(Q,K) \Bigr].
\end{aligned}
\ee
Using the Fourier transform of the spectral functions
\be
\tilde\Gamma_{\ell'\ell}^{I}(\varepsilon_\gamma + N\omega)
= \frac{1}{2\pi} \int_{-\infty}^\infty \mathrm{d}\tau\, 
\Gamma_{\ell'\ell}^{I}(\tau)\,e^{-i(\varepsilon_\gamma + N\omega)\tau} ,
\ee
neglecting the principal-value (Lamb-shift) contribution as in Sec.~\ref{sec:open_qs}, and introducing the shorthand notation $\Delta^{N}_{\gamma }=\varepsilon_\gamma + N\omega$, we finally arrive at
\begin{equation}
\begin{split}
& \braa{z_{\ell'}(t)}
= \pi \sum_\ell^{4L} \sum_\gamma^{2L} \sum_{Q,K \in \mathbb{Z}}
   e^{-i(K-Q) \omega t} \braa{s_\ell}\\ 
   & \times \Bigl[\tilde{\Gamma}_{{\ell'}\ell}^I(-\Delta^{Q}_{\gamma})
   \Pi_\gamma(K,Q) + 
   \tilde{\Gamma}_{{\ell'}\ell}^I(+\Delta^{K}_{\gamma})
   \Pi_\gamma^*(Q,K) \Bigr], 
\end{split}
\label{eq:app_zdash}
\end{equation}
where the time dependence is now explicit. This is Eq.~(\ref{eq:zdash}) of the main text, below which its physical interpretation is discussed.

\section{Correlations of Majorana Fermions}\label{sec:app_correlations}

\begin{figure*}
    \centering
    \includegraphics[width=0.9\textwidth]{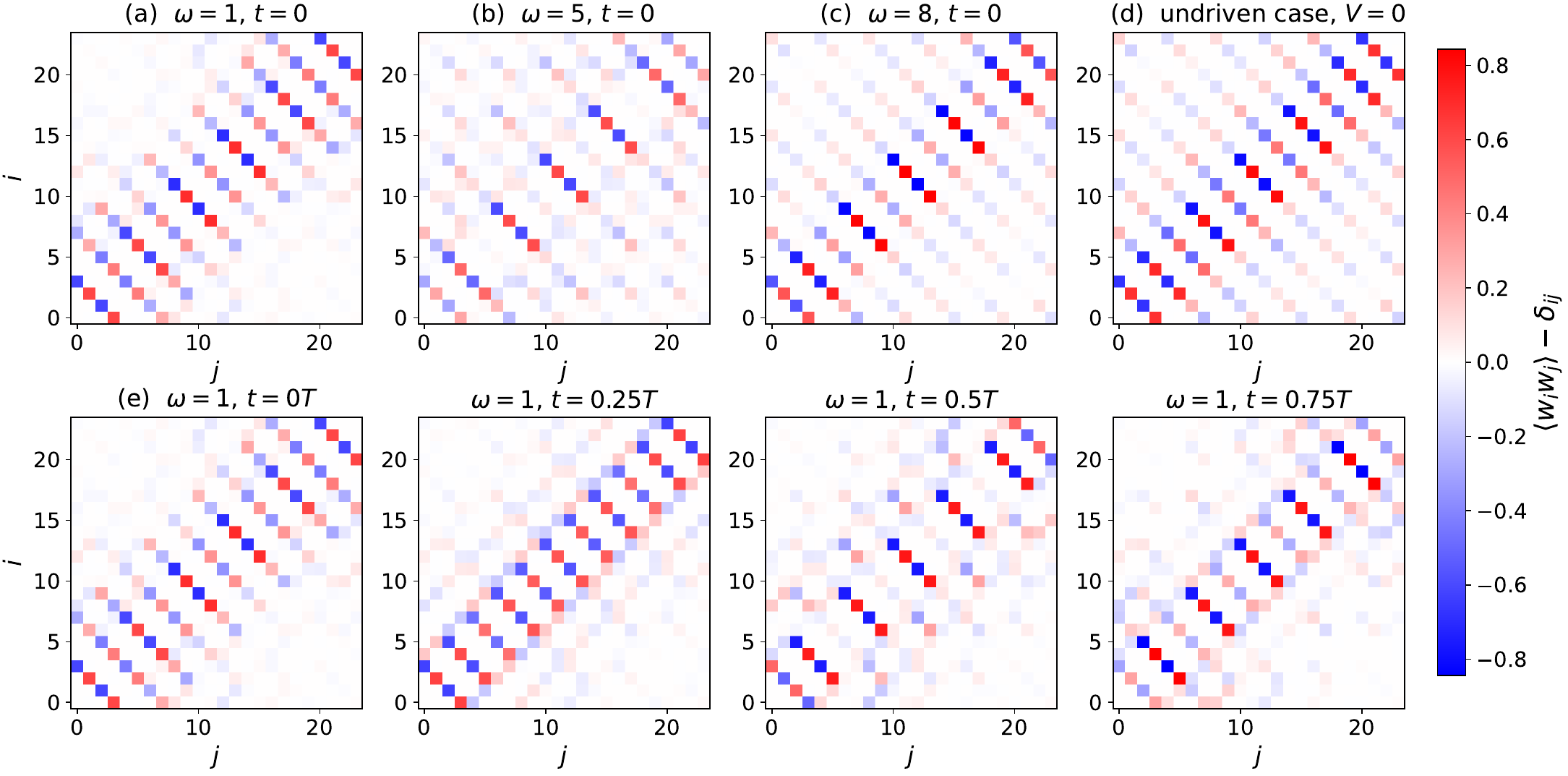}
    \caption{Behaviour of the steady-state correlations between Majorana sites across a driven and undriven dissipative SSH chain. 
    (a)--(c) Correlation matrix $\Im[W_{\text{NESS}}(0)-\mathds{1}] = - C_{\text{NESS}}(0)$ evaluated at time $t=0$  for a chain of length $L = 6$, and same remaining parameters as in Fig.~\ref{img:test1}, for $\omega = 1$ (a), $\omega = 5$ (b), and $\omega = 8$ (c). (d) $W_{\text{NESS}}(t)-\mathds{1}$ for the undriven case $V=0$.
    (e) $W_{\text{NESS}}(t)-\mathds{1}$ for $\omega = 1$ and $t \in \{0, 0.25T, 0.5T, 0.75T\}$, illustrating the effect of micromotion and indicating that analysing the system at only stroboscopic times can miss micromotion contributions. The NESS remains structured and correlated.}
    \label{fig:covallFBZ_app} %
\end{figure*}

In Fig.~\ref{fig:covallFBZ_app}(a)--(c), we show the correlations $W_{jk} \equiv \langle w_j w_k \rangle  $ between Majorana quasi-particles at different sites $w_j, w_k$ by plotting the correlation matrix $ \Im[W_{\text{NESS}}(0)-\mathds{1}] = - C_{\text{NESS}}(0)$ (noting that $W_{\text{NESS}}(0)-\mathds{1} = - iC_{\text{NESS}}(0)$ is purely imaginary)
at driving frequencies (a) $\omega = 1$, (b) $\omega = 5$, and (c) $\omega = 8$.  Here, we show the correlations at micromotion time $t=0$ in the steady state, i.e.~at late physical time, when the system is relaxed and the remaining dynamics is solely due to the micromotion. 
We observe that different short- and long-range correlations between Majorana quasi-particles from different (even non-neighbouring) unit cells can be engineered through the periodic modulation, as seen in the different structures observed in Fig.~\ref{fig:covallFBZ_app}(a)--(c).  
For high frequencies, Fig.~\ref{fig:covallFBZ_app}(c), the full correlation matrix shows its dominating correlations in a distinct \mbox{(sign-)alternating} pattern in proximity to the main diagonal, which implies that short-range correlations (nearest and next-nearest neighbour) rather than long-range correlations are dominantly present between Majorana pairs. Here, the system approaches a time-averaged static high-frequency limit in which $-C_{\text{NESS}}(t)$ resembles (though is not identical to) the static case, shown in Fig.~\ref{fig:covallFBZ_app}(d), consistent with our expectation from Floquet theory \cite{Eckardt_2015,EckardtColl}.  
Moreover, in Fig.~\ref{fig:covallFBZ_app}(e)  we show the micromotion of $- C_{\text{NESS}}(t)$ for fixed $\omega = 1$ and $t \in \{0, 0.25T, 0.5T, 0.75T\}$. We observe that it substantially modifies the covariances, indicating that analysing the system at only stroboscopic times can miss micromotion contributions. The figure demonstrates that the NESS remains structured and correlated.

\section{\texorpdfstring{Independence of $\mathbb{X}(t)$ from $\beta$ and $\mu$ for the Considered Driven-Dissipative System}{Independence of X(t) from beta and mu for the Considered Driven-Dissipative System}} \label{sec:app_independence}

In the following, we show that for our system--bath correlation functions the operator $\mathbb{X}(t)$ is blind to the chemical potentials $\mu_{A,B}$ and inverse temperatures $\beta_{A,B}$ of the baths. Recall that $\mathbb X(t)
= 4\bigl( i  \mathbb H_\mathrm{Majo}(t) + \mathbb M_r(t) \bigr)$, where ${\mathbb M_r}(t)=(\mathbb M(t)+\mathbb M^*(t))/2$. We only need to show that ${\mathbb M_r}(t)$  is independent of $\mu_{A,B}$ and $\beta_{A,B}$ since $\mathbb H_\mathrm{Majo}(t)$ is the system Hamiltonian and thus independent of bath parameters. Recall that
\begin{equation}
\mathbb M(t) = \sum_{\ell'} \kett{s_{\ell'}} \braa{z_{\ell'}(t)} , 
\end{equation}
where the bath operator $\braa{z_{\ell'}(t)}$ decomposes as
\begin{equation} 
\begin{split}
& \braa{z_{\ell'}(t)}
= \pi \sum_\ell^{4L} \sum_\gamma^{2L} \sum_{Q,K \in \mathbb{Z}}
   e^{-i(K-Q) \omega t} \\ 
   & \times \Bigl[\tilde{\Gamma}_{{\ell'}\ell}^I(-\Delta^{Q}_{\gamma})
   \braa{s_\ell}\Pi_\gamma(K,Q) \\
   & \quad + \tilde{\Gamma}_{{\ell'}\ell}^I(+\Delta^{K}_{\gamma})
   \braa{s_\ell}\Pi_\gamma^*(Q,K) \Bigr] .
\end{split}
\end{equation}
Plugging $\braa{z_{\ell'}(t)}$ into $\mathbb M(t)$ yields
\begin{align}
\mathbb M(t) &= \pi \sum_{\ell',\ell} \sum_\gamma \sum_{Q,K \in \mathbb{Z}}
   e^{-i(K-Q) \omega t} \nonumber \\
   & \times \Bigl[\tilde{\Gamma}_{{\ell'}\ell}^I(-\Delta^{Q}_{\gamma})
   \, \kett{s_{\ell'}}\braa{s_\ell}\Pi_\gamma(K,Q) \nonumber \\
   & \hspace{4mm} + \tilde{\Gamma}_{{\ell'}\ell}^I(+\Delta^{K}_{\gamma})
   \, \kett{s_{\ell'}}\braa{s_\ell}\Pi_\gamma^*(Q,K) \Bigr] .
\end{align}
Taking the complex conjugate of $\mathbb M(t)$ and using that the
$\kett{s_\ell}$ are real together with
$[\Pi_\gamma(K,Q)]^* = \Pi^*_\gamma(K,Q)$, we obtain
\begin{align}
\mathbb M^*(t) &= \pi \sum_{\ell',\ell} \sum_\gamma \sum_{Q,K \in \mathbb{Z}}
   e^{+i(K-Q) \omega t} \nonumber \\
   & \times \Bigl[\bigl[\tilde{\Gamma}_{{\ell'}\ell}^I(-\Delta^{Q}_{\gamma})\bigr]^*
   \, \kett{s_{\ell'}}\braa{s_\ell}\Pi^*_\gamma(K,Q) \nonumber \\
   & \hspace{4mm} + \bigl[\tilde{\Gamma}_{{\ell'}\ell}^I(+\Delta^{K}_{\gamma})\bigr]^*
   \, \kett{s_{\ell'}}\braa{s_\ell}\Pi_\gamma(Q,K) \Bigr] .
\end{align}
Since $Q$ and $K$ are summed over independently, we may relabel
$K \leftrightarrow Q$, which returns the phase factor to
$e^{-i(K-Q)\omega t}$ and brings each term onto the same projector as one of
the two terms in $\mathbb M(t)$, i.e.
\begin{align}
\mathbb M^*(t) &= \pi \sum_{\ell',\ell} \sum_\gamma \sum_{Q,K \in \mathbb{Z}}
   e^{-i(K-Q) \omega t} \nonumber \\
   & \times \Bigl[\bigl[\tilde{\Gamma}_{{\ell'}\ell}^I(-\Delta^{K}_{\gamma})\bigr]^*
   \, \kett{s_{\ell'}}\braa{s_\ell}\Pi^*_\gamma(Q,K) \nonumber \\
   & \hspace{4mm} + \bigl[\tilde{\Gamma}_{{\ell'}\ell}^I(+\Delta^{Q}_{\gamma})\bigr]^*
   \, \kett{s_{\ell'}}\braa{s_\ell}\Pi_\gamma(K,Q) \Bigr] .
\end{align}
Adding the two expressions, the spectral functions combine pairwise on each
projector and we find
\begin{align}
& \mathbb M(t) + \mathbb M^*(t)
 = \pi \sum_{\ell',\ell} \sum_\gamma \sum_{Q,K \in \mathbb{Z}}
    e^{-i(K-Q) \omega t} \, \kett{s_{\ell'}}\braa{s_\ell} \nonumber \\
 & \times \Bigl[
   \Bigl(\tilde{\Gamma}_{{\ell'}\ell}^I(-\Delta^{Q}_{\gamma})
   + \bigl[\tilde{\Gamma}_{{\ell'}\ell}^I(+\Delta^{Q}_{\gamma})\bigr]^*\Bigr)
   \Pi_\gamma(K,Q) \nonumber \\
 & \hspace{4mm} + \Bigl(\tilde{\Gamma}_{{\ell'}\ell}^I(+\Delta^{K}_{\gamma})
   + \bigl[\tilde{\Gamma}_{{\ell'}\ell}^I(-\Delta^{K}_{\gamma})\bigr]^*\Bigr)
   \Pi^*_\gamma(Q,K) \Bigr] .
\end{align}
Note that both brackets are of the form
$\tilde{\Gamma}^I(\Omega) + \bigl[\tilde{\Gamma}^I(-\Omega)\bigr]^*$, with
$\Omega = -\Delta^{Q}_{\gamma}$ and $\Omega = +\Delta^{K}_{\gamma}$
respectively.
Now recall that 
\begin{equation}
\tilde\Gamma(\Omega)
= \mathrm{diag}\bigl(\tilde\Gamma_A(\Omega), \tilde\Gamma_B(\Omega),
\tilde\Gamma_A(\Omega), \tilde\Gamma_B(\Omega), \dots\bigr),
\end{equation}
with two-by-two matrices
\begin{equation}
\tilde\Gamma_I(\Omega)
= \kappa^2 g_I(\Omega)\bigl[f_+(\Omega)
\mathds{1} - f_-(\Omega)\sigma_y\bigr],
\end{equation} 
and
$
f_{\pm}(\Omega) \equiv n_{I}(\Omega) \pm \big(1 - n_{I}(-\Omega)\big) 
$.
The Fermi--Dirac distribution of reservoir $\mathcal{R}_I$ is $n_{I}(\Omega) = (e^{\beta_{I}(\Omega-\mu_{I})}+1)^{-1}$.
Using the identities
\begin{equation}
f_+(\Omega) + f_+(-\Omega) = 2 , \qquad f_-(-\Omega) = f_-(\Omega) ,
\end{equation}
and the fact that  $g_I(\Omega)$ is even, i.e.\ $g_I(-\Omega)=g_I(\Omega)$, one finds
\begin{equation}
\tilde{\Gamma}^I_{\ell'\ell}(\Omega) + \bigl[\tilde{\Gamma}^I_{\ell'\ell}(-\Omega)\bigr]^*
= 2 \kappa^2 g   \delta_{\ell'\ell} ,
\end{equation}
with $g_I(\Omega)\equiv g=\mathrm{const.}$, such that the Fermi functions cancel out. Combining all results, we find
\begin{align}
 \mathbb M_r(t) 
& = \pi  \kappa^2 g \sum_{\ell} \sum_\gamma \sum_{Q,K \in \mathbb{Z}}
    e^{-i(K-Q) \omega t} \, \kett{s_{\ell}}\braa{s_\ell} \nonumber \\
 & \times \Bigl[
   \Pi_\gamma(K,Q) +  \Pi^*_\gamma(Q,K) \Bigr] .
\end{align}
Since the Floquet modes $\kett{\Phi_\gamma(t)}$ together with their conjugate partners
$\kett{\Phi^*_\gamma(t)}$ form a complete basis, i.e.\
$\sum_\gamma \bigl[\kett{\Phi_\gamma(t)}\braa{\Phi^*_\gamma(t)} + \kett{\Phi^*_\gamma(t)}\braa{\Phi_\gamma(t)}\bigr] = \mathds{1}$,
summing over the Fourier components we find
\be
\begin{split} 
&\sum_\gamma^{2L} \sum_{Q,K \in \mathbb{Z}}
    e^{-i(K-Q) \omega t} \\
&\quad\times \Bigl[
   \Pi_\gamma(K,Q) +  \Pi^*_\gamma(Q,K) \Bigr] = \mathds{1} .
\end{split}
\ee
Thus, the real part of the bath matrix finally reads
\be
 \mathbb M_r(t) =  \mathbb M_r = \pi  \kappa^2 g \sum_{\ell} \kett{s_{\ell}}\braa{s_\ell} ,
\ee
which yields 
\be
\mathbb X(t)
= 4\Bigl[ i  \mathbb H_\mathrm{Majo}(t) + \pi  \kappa^2 g \sum_{\ell} \kett{s_{\ell}}\braa{s_\ell} \Bigr] ,
\ee
which is fully independent of $\beta_{A,B}$ and $\mu_{A,B}$. Note that this result need not hold for arbitrary system--bath correlation functions. The derivation also leverages that $g_I(\Omega)$ is even and that Lamb shifts are neglected. 

For the system--bath coupling $s_{\ell,j} = \delta_{\ell,j}$ considered in this work, we have 
$
\sum_{\ell}^{4L} \kett{s_{\ell}}\braa{s_\ell} = \mathds{1} 
$,
such that the real part of the bath matrix reads as 
$
 \mathbb M_r =  \pi  \kappa^2 g \mathds{1} 
$,
and we finally find 
\be
\mathbb X(t)
 = 4\Bigl[ i  \mathbb H_\mathrm{Majo}(t) + \pi  \kappa^2 g \mathds{1}  \Bigr] ,
\ee
which has decay rates $\Re (\beta_j)=4\pi\kappa^2 g$.
\section{\texorpdfstring{Quantisation of $\phi_{\mathrm{EGP}}(t)$ at All Times in the Symmetric Regime}{Quantisation of the EGP at All Times in the Symmetric Regime}} \label{sec:app_quantisation}
\begin{figure}
    \centering
    \includegraphics[width=0.47\textwidth]{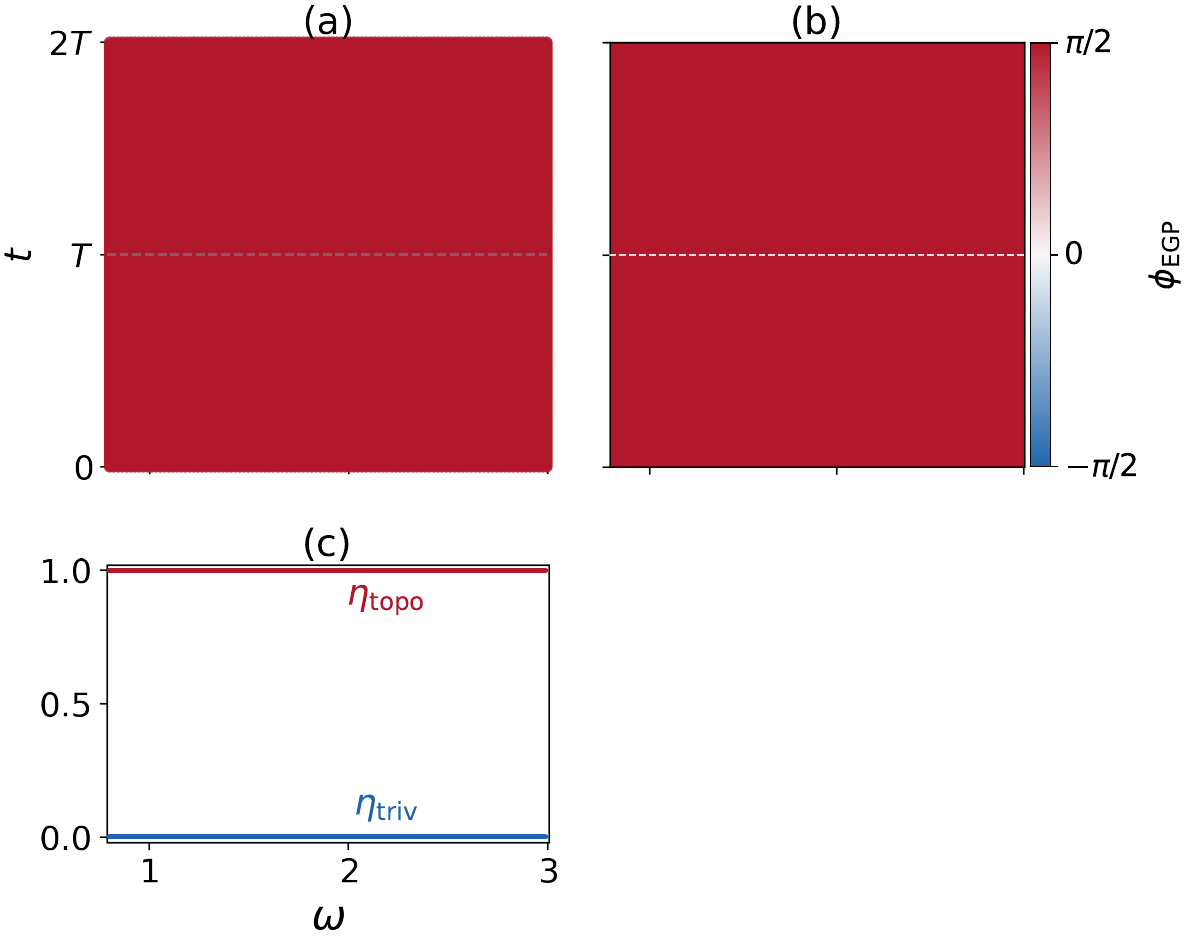}
    \caption{Erased intra-period topological switching with a static dissipator. Same diagnostics and parameters as in Fig.~\ref{img:test3}(a)--(c), with the difference that the impact of the drive on the dissipator is neglected and we instead use the bath-operator construction given by Eq.~(\ref{eq:z_spec_stat}). The intra-period switching physics is erased, and the topologically nontrivial and trivial fractions are fixed to values $(\eta_{\mathrm{triv}},\eta_{\mathrm{topo}}) = (0,1) \ \forall \omega$ (cf. (c)). The figure demonstrates that the correct treatment of the dissipator within the Floquet--Born--Markov approach is essential for capturing the intra-period topological switching.}
    \label{img:test4}
\end{figure}
Here we give the details of the derivation in Sec.~\ref{sec:res_sym}. Suppose the bath temperatures $\beta$ are finite and consider the following transformation~\cite{Molignini_2023}
\be
f_{n,A} = -\tilde{f}_{n,A}^{\dagger}, \qquad f_{n,B} = \tilde{f}_{n,B}^{\dagger} , 
\ee
which relies on the chiral symmetry of the model, as it acts differently on the A and B sublattices. By defining 
$
\Lambda^{\mathrm{cell}} = \operatorname{diag}(-1,+1,+1,-1) 
$, 
the transformation can be written in terms of Majorana operators $w_j$ as 
\be
\begin{bmatrix}
\tilde{w}_{2n-1,A} \\
\tilde{w}_{2n,A} \\
\tilde{w}_{2n-1,B} \\
\tilde{w}_{2n,B}
\end{bmatrix}
=
\Lambda^{\mathrm{cell}}
\begin{bmatrix}
w_{2n-1,A} \\
w_{2n,A} \\
w_{2n-1,B} \\
w_{2n,B}
\end{bmatrix}.
\ee
The transformation across the whole chain is given by
\be
\Lambda = \bigoplus_{j=1}^L \Lambda^{\mathrm{cell}} . 
\ee
Regardless of $\mu$ and $\beta$, at all times $t$ we find 
\be
\Lambda \mathbb{X}(t) \Lambda = \mathbb{X}(t) , \qquad \Lambda \mathcal{K}_{1,2} \Lambda = - \mathcal{K}_{1,2} .
\ee
Further, if $\mu=0$, we also have
\be
\Lambda \mathbb{Y}(t) \Lambda = \mathbb{Y}(t) , 
\ee
which, in turn, results in the symmetry being inherited by the steady-state covariance matrix $C_{\text{NESS}}(t)$ \cite{Bardyn_2013}, 
\be
\Lambda C_{\text{NESS}}(t) \Lambda = C_{\text{NESS}}(t) ,
\ee
and its inverse $C^{-1}_{\mathrm{NESS}}(t)$. We assume $C_{\text{NESS}}(t)$ to be symmetric under $\Lambda$ for the following. 
Consider the factors in $\phi_{\mathrm{EGP}}(t)=\arg(Z(t))$, with
\begin{equation}
Z(t) = c\,\Omega\,\mathrm{Pf}\bigl(C_{\mathrm{NESS}}(t)\bigr)\Bigl[ \mathrm{Pf}\bigl(\mathcal{N}_{1}(t)\bigr)
- \mathrm{Pf}\bigl(\mathcal{N}_{2}(t)\bigr)
\Bigr],
\end{equation} 
where we have defined 
$
\mathcal{N}_{1,2}(t) = \mathcal{K}_{1,2} - C^{-1}_{\mathrm{NESS}}(t)  
$. 
Importantly, we find that
\be
\mathcal{N}_{1,2}^*(t) = - \mathcal{K}_{1,2} - C^{-1}_{\mathrm{NESS}}(t) = \Lambda \mathcal{N}_{1,2}(t) \Lambda ,  \label{eq:app_ident}
\ee
i.e.\ complex conjugation has the same effect on $\mathcal{N}_{1,2}$ as the chosen transformation $\Lambda$. Taking the Pfaffian on both sides
\be
\mathrm{Pf}\bigl(\mathcal{N}_{1,2}^*(t)\bigr) =  \mathrm{Pf}\bigl(\Lambda \mathcal{N}_{1,2}(t) \Lambda \bigr), 
\ee
and using known Pfaffian identities ($\mathrm{Pf}(A^*) = \mathrm{Pf}(A)^*$ and $\mathrm{Pf}(RAR^T)=\operatorname{det}(R)\,\mathrm{Pf}(A)$ for a square matrix $R$), and noting that since $\det(\Lambda^{\mathrm{cell}})=+1$, one has $\det(\Lambda)=+1$, we find 
\begin{align}
&\mathrm{Pf}\bigl(\mathcal{N}_{1,2}(t)\bigr)^*
= \operatorname{det}(\Lambda)\,
\mathrm{Pf}\bigl(\mathcal{N}_{1,2}(t)\bigr)
= +1\,\mathrm{Pf}\bigl(\mathcal{N}_{1,2}(t)\bigr)
\nonumber\\
&\Longrightarrow\quad
\mathrm{Pf}\bigl(\mathcal{N}_{1,2}(t)\bigr) \in \mathbb{R},  \quad \forall t \in [0,T] .
\end{align}
Further, since $C_{\mathrm{NESS}}(t) \in \mathbb{R}$, we have $\mathrm{Pf}\bigl(C_{\mathrm{NESS}}(t)\bigr) \in \mathbb{R}$, as well as $\Omega \in \mathbb{R}$, such that so far all the factors except $c$ are real. Consider 
\be
c = \exp{ \Bigl(\frac{i \pi}{2}(2L+3) \Bigr)} . 
\ee
Since $2L+3$ is always odd we find that $c^* = -c$. Combining all results, we thus find
\be
Z^*(t) = - Z(t) \Longrightarrow  Z(t) = \left|Z(t)\right|
 e^{\pm i \pi/2} , 
\ee
which implies the quantisation of the EGP 
\be
\phi_{\mathrm{EGP}}(t) = \frac{\pi}{2}
\ \mathrm{mod}\,\pi \quad \forall t \in [0,T],
\ee
provided $Z(t) \neq 0$.
Note that having both $\mu\neq0$ and $\beta\neq0$ breaks the symmetry and the quantisation of the EGP, as it breaks the identification with the complex conjugate in Eq.~(\ref{eq:app_ident}).

\section{\texorpdfstring{Static Dissipator}{Static Dissipator}} \label{sec:stat_diss}

Here we show the full diagnostics for the static-dissipator case discussed in Sec.~\ref{sec:res_sym}. Fig.~\ref{img:test4} shows the same quantities as Fig.~\ref{img:test3}(a)--(c), but with the impact of the drive on the dissipator neglected, using the bath-operator construction of Eq.~(\ref{eq:z_spec_stat}). In contrast to the Floquet--Born--Markov treatment, the purity gap never closes, and $0$-purity edge states are present at all points $(\omega,t)$ in (a). Correspondingly, $\phi_\mathrm{EGP}=+\pi/2$ throughout (b), so that $\eta_\mathrm{topo}=1$ and $\eta_\mathrm{triv}=0$ for all $\omega$ in (c).
\begin{figure}
    \centering
    \includegraphics[width=0.48\textwidth]{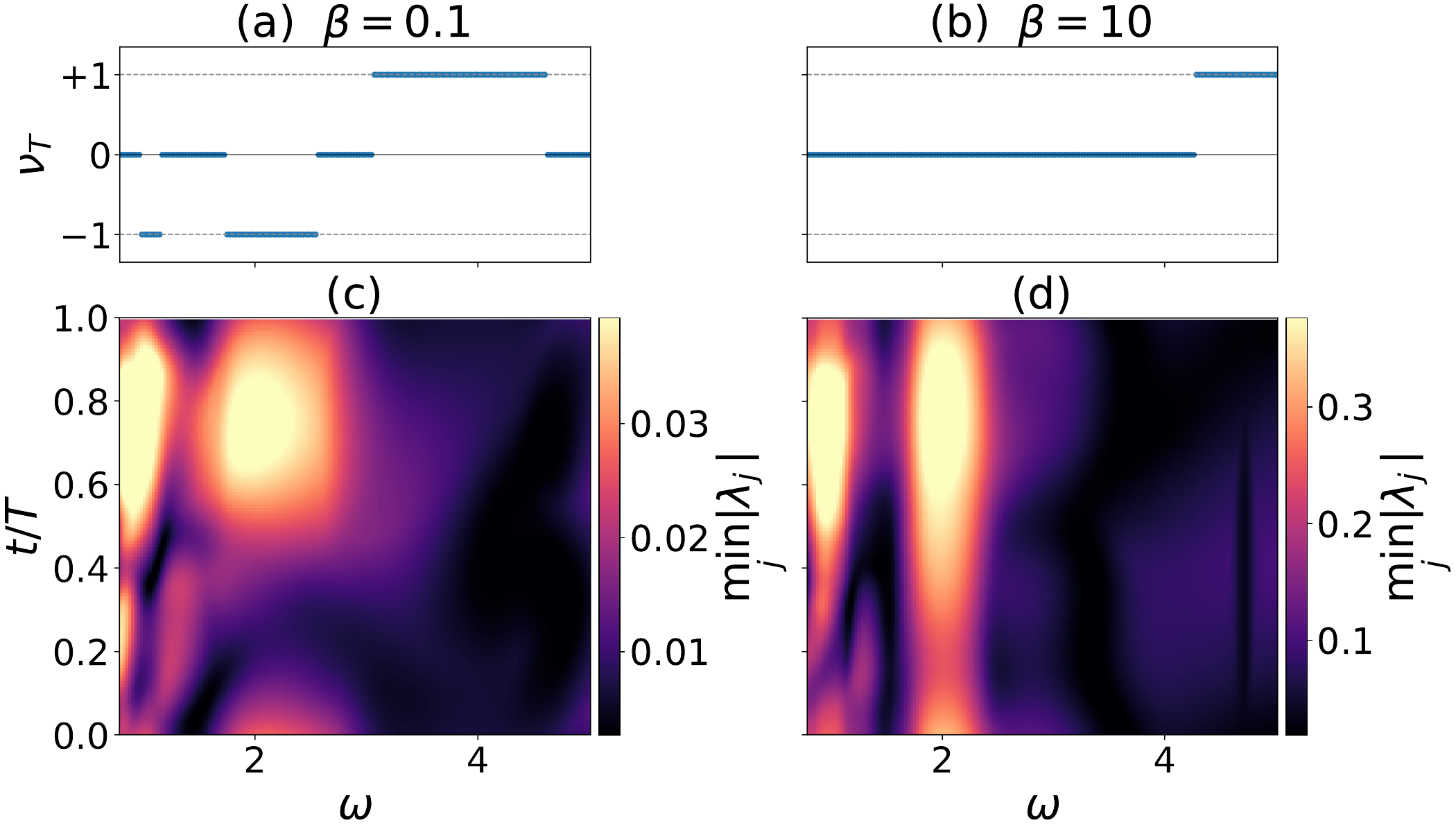}
    \caption{Decoupling of the winding number $\nu_T$ and $0$-purity states in the purity spectrum $\lambda_j$, with $L=10$ and same remaining parameters as in Fig.~\ref{img:test51}.~(a) and (b) show $\nu_T$ against $\omega$ for $\beta=0.1$ and $\beta=10$, respectively.~(c) and (d) magnitude of the lowest purity band $\min_j|\lambda_j|$ (OBC) over $(\omega,t)$ for $\beta=0.1$ and $\beta=10$, respectively.}
    \label{img:windingpurity}
\end{figure}
\section{\texorpdfstring{Decoupling of the Winding Number $\nu_T$ and $0$-Purity States in the Purity Spectrum $\lambda_j$}{Decoupling of the Winding Number and 0-Purity States in the Purity Spectrum}} \label{sec:nu_purity} 

Since the purity spectrum characterises $C_{\mathrm{NESS}}(t)$ at a single
instant $t$, whereas $\nu_T$ depends on the full closed trajectory of $Z(t)$
over the period, we expect $\nu_T$ and the existence of $0$-purity states in the OBC steady-state purity spectrum to not be directly correlated. We show this in Fig.~\ref{img:windingpurity} for $L=10$ and the same remaining parameters as in Fig.~\ref{img:test51}. In (a) and (b) we show $\nu_T$ against $\omega$ for $\beta=0.1$ and $\beta=10$, respectively. In (c) and (d) we plot the magnitude of the lowest purity band $\min_j|\lambda_j|$ (OBC) over $(\omega,t)$ for $\beta=0.1$ and $\beta=10$, respectively. No direct connection between nontrivial values of $\nu_T$ and $0$-purity states in the purity spectra is visible. We find no bulk-boundary correspondence between $\nu_T$ and the instantaneous OBC purity spectrum, as well as the OBC quasipurity spectrum, in our numerics.

\bibliography{apssamp}%

\begin{thebibliography}{67}%
\makeatletter
\providecommand \@ifxundefined [1]{%
 \@ifx{#1\undefined}
}%
\providecommand \@ifnum [1]{%
 \ifnum #1\expandafter \@firstoftwo
 \else \expandafter \@secondoftwo
 \fi
}%
\providecommand \@ifx [1]{%
 \ifx #1\expandafter \@firstoftwo
 \else \expandafter \@secondoftwo
 \fi
}%
\providecommand \natexlab [1]{#1}%
\providecommand \enquote  [1]{``#1''}%
\providecommand \bibnamefont  [1]{#1}%
\providecommand \bibfnamefont [1]{#1}%
\providecommand \citenamefont [1]{#1}%
\providecommand \href@noop [0]{\@secondoftwo}%
\providecommand \href [0]{\begingroup \@sanitize@url \@href}%
\providecommand \@href[1]{\@@startlink{#1}\@@href}%
\providecommand \@@href[1]{\endgroup#1\@@endlink}%
\providecommand \@sanitize@url [0]{\catcode `\\12\catcode `\$12\catcode `\&12\catcode `\#12\catcode `\^12\catcode `\_12\catcode `\%12\relax}%
\providecommand \@@startlink[1]{}%
\providecommand \@@endlink[0]{}%
\providecommand \url  [0]{\begingroup\@sanitize@url \@url }%
\providecommand \@url [1]{\endgroup\@href {#1}{\urlprefix }}%
\providecommand \urlprefix  [0]{URL }%
\providecommand \Eprint [0]{\href }%
\providecommand \doibase [0]{https://doi.org/}%
\providecommand \selectlanguage [0]{\@gobble}%
\providecommand \bibinfo  [0]{\@secondoftwo}%
\providecommand \bibfield  [0]{\@secondoftwo}%
\providecommand \translation [1]{[#1]}%
\providecommand \BibitemOpen [0]{}%
\providecommand \bibitemStop [0]{}%
\providecommand \bibitemNoStop [0]{.\EOS\space}%
\providecommand \EOS [0]{\spacefactor3000\relax}%
\providecommand \BibitemShut  [1]{\csname bibitem#1\endcsname}%
\let\auto@bib@innerbib\@empty
\bibitem [{\citenamefont {Ozawa}\ \emph {et~al.}(2019)\citenamefont {Ozawa}, \citenamefont {Price}, \citenamefont {Amo}, \citenamefont {Goldman}, \citenamefont {Hafezi}, \citenamefont {Lu}, \citenamefont {Rechtsman}, \citenamefont {Schuster}, \citenamefont {Simon}, \citenamefont {Zilberberg},\ and\ \citenamefont {Carusotto}}]{Ozawa2019}%
  \BibitemOpen
  \bibfield  {author} {\bibinfo {author} {\bibfnamefont {T.}~\bibnamefont {Ozawa}}, \bibinfo {author} {\bibfnamefont {H.~M.}\ \bibnamefont {Price}}, \bibinfo {author} {\bibfnamefont {A.}~\bibnamefont {Amo}}, \bibinfo {author} {\bibfnamefont {N.}~\bibnamefont {Goldman}}, \bibinfo {author} {\bibfnamefont {M.}~\bibnamefont {Hafezi}}, \bibinfo {author} {\bibfnamefont {L.}~\bibnamefont {Lu}}, \bibinfo {author} {\bibfnamefont {M.~C.}\ \bibnamefont {Rechtsman}}, \bibinfo {author} {\bibfnamefont {D.}~\bibnamefont {Schuster}}, \bibinfo {author} {\bibfnamefont {J.}~\bibnamefont {Simon}}, \bibinfo {author} {\bibfnamefont {O.}~\bibnamefont {Zilberberg}},\ and\ \bibinfo {author} {\bibfnamefont {I.}~\bibnamefont {Carusotto}},\ }\bibfield  {title} {\bibinfo {title} {Topological photonics},\ }\href {https://doi.org/10.1103/RevModPhys.91.015006} {\bibfield  {journal} {\bibinfo  {journal} {Rev. Mod. Phys.}\ }\textbf {\bibinfo {volume} {91}},\ \bibinfo {pages} {015006} (\bibinfo {year} {2019})}\BibitemShut {NoStop}%
\bibitem [{\citenamefont {Bernevig}\ \emph {et~al.}(2025)\citenamefont {Bernevig}, \citenamefont {Fu}, \citenamefont {Ju}, \citenamefont {MacDonald}, \citenamefont {Mak},\ and\ \citenamefont {Shan}}]{Bernevig2025}%
  \BibitemOpen
  \bibfield  {author} {\bibinfo {author} {\bibfnamefont {B.~A.}\ \bibnamefont {Bernevig}}, \bibinfo {author} {\bibfnamefont {L.}~\bibnamefont {Fu}}, \bibinfo {author} {\bibfnamefont {L.}~\bibnamefont {Ju}}, \bibinfo {author} {\bibfnamefont {A.~H.}\ \bibnamefont {MacDonald}}, \bibinfo {author} {\bibfnamefont {K.~F.}\ \bibnamefont {Mak}},\ and\ \bibinfo {author} {\bibfnamefont {J.}~\bibnamefont {Shan}},\ }\bibfield  {title} {\bibinfo {title} {Fractional quantization in insulators from {H}all to {C}hern},\ }\href {https://doi.org/10.1038/s41567-025-03072-8} {\bibfield  {journal} {\bibinfo  {journal} {Nat. Phys.}\ }\textbf {\bibinfo {volume} {21}},\ \bibinfo {pages} {1702} (\bibinfo {year} {2025})}\BibitemShut {NoStop}%
\bibitem [{\citenamefont {Altland}\ and\ \citenamefont {Simons}(2010)}]{Altland2010}%
  \BibitemOpen
  \bibfield  {author} {\bibinfo {author} {\bibfnamefont {A.}~\bibnamefont {Altland}}\ and\ \bibinfo {author} {\bibfnamefont {B.~D.}\ \bibnamefont {Simons}},\ }\href {https://doi.org/10.1017/CBO9780511789984} {\emph {\bibinfo {title} {Condensed Matter Field Theory}}},\ \bibinfo {edition} {2nd}\ ed.\ (\bibinfo  {publisher} {Cambridge University Press},\ \bibinfo {address} {Cambridge},\ \bibinfo {year} {2010})\BibitemShut {NoStop}%
\bibitem [{\citenamefont {Batra}\ and\ \citenamefont {Sheet}(2020)}]{Batra2020}%
  \BibitemOpen
  \bibfield  {author} {\bibinfo {author} {\bibfnamefont {N.}~\bibnamefont {Batra}}\ and\ \bibinfo {author} {\bibfnamefont {G.}~\bibnamefont {Sheet}},\ }\bibfield  {title} {\bibinfo {title} {Physics with coffee and doughnuts},\ }\href {https://doi.org/10.1007/s12045-020-0995-x} {\bibfield  {journal} {\bibinfo  {journal} {Resonance}\ }\textbf {\bibinfo {volume} {25}},\ \bibinfo {pages} {765} (\bibinfo {year} {2020})}\BibitemShut {NoStop}%
\bibitem [{\citenamefont {Zak}(1989)}]{Zak1989}%
  \BibitemOpen
  \bibfield  {author} {\bibinfo {author} {\bibfnamefont {J.}~\bibnamefont {Zak}},\ }\bibfield  {title} {\bibinfo {title} {{B}erry's phase for energy bands in solids},\ }\href {https://doi.org/10.1103/PhysRevLett.62.2747} {\bibfield  {journal} {\bibinfo  {journal} {Phys. Rev. Lett.}\ }\textbf {\bibinfo {volume} {62}},\ \bibinfo {pages} {2747} (\bibinfo {year} {1989})}\BibitemShut {NoStop}%
\bibitem [{\citenamefont {Delplace}\ \emph {et~al.}(2011)\citenamefont {Delplace}, \citenamefont {Ullmo},\ and\ \citenamefont {Montambaux}}]{Delplace2011}%
  \BibitemOpen
  \bibfield  {author} {\bibinfo {author} {\bibfnamefont {P.}~\bibnamefont {Delplace}}, \bibinfo {author} {\bibfnamefont {D.}~\bibnamefont {Ullmo}},\ and\ \bibinfo {author} {\bibfnamefont {G.}~\bibnamefont {Montambaux}},\ }\bibfield  {title} {\bibinfo {title} {{Z}ak phase and the existence of edge states in graphene},\ }\href {https://doi.org/10.1103/PhysRevB.84.195452} {\bibfield  {journal} {\bibinfo  {journal} {Phys. Rev. B}\ }\textbf {\bibinfo {volume} {84}},\ \bibinfo {pages} {195452} (\bibinfo {year} {2011})}\BibitemShut {NoStop}%
\bibitem [{\citenamefont {Aihara}\ \emph {et~al.}(2020)\citenamefont {Aihara}, \citenamefont {Hirayama},\ and\ \citenamefont {Murakami}}]{Aihara2020}%
  \BibitemOpen
  \bibfield  {author} {\bibinfo {author} {\bibfnamefont {Y.}~\bibnamefont {Aihara}}, \bibinfo {author} {\bibfnamefont {M.}~\bibnamefont {Hirayama}},\ and\ \bibinfo {author} {\bibfnamefont {S.}~\bibnamefont {Murakami}},\ }\bibfield  {title} {\bibinfo {title} {Anomalous dielectric response in insulators with the $\pi$ {Z}ak phase},\ }\href {https://doi.org/10.1103/PhysRevResearch.2.033224} {\bibfield  {journal} {\bibinfo  {journal} {Phys. Rev. Res.}\ }\textbf {\bibinfo {volume} {2}},\ \bibinfo {pages} {033224} (\bibinfo {year} {2020})}\BibitemShut {NoStop}%
\bibitem [{\citenamefont {Xiao}\ \emph {et~al.}(2014)\citenamefont {Xiao}, \citenamefont {Zhang},\ and\ \citenamefont {Chan}}]{Xiao2014}%
  \BibitemOpen
  \bibfield  {author} {\bibinfo {author} {\bibfnamefont {M.}~\bibnamefont {Xiao}}, \bibinfo {author} {\bibfnamefont {Z.~Q.}\ \bibnamefont {Zhang}},\ and\ \bibinfo {author} {\bibfnamefont {C.~T.}\ \bibnamefont {Chan}},\ }\bibfield  {title} {\bibinfo {title} {Surface impedance and bulk band geometric phases in one-dimensional systems},\ }\href {https://doi.org/10.1103/PhysRevX.4.021017} {\bibfield  {journal} {\bibinfo  {journal} {Phys. Rev. X}\ }\textbf {\bibinfo {volume} {4}},\ \bibinfo {pages} {021017} (\bibinfo {year} {2014})}\BibitemShut {NoStop}%
\bibitem [{\citenamefont {Wang}\ \emph {et~al.}(2016)\citenamefont {Wang}, \citenamefont {Xiao}, \citenamefont {Liu}, \citenamefont {Zhu},\ and\ \citenamefont {Chan}}]{Wang2016}%
  \BibitemOpen
  \bibfield  {author} {\bibinfo {author} {\bibfnamefont {Q.}~\bibnamefont {Wang}}, \bibinfo {author} {\bibfnamefont {M.}~\bibnamefont {Xiao}}, \bibinfo {author} {\bibfnamefont {H.}~\bibnamefont {Liu}}, \bibinfo {author} {\bibfnamefont {S.}~\bibnamefont {Zhu}},\ and\ \bibinfo {author} {\bibfnamefont {C.~T.}\ \bibnamefont {Chan}},\ }\bibfield  {title} {\bibinfo {title} {Measurement of the {Z}ak phase of photonic bands through the interface states of a metasurface/photonic crystal},\ }\href {https://doi.org/10.1103/PhysRevB.93.041415} {\bibfield  {journal} {\bibinfo  {journal} {Phys. Rev. B}\ }\textbf {\bibinfo {volume} {93}},\ \bibinfo {pages} {041415} (\bibinfo {year} {2016})}\BibitemShut {NoStop}%
\bibitem [{\citenamefont {Saei Ghareh~Naz}\ \emph {et~al.}(2018)\citenamefont {Saei Ghareh~Naz}, \citenamefont {Fulga}, \citenamefont {Ma}, \citenamefont {Schmidt},\ and\ \citenamefont {van~den Brink}}]{SaeiGharehNaz2018}%
  \BibitemOpen
  \bibfield  {author} {\bibinfo {author} {\bibfnamefont {E.}~\bibnamefont {Saei Ghareh~Naz}}, \bibinfo {author} {\bibfnamefont {I.~C.}\ \bibnamefont {Fulga}}, \bibinfo {author} {\bibfnamefont {L.}~\bibnamefont {Ma}}, \bibinfo {author} {\bibfnamefont {O.~G.}\ \bibnamefont {Schmidt}},\ and\ \bibinfo {author} {\bibfnamefont {J.}~\bibnamefont {van~den Brink}},\ }\bibfield  {title} {\bibinfo {title} {Topological phase transition in a stretchable photonic crystal},\ }\href {https://doi.org/10.1103/PhysRevA.98.033830} {\bibfield  {journal} {\bibinfo  {journal} {Phys. Rev. A}\ }\textbf {\bibinfo {volume} {98}},\ \bibinfo {pages} {033830} (\bibinfo {year} {2018})}\BibitemShut {NoStop}%
\bibitem [{\citenamefont {Malkova}\ \emph {et~al.}(2009)\citenamefont {Malkova}, \citenamefont {Hromada}, \citenamefont {Wang}, \citenamefont {Bryant},\ and\ \citenamefont {Chen}}]{Malkova2009}%
  \BibitemOpen
  \bibfield  {author} {\bibinfo {author} {\bibfnamefont {N.}~\bibnamefont {Malkova}}, \bibinfo {author} {\bibfnamefont {I.}~\bibnamefont {Hromada}}, \bibinfo {author} {\bibfnamefont {X.}~\bibnamefont {Wang}}, \bibinfo {author} {\bibfnamefont {G.}~\bibnamefont {Bryant}},\ and\ \bibinfo {author} {\bibfnamefont {Z.}~\bibnamefont {Chen}},\ }\bibfield  {title} {\bibinfo {title} {Observation of optical {S}hockley-like surface states in photonic superlattices},\ }\href {https://doi.org/10.1364/OL.34.001633} {\bibfield  {journal} {\bibinfo  {journal} {Opt. Lett.}\ }\textbf {\bibinfo {volume} {34}},\ \bibinfo {pages} {1633} (\bibinfo {year} {2009})}\BibitemShut {NoStop}%
\bibitem [{\citenamefont {Queralt{\'o}}\ \emph {et~al.}(2020)\citenamefont {Queralt{\'o}}, \citenamefont {Kremer}, \citenamefont {Maczewsky}, \citenamefont {Heinrich}, \citenamefont {Mompart}, \citenamefont {Ahufinger},\ and\ \citenamefont {Szameit}}]{Queralto2020}%
  \BibitemOpen
  \bibfield  {author} {\bibinfo {author} {\bibfnamefont {G.}~\bibnamefont {Queralt{\'o}}}, \bibinfo {author} {\bibfnamefont {M.}~\bibnamefont {Kremer}}, \bibinfo {author} {\bibfnamefont {L.~J.}\ \bibnamefont {Maczewsky}}, \bibinfo {author} {\bibfnamefont {M.}~\bibnamefont {Heinrich}}, \bibinfo {author} {\bibfnamefont {J.}~\bibnamefont {Mompart}}, \bibinfo {author} {\bibfnamefont {V.}~\bibnamefont {Ahufinger}},\ and\ \bibinfo {author} {\bibfnamefont {A.}~\bibnamefont {Szameit}},\ }\bibfield  {title} {\bibinfo {title} {Topological state engineering via supersymmetric transformations},\ }\href {https://doi.org/10.1038/s42005-020-0316-4} {\bibfield  {journal} {\bibinfo  {journal} {Commun. Phys.}\ }\textbf {\bibinfo {volume} {3}},\ \bibinfo {pages} {49} (\bibinfo {year} {2020})}\BibitemShut {NoStop}%
\bibitem [{\citenamefont {Lee}\ \emph {et~al.}(2018)\citenamefont {Lee}, \citenamefont {Imhof}, \citenamefont {Berger}, \citenamefont {Bayer}, \citenamefont {Brehm}, \citenamefont {Molenkamp}, \citenamefont {Kiessling},\ and\ \citenamefont {Thomale}}]{Lee2018}%
  \BibitemOpen
  \bibfield  {author} {\bibinfo {author} {\bibfnamefont {C.~H.}\ \bibnamefont {Lee}}, \bibinfo {author} {\bibfnamefont {S.}~\bibnamefont {Imhof}}, \bibinfo {author} {\bibfnamefont {C.}~\bibnamefont {Berger}}, \bibinfo {author} {\bibfnamefont {F.}~\bibnamefont {Bayer}}, \bibinfo {author} {\bibfnamefont {J.}~\bibnamefont {Brehm}}, \bibinfo {author} {\bibfnamefont {L.~W.}\ \bibnamefont {Molenkamp}}, \bibinfo {author} {\bibfnamefont {T.}~\bibnamefont {Kiessling}},\ and\ \bibinfo {author} {\bibfnamefont {R.}~\bibnamefont {Thomale}},\ }\bibfield  {title} {\bibinfo {title} {Topolectrical circuits},\ }\href {https://doi.org/10.1038/s42005-018-0035-2} {\bibfield  {journal} {\bibinfo  {journal} {Commun. Phys.}\ }\textbf {\bibinfo {volume} {1}},\ \bibinfo {pages} {39} (\bibinfo {year} {2018})}\BibitemShut {NoStop}%
\bibitem [{\citenamefont {Atala}\ \emph {et~al.}(2013)\citenamefont {Atala}, \citenamefont {Aidelsburger}, \citenamefont {Barreiro}, \citenamefont {Abanin}, \citenamefont {Kitagawa}, \citenamefont {Demler},\ and\ \citenamefont {Bloch}}]{Atala2013}%
  \BibitemOpen
  \bibfield  {author} {\bibinfo {author} {\bibfnamefont {M.}~\bibnamefont {Atala}}, \bibinfo {author} {\bibfnamefont {M.}~\bibnamefont {Aidelsburger}}, \bibinfo {author} {\bibfnamefont {J.~T.}\ \bibnamefont {Barreiro}}, \bibinfo {author} {\bibfnamefont {D.}~\bibnamefont {Abanin}}, \bibinfo {author} {\bibfnamefont {T.}~\bibnamefont {Kitagawa}}, \bibinfo {author} {\bibfnamefont {E.}~\bibnamefont {Demler}},\ and\ \bibinfo {author} {\bibfnamefont {I.}~\bibnamefont {Bloch}},\ }\bibfield  {title} {\bibinfo {title} {Direct measurement of the {Z}ak phase in topological {B}loch bands},\ }\href {https://doi.org/10.1038/nphys2790} {\bibfield  {journal} {\bibinfo  {journal} {Nat. Phys.}\ }\textbf {\bibinfo {volume} {9}},\ \bibinfo {pages} {795} (\bibinfo {year} {2013})}\BibitemShut {NoStop}%
\bibitem [{\citenamefont {St-Jean}\ \emph {et~al.}(2017)\citenamefont {St-Jean}, \citenamefont {Goblot}, \citenamefont {Galopin}, \citenamefont {Lema{\^i}tre}, \citenamefont {Ozawa}, \citenamefont {Le~Gratiet}, \citenamefont {Sagnes}, \citenamefont {Bloch},\ and\ \citenamefont {Amo}}]{St-Jean2017}%
  \BibitemOpen
  \bibfield  {author} {\bibinfo {author} {\bibfnamefont {P.}~\bibnamefont {St-Jean}}, \bibinfo {author} {\bibfnamefont {V.}~\bibnamefont {Goblot}}, \bibinfo {author} {\bibfnamefont {E.}~\bibnamefont {Galopin}}, \bibinfo {author} {\bibfnamefont {A.}~\bibnamefont {Lema{\^i}tre}}, \bibinfo {author} {\bibfnamefont {T.}~\bibnamefont {Ozawa}}, \bibinfo {author} {\bibfnamefont {L.}~\bibnamefont {Le~Gratiet}}, \bibinfo {author} {\bibfnamefont {I.}~\bibnamefont {Sagnes}}, \bibinfo {author} {\bibfnamefont {J.}~\bibnamefont {Bloch}},\ and\ \bibinfo {author} {\bibfnamefont {A.}~\bibnamefont {Amo}},\ }\bibfield  {title} {\bibinfo {title} {Lasing in topological edge states of a one-dimensional lattice},\ }\href {https://doi.org/10.1038/s41566-017-0006-2} {\bibfield  {journal} {\bibinfo  {journal} {Nat. Photonics}\ }\textbf {\bibinfo {volume} {11}},\ \bibinfo {pages} {651} (\bibinfo {year} {2017})}\BibitemShut {NoStop}%
\bibitem [{\citenamefont {Solnyshkov}\ \emph {et~al.}(2016)\citenamefont {Solnyshkov}, \citenamefont {Nalitov},\ and\ \citenamefont {Malpuech}}]{Solnyshkov2016}%
  \BibitemOpen
  \bibfield  {author} {\bibinfo {author} {\bibfnamefont {D.~D.}\ \bibnamefont {Solnyshkov}}, \bibinfo {author} {\bibfnamefont {A.~V.}\ \bibnamefont {Nalitov}},\ and\ \bibinfo {author} {\bibfnamefont {G.}~\bibnamefont {Malpuech}},\ }\bibfield  {title} {\bibinfo {title} {{K}ibble-{Z}urek mechanism in topologically nontrivial zigzag chains of polariton micropillars},\ }\href {https://doi.org/10.1103/PhysRevLett.116.046402} {\bibfield  {journal} {\bibinfo  {journal} {Phys. Rev. Lett.}\ }\textbf {\bibinfo {volume} {116}},\ \bibinfo {pages} {046402} (\bibinfo {year} {2016})}\BibitemShut {NoStop}%
\bibitem [{\citenamefont {Yu}\ \emph {et~al.}(2026)\citenamefont {Yu}, \citenamefont {Jin}, \citenamefont {Ren}, \citenamefont {Mandal}, \citenamefont {Zhang},\ and\ \citenamefont {Su}}]{Yu2025}%
  \BibitemOpen
  \bibfield  {author} {\bibinfo {author} {\bibfnamefont {Z.}~\bibnamefont {Yu}}, \bibinfo {author} {\bibfnamefont {F.}~\bibnamefont {Jin}}, \bibinfo {author} {\bibfnamefont {J.}~\bibnamefont {Ren}}, \bibinfo {author} {\bibfnamefont {S.}~\bibnamefont {Mandal}}, \bibinfo {author} {\bibfnamefont {B.}~\bibnamefont {Zhang}},\ and\ \bibinfo {author} {\bibfnamefont {R.}~\bibnamefont {Su}},\ }\bibfield  {title} {\bibinfo {title} {Topological exciton-polaritons with negative coupling},\ }\href {https://doi.org/10.1038/s41467-025-68025-4} {\bibfield  {journal} {\bibinfo  {journal} {Nat. Commun.}\ }\textbf {\bibinfo {volume} {17}},\ \bibinfo {pages} {1269} (\bibinfo {year} {2026})}\BibitemShut {NoStop}%
\bibitem [{\citenamefont {Eckardt}(2017)}]{EckardtColl}%
  \BibitemOpen
  \bibfield  {author} {\bibinfo {author} {\bibfnamefont {A.}~\bibnamefont {Eckardt}},\ }\bibfield  {title} {\bibinfo {title} {Colloquium: Atomic quantum gases in periodically driven optical lattices},\ }\href {https://doi.org/10.1103/RevModPhys.89.011004} {\bibfield  {journal} {\bibinfo  {journal} {Rev. Mod. Phys.}\ }\textbf {\bibinfo {volume} {89}},\ \bibinfo {pages} {011004} (\bibinfo {year} {2017})}\BibitemShut {NoStop}%
\bibitem [{\citenamefont {Eckardt}\ and\ \citenamefont {Anisimovas}(2015)}]{Eckardt_2015}%
  \BibitemOpen
  \bibfield  {author} {\bibinfo {author} {\bibfnamefont {A.}~\bibnamefont {Eckardt}}\ and\ \bibinfo {author} {\bibfnamefont {E.}~\bibnamefont {Anisimovas}},\ }\bibfield  {title} {\bibinfo {title} {High-frequency approximation for periodically driven quantum systems from a {F}loquet-space perspective},\ }\href {https://doi.org/10.1088/1367-2630/17/9/093039} {\bibfield  {journal} {\bibinfo  {journal} {New J. Phys.}\ }\textbf {\bibinfo {volume} {17}},\ \bibinfo {pages} {093039} (\bibinfo {year} {2015})}\BibitemShut {NoStop}%
\bibitem [{\citenamefont {Nathan}\ and\ \citenamefont {Rudner}(2015)}]{Nathan_2015}%
  \BibitemOpen
  \bibfield  {author} {\bibinfo {author} {\bibfnamefont {F.}~\bibnamefont {Nathan}}\ and\ \bibinfo {author} {\bibfnamefont {M.~S.}\ \bibnamefont {Rudner}},\ }\bibfield  {title} {\bibinfo {title} {Topological singularities and the general classification of {F}loquet--{B}loch systems},\ }\href {https://doi.org/10.1088/1367-2630/17/12/125014} {\bibfield  {journal} {\bibinfo  {journal} {New J. Phys.}\ }\textbf {\bibinfo {volume} {17}},\ \bibinfo {pages} {125014} (\bibinfo {year} {2015})}\BibitemShut {NoStop}%
\bibitem [{\citenamefont {Kitagawa}\ \emph {et~al.}(2010)\citenamefont {Kitagawa}, \citenamefont {Berg}, \citenamefont {Rudner},\ and\ \citenamefont {Demler}}]{Kitagawa_2010}%
  \BibitemOpen
  \bibfield  {author} {\bibinfo {author} {\bibfnamefont {T.}~\bibnamefont {Kitagawa}}, \bibinfo {author} {\bibfnamefont {E.}~\bibnamefont {Berg}}, \bibinfo {author} {\bibfnamefont {M.}~\bibnamefont {Rudner}},\ and\ \bibinfo {author} {\bibfnamefont {E.}~\bibnamefont {Demler}},\ }\bibfield  {title} {\bibinfo {title} {Topological characterization of periodically driven quantum systems},\ }\href {https://doi.org/10.1103/PhysRevB.82.235114} {\bibfield  {journal} {\bibinfo  {journal} {Phys. Rev. B}\ }\textbf {\bibinfo {volume} {82}},\ \bibinfo {pages} {235114} (\bibinfo {year} {2010})}\BibitemShut {NoStop}%
\bibitem [{\citenamefont {Wintersperger}\ \emph {et~al.}(2020)\citenamefont {Wintersperger}, \citenamefont {Braun}, \citenamefont {{\"U}nal}, \citenamefont {Eckardt}, \citenamefont {Di~Liberto}, \citenamefont {Goldman}, \citenamefont {Bloch},\ and\ \citenamefont {Aidelsburger}}]{Wintersperger2020a}%
  \BibitemOpen
  \bibfield  {author} {\bibinfo {author} {\bibfnamefont {K.}~\bibnamefont {Wintersperger}}, \bibinfo {author} {\bibfnamefont {C.}~\bibnamefont {Braun}}, \bibinfo {author} {\bibfnamefont {F.~N.}\ \bibnamefont {{\"U}nal}}, \bibinfo {author} {\bibfnamefont {A.}~\bibnamefont {Eckardt}}, \bibinfo {author} {\bibfnamefont {M.}~\bibnamefont {Di~Liberto}}, \bibinfo {author} {\bibfnamefont {N.}~\bibnamefont {Goldman}}, \bibinfo {author} {\bibfnamefont {I.}~\bibnamefont {Bloch}},\ and\ \bibinfo {author} {\bibfnamefont {M.}~\bibnamefont {Aidelsburger}},\ }\bibfield  {title} {\bibinfo {title} {Realization of an anomalous {F}loquet topological system with ultracold atoms},\ }\href@noop {} {\bibfield  {journal} {\bibinfo  {journal} {Nat. Phys.}\ }\textbf {\bibinfo {volume} {16}},\ \bibinfo {pages} {1058} (\bibinfo {year} {2020})}\BibitemShut {NoStop}%
\bibitem [{\citenamefont {Schnell}\ \emph {et~al.}(2024)\citenamefont {Schnell}, \citenamefont {Weitenberg},\ and\ \citenamefont {Eckardt}}]{Schnell2024a}%
  \BibitemOpen
  \bibfield  {author} {\bibinfo {author} {\bibfnamefont {A.}~\bibnamefont {Schnell}}, \bibinfo {author} {\bibfnamefont {C.}~\bibnamefont {Weitenberg}},\ and\ \bibinfo {author} {\bibfnamefont {A.}~\bibnamefont {Eckardt}},\ }\bibfield  {title} {\bibinfo {title} {Dissipative preparation of a {F}loquet topological insulator in an optical lattice via bath engineering},\ }\href {https://doi.org/10.21468/SciPostPhys.17.2.052} {\bibfield  {journal} {\bibinfo  {journal} {SciPost Phys.}\ }\textbf {\bibinfo {volume} {17}},\ \bibinfo {pages} {052} (\bibinfo {year} {2024})}\BibitemShut {NoStop}%
\bibitem [{\citenamefont {Rudner}\ \emph {et~al.}(2013)\citenamefont {Rudner}, \citenamefont {Lindner}, \citenamefont {Berg},\ and\ \citenamefont {Levin}}]{Rudner_2013}%
  \BibitemOpen
  \bibfield  {author} {\bibinfo {author} {\bibfnamefont {M.~S.}\ \bibnamefont {Rudner}}, \bibinfo {author} {\bibfnamefont {N.~H.}\ \bibnamefont {Lindner}}, \bibinfo {author} {\bibfnamefont {E.}~\bibnamefont {Berg}},\ and\ \bibinfo {author} {\bibfnamefont {M.}~\bibnamefont {Levin}},\ }\bibfield  {title} {\bibinfo {title} {Anomalous edge states and the bulk-edge correspondence for periodically driven two-dimensional systems},\ }\href {https://doi.org/10.1103/PhysRevX.3.031005} {\bibfield  {journal} {\bibinfo  {journal} {Phys. Rev. X}\ }\textbf {\bibinfo {volume} {3}},\ \bibinfo {pages} {031005} (\bibinfo {year} {2013})}\BibitemShut {NoStop}%
\bibitem [{\citenamefont {Peralta~Gavensky}\ \emph {et~al.}(2025)\citenamefont {Peralta~Gavensky}, \citenamefont {Usaj},\ and\ \citenamefont {Goldman}}]{Gavensky2025}%
  \BibitemOpen
  \bibfield  {author} {\bibinfo {author} {\bibfnamefont {L.}~\bibnamefont {Peralta~Gavensky}}, \bibinfo {author} {\bibfnamefont {G.}~\bibnamefont {Usaj}},\ and\ \bibinfo {author} {\bibfnamefont {N.}~\bibnamefont {Goldman}},\ }\bibfield  {title} {\bibinfo {title} {{S}t{\v{r}}eda formula for {F}loquet systems: Topological invariants and quantized anomalies from {C}es{\`a}ro summation},\ }\href {https://doi.org/10.1103/b3pw-my97} {\bibfield  {journal} {\bibinfo  {journal} {Phys. Rev. X}\ }\textbf {\bibinfo {volume} {15}},\ \bibinfo {pages} {031067} (\bibinfo {year} {2025})}\BibitemShut {NoStop}%
\bibitem [{\citenamefont {Breuer}\ and\ \citenamefont {Petruccione}(2002)}]{breuer2002theory}%
  \BibitemOpen
  \bibfield  {author} {\bibinfo {author} {\bibfnamefont {H.-P.}\ \bibnamefont {Breuer}}\ and\ \bibinfo {author} {\bibfnamefont {F.}~\bibnamefont {Petruccione}},\ }\href@noop {} {\emph {\bibinfo {title} {The Theory of Open Quantum Systems}}}\ (\bibinfo  {publisher} {Oxford University Press},\ \bibinfo {address} {Oxford},\ \bibinfo {year} {2002})\BibitemShut {NoStop}%
\bibitem [{\citenamefont {Weiss}(2012)}]{Weiss2012}%
  \BibitemOpen
  \bibfield  {author} {\bibinfo {author} {\bibfnamefont {U.}~\bibnamefont {Weiss}},\ }\href {https://doi.org/10.1142/8334} {\emph {\bibinfo {title} {Quantum Dissipative Systems}}},\ \bibinfo {edition} {4th}\ ed.\ (\bibinfo  {publisher} {World Scientific},\ \bibinfo {address} {Singapore},\ \bibinfo {year} {2012})\BibitemShut {NoStop}%
\bibitem [{\citenamefont {Wawer}\ and\ \citenamefont {Fleischhauer}(2021{\natexlab{a}})}]{Wawer2021}%
  \BibitemOpen
  \bibfield  {author} {\bibinfo {author} {\bibfnamefont {L.}~\bibnamefont {Wawer}}\ and\ \bibinfo {author} {\bibfnamefont {M.}~\bibnamefont {Fleischhauer}},\ }\bibfield  {title} {\bibinfo {title} {Chern number and {B}erry curvature for {G}aussian mixed states of fermions},\ }\href {https://doi.org/10.1103/PhysRevB.104.094104} {\bibfield  {journal} {\bibinfo  {journal} {Phys. Rev. B}\ }\textbf {\bibinfo {volume} {104}},\ \bibinfo {pages} {094104} (\bibinfo {year} {2021}{\natexlab{a}})}\BibitemShut {NoStop}%
\bibitem [{\citenamefont {Hannukainen}\ \emph {et~al.}(2022)\citenamefont {Hannukainen}, \citenamefont {Mart{\'i}nez}, \citenamefont {Bardarson},\ and\ \citenamefont {Kvorning}}]{Hannukainen2022}%
  \BibitemOpen
  \bibfield  {author} {\bibinfo {author} {\bibfnamefont {J.~D.}\ \bibnamefont {Hannukainen}}, \bibinfo {author} {\bibfnamefont {M.~F.}\ \bibnamefont {Mart{\'i}nez}}, \bibinfo {author} {\bibfnamefont {J.~H.}\ \bibnamefont {Bardarson}},\ and\ \bibinfo {author} {\bibfnamefont {T.~K.}\ \bibnamefont {Kvorning}},\ }\bibfield  {title} {\bibinfo {title} {Local topological markers in odd spatial dimensions and their application to amorphous topological matter},\ }\href {https://doi.org/10.1103/PhysRevLett.129.277601} {\bibfield  {journal} {\bibinfo  {journal} {Phys. Rev. Lett.}\ }\textbf {\bibinfo {volume} {129}},\ \bibinfo {pages} {277601} (\bibinfo {year} {2022})}\BibitemShut {NoStop}%
\bibitem [{\citenamefont {Tran}\ \emph {et~al.}(2017)\citenamefont {Tran}, \citenamefont {Dauphin}, \citenamefont {Grushin}, \citenamefont {Zoller},\ and\ \citenamefont {Goldman}}]{Tran2017}%
  \BibitemOpen
  \bibfield  {author} {\bibinfo {author} {\bibfnamefont {D.~T.}\ \bibnamefont {Tran}}, \bibinfo {author} {\bibfnamefont {A.}~\bibnamefont {Dauphin}}, \bibinfo {author} {\bibfnamefont {A.~G.}\ \bibnamefont {Grushin}}, \bibinfo {author} {\bibfnamefont {P.}~\bibnamefont {Zoller}},\ and\ \bibinfo {author} {\bibfnamefont {N.}~\bibnamefont {Goldman}},\ }\bibfield  {title} {\bibinfo {title} {Probing topology by ``heating'': Quantized circular dichroism in ultracold atoms},\ }\href {https://doi.org/10.1126/sciadv.1701207} {\bibfield  {journal} {\bibinfo  {journal} {Sci. Adv.}\ }\textbf {\bibinfo {volume} {3}},\ \bibinfo {pages} {e1701207} (\bibinfo {year} {2017})}\BibitemShut {NoStop}%
\bibitem [{\citenamefont {Yang}\ \emph {et~al.}(2026{\natexlab{a}})\citenamefont {Yang}, \citenamefont {Han}, \citenamefont {Ning}, \citenamefont {Wu}, \citenamefont {Yang},\ and\ \citenamefont {Zheng}}]{yang2026}%
  \BibitemOpen
  \bibfield  {author} {\bibinfo {author} {\bibfnamefont {S.-B.}\ \bibnamefont {Yang}}, \bibinfo {author} {\bibfnamefont {P.-R.}\ \bibnamefont {Han}}, \bibinfo {author} {\bibfnamefont {W.}~\bibnamefont {Ning}}, \bibinfo {author} {\bibfnamefont {F.}~\bibnamefont {Wu}}, \bibinfo {author} {\bibfnamefont {Z.-B.}\ \bibnamefont {Yang}},\ and\ \bibinfo {author} {\bibfnamefont {S.-B.}\ \bibnamefont {Zheng}},\ }\href {https://arxiv.org/abs/2602.10831} {\bibinfo {title} {Mixed-state topology in non-{H}ermitian systems}} (\bibinfo {year} {2026}{\natexlab{a}}),\ \Eprint {https://arxiv.org/abs/2602.10831} {arXiv:2602.10831 [quant-ph]} \BibitemShut {NoStop}%
\bibitem [{\citenamefont {Yang}\ \emph {et~al.}(2026{\natexlab{b}})\citenamefont {Yang}, \citenamefont {Shi},\ and\ \citenamefont {Lee}}]{Yang2025}%
  \BibitemOpen
  \bibfield  {author} {\bibinfo {author} {\bibfnamefont {T.-H.}\ \bibnamefont {Yang}}, \bibinfo {author} {\bibfnamefont {B.}~\bibnamefont {Shi}},\ and\ \bibinfo {author} {\bibfnamefont {J.~Y.}\ \bibnamefont {Lee}},\ }\bibfield  {title} {\bibinfo {title} {Topological mixed states: Phases of matter from axiomatic approaches},\ }\href {https://doi.org/10.1103/3bpt-f9pd} {\bibfield  {journal} {\bibinfo  {journal} {Phys. Rev. X}\ }\textbf {\bibinfo {volume} {16}},\ \bibinfo {pages} {041002} (\bibinfo {year} {2026}{\natexlab{b}})}\BibitemShut {NoStop}%
\bibitem [{\citenamefont {Uhlmann}(1986)}]{Uhlmann_1986}%
  \BibitemOpen
  \bibfield  {author} {\bibinfo {author} {\bibfnamefont {A.}~\bibnamefont {Uhlmann}},\ }\bibfield  {title} {\bibinfo {title} {Parallel transport and ``quantum holonomy'' along density operators},\ }\href {https://doi.org/10.1016/0034-4877(86)90055-8} {\bibfield  {journal} {\bibinfo  {journal} {Rep. Math. Phys.}\ }\textbf {\bibinfo {volume} {24}},\ \bibinfo {pages} {229} (\bibinfo {year} {1986})}\BibitemShut {NoStop}%
\bibitem [{\citenamefont {Uhlmann}(1991)}]{Uhlmann_1991}%
  \BibitemOpen
  \bibfield  {author} {\bibinfo {author} {\bibfnamefont {A.}~\bibnamefont {Uhlmann}},\ }\bibfield  {title} {\bibinfo {title} {A gauge field governing parallel transport along mixed states},\ }\href {https://doi.org/10.1007/BF00420373} {\bibfield  {journal} {\bibinfo  {journal} {Lett. Math. Phys.}\ }\textbf {\bibinfo {volume} {21}},\ \bibinfo {pages} {229} (\bibinfo {year} {1991})}\BibitemShut {NoStop}%
\bibitem [{\citenamefont {Uhlmann}(1989)}]{Uhlmann_1989}%
  \BibitemOpen
  \bibfield  {author} {\bibinfo {author} {\bibfnamefont {A.}~\bibnamefont {Uhlmann}},\ }\bibfield  {title} {\bibinfo {title} {On {B}erry phases along mixtures of states},\ }\href {https://doi.org/10.1002/andp.19895010108} {\bibfield  {journal} {\bibinfo  {journal} {Ann. Phys. (Berlin)}\ }\textbf {\bibinfo {volume} {501}},\ \bibinfo {pages} {63} (\bibinfo {year} {1989})}\BibitemShut {NoStop}%
\bibitem [{\citenamefont {Viyuela}\ \emph {et~al.}(2014)\citenamefont {Viyuela}, \citenamefont {Rivas},\ and\ \citenamefont {Martin-Delgado}}]{Viyuela_2014}%
  \BibitemOpen
  \bibfield  {author} {\bibinfo {author} {\bibfnamefont {O.}~\bibnamefont {Viyuela}}, \bibinfo {author} {\bibfnamefont {A.}~\bibnamefont {Rivas}},\ and\ \bibinfo {author} {\bibfnamefont {M.~A.}\ \bibnamefont {Martin-Delgado}},\ }\bibfield  {title} {\bibinfo {title} {{U}hlmann phase as a topological measure for one-dimensional fermion systems},\ }\href {https://doi.org/10.1103/PhysRevLett.112.130401} {\bibfield  {journal} {\bibinfo  {journal} {Phys. Rev. Lett.}\ }\textbf {\bibinfo {volume} {112}},\ \bibinfo {pages} {130401} (\bibinfo {year} {2014})}\BibitemShut {NoStop}%
\bibitem [{\citenamefont {Hou}\ \emph {et~al.}(2023)\citenamefont {Hou}, \citenamefont {Wang}, \citenamefont {Zhou}, \citenamefont {Guo},\ and\ \citenamefont {Chien}}]{Hou_2023}%
  \BibitemOpen
  \bibfield  {author} {\bibinfo {author} {\bibfnamefont {X.-Y.}\ \bibnamefont {Hou}}, \bibinfo {author} {\bibfnamefont {X.}~\bibnamefont {Wang}}, \bibinfo {author} {\bibfnamefont {Z.}~\bibnamefont {Zhou}}, \bibinfo {author} {\bibfnamefont {H.}~\bibnamefont {Guo}},\ and\ \bibinfo {author} {\bibfnamefont {C.-C.}\ \bibnamefont {Chien}},\ }\bibfield  {title} {\bibinfo {title} {Geometric phases of mixed quantum states: A comparative study of interferometric and {U}hlmann phases},\ }\href {https://doi.org/10.1103/PhysRevB.107.165415} {\bibfield  {journal} {\bibinfo  {journal} {Phys. Rev. B}\ }\textbf {\bibinfo {volume} {107}},\ \bibinfo {pages} {165415} (\bibinfo {year} {2023})}\BibitemShut {NoStop}%
\bibitem [{\citenamefont {Bardyn}\ \emph {et~al.}(2018)\citenamefont {Bardyn}, \citenamefont {Wawer}, \citenamefont {Altland}, \citenamefont {Fleischhauer},\ and\ \citenamefont {Diehl}}]{Bardyn_2018}%
  \BibitemOpen
  \bibfield  {author} {\bibinfo {author} {\bibfnamefont {C.-E.}\ \bibnamefont {Bardyn}}, \bibinfo {author} {\bibfnamefont {L.}~\bibnamefont {Wawer}}, \bibinfo {author} {\bibfnamefont {A.}~\bibnamefont {Altland}}, \bibinfo {author} {\bibfnamefont {M.}~\bibnamefont {Fleischhauer}},\ and\ \bibinfo {author} {\bibfnamefont {S.}~\bibnamefont {Diehl}},\ }\bibfield  {title} {\bibinfo {title} {Probing the topology of density matrices},\ }\href {https://doi.org/10.1103/PhysRevX.8.011035} {\bibfield  {journal} {\bibinfo  {journal} {Phys. Rev. X}\ }\textbf {\bibinfo {volume} {8}},\ \bibinfo {pages} {011035} (\bibinfo {year} {2018})}\BibitemShut {NoStop}%
\bibitem [{\citenamefont {Unanyan}\ \emph {et~al.}(2020)\citenamefont {Unanyan}, \citenamefont {Kiefer-Emmanouilidis},\ and\ \citenamefont {Fleischhauer}}]{Unanyan2020}%
  \BibitemOpen
  \bibfield  {author} {\bibinfo {author} {\bibfnamefont {R.}~\bibnamefont {Unanyan}}, \bibinfo {author} {\bibfnamefont {M.}~\bibnamefont {Kiefer-Emmanouilidis}},\ and\ \bibinfo {author} {\bibfnamefont {M.}~\bibnamefont {Fleischhauer}},\ }\bibfield  {title} {\bibinfo {title} {Finite-temperature topological invariant for interacting systems},\ }\href {https://doi.org/10.1103/PhysRevLett.125.215701} {\bibfield  {journal} {\bibinfo  {journal} {Phys. Rev. Lett.}\ }\textbf {\bibinfo {volume} {125}},\ \bibinfo {pages} {215701} (\bibinfo {year} {2020})}\BibitemShut {NoStop}%
\bibitem [{\citenamefont {Molignini}\ and\ \citenamefont {Cooper}(2023)}]{Molignini_2023}%
  \BibitemOpen
  \bibfield  {author} {\bibinfo {author} {\bibfnamefont {P.}~\bibnamefont {Molignini}}\ and\ \bibinfo {author} {\bibfnamefont {N.~R.}\ \bibnamefont {Cooper}},\ }\bibfield  {title} {\bibinfo {title} {Topological phase transitions at finite temperature},\ }\href {https://doi.org/10.1103/PhysRevResearch.5.023004} {\bibfield  {journal} {\bibinfo  {journal} {Phys. Rev. Res.}\ }\textbf {\bibinfo {volume} {5}},\ \bibinfo {pages} {023004} (\bibinfo {year} {2023})}\BibitemShut {NoStop}%
\bibitem [{\citenamefont {Huang}\ \emph {et~al.}(2022)\citenamefont {Huang}, \citenamefont {Sun},\ and\ \citenamefont {Diehl}}]{Huang2022}%
  \BibitemOpen
  \bibfield  {author} {\bibinfo {author} {\bibfnamefont {Z.-M.}\ \bibnamefont {Huang}}, \bibinfo {author} {\bibfnamefont {X.-Q.}\ \bibnamefont {Sun}},\ and\ \bibinfo {author} {\bibfnamefont {S.}~\bibnamefont {Diehl}},\ }\bibfield  {title} {\bibinfo {title} {Topological gauge theory for mixed {D}irac stationary states in all dimensions},\ }\href {https://doi.org/10.1103/PhysRevB.106.245204} {\bibfield  {journal} {\bibinfo  {journal} {Phys. Rev. B}\ }\textbf {\bibinfo {volume} {106}},\ \bibinfo {pages} {245204} (\bibinfo {year} {2022})}\BibitemShut {NoStop}%
\bibitem [{\citenamefont {Mink}\ \emph {et~al.}(2019)\citenamefont {Mink}, \citenamefont {Fleischhauer},\ and\ \citenamefont {Unanyan}}]{Mink2019}%
  \BibitemOpen
  \bibfield  {author} {\bibinfo {author} {\bibfnamefont {C.~D.}\ \bibnamefont {Mink}}, \bibinfo {author} {\bibfnamefont {M.}~\bibnamefont {Fleischhauer}},\ and\ \bibinfo {author} {\bibfnamefont {R.}~\bibnamefont {Unanyan}},\ }\bibfield  {title} {\bibinfo {title} {Absence of topology in {G}aussian mixed states of bosons},\ }\href {https://doi.org/10.1103/PhysRevB.100.014305} {\bibfield  {journal} {\bibinfo  {journal} {Phys. Rev. B}\ }\textbf {\bibinfo {volume} {100}},\ \bibinfo {pages} {014305} (\bibinfo {year} {2019})}\BibitemShut {NoStop}%
\bibitem [{\citenamefont {Hannukainen}\ and\ \citenamefont {Cooper}(2026)}]{hannukainen2026}%
  \BibitemOpen
  \bibfield  {author} {\bibinfo {author} {\bibfnamefont {J.~D.}\ \bibnamefont {Hannukainen}}\ and\ \bibinfo {author} {\bibfnamefont {N.~R.}\ \bibnamefont {Cooper}},\ }\bibfield  {title} {\bibinfo {title} {Characterizing topology at nonzero temperature: Topological invariants and indicators in the extended {S}u-{S}chrieffer-{H}eeger model},\ }\href {https://doi.org/10.1103/pwkc-lqq6} {\bibfield  {journal} {\bibinfo  {journal} {Phys. Rev. Res.}\ }\textbf {\bibinfo {volume} {8}},\ \bibinfo {pages} {033080} (\bibinfo {year} {2026})}\BibitemShut {NoStop}%
\bibitem [{\citenamefont {Huang}\ and\ \citenamefont {Diehl}(2025)}]{Huang2025}%
  \BibitemOpen
  \bibfield  {author} {\bibinfo {author} {\bibfnamefont {Z.-M.}\ \bibnamefont {Huang}}\ and\ \bibinfo {author} {\bibfnamefont {S.}~\bibnamefont {Diehl}},\ }\bibfield  {title} {\bibinfo {title} {Interaction-induced topological phase transition at finite temperature},\ }\href {https://doi.org/10.1103/PhysRevLett.134.053002} {\bibfield  {journal} {\bibinfo  {journal} {Phys. Rev. Lett.}\ }\textbf {\bibinfo {volume} {134}},\ \bibinfo {pages} {053002} (\bibinfo {year} {2025})}\BibitemShut {NoStop}%
\bibitem [{\citenamefont {Wawer}\ and\ \citenamefont {Fleischhauer}(2021{\natexlab{b}})}]{Wawer2021_2}%
  \BibitemOpen
  \bibfield  {author} {\bibinfo {author} {\bibfnamefont {L.}~\bibnamefont {Wawer}}\ and\ \bibinfo {author} {\bibfnamefont {M.}~\bibnamefont {Fleischhauer}},\ }\bibfield  {title} {\bibinfo {title} {{$\mathbb{Z}_2$} topological invariants for mixed states of fermions in time-reversal invariant band structures},\ }\href {https://doi.org/10.1103/PhysRevB.104.214107} {\bibfield  {journal} {\bibinfo  {journal} {Phys. Rev. B}\ }\textbf {\bibinfo {volume} {104}},\ \bibinfo {pages} {214107} (\bibinfo {year} {2021}{\natexlab{b}})}\BibitemShut {NoStop}%
\bibitem [{\citenamefont {Asb\'oth}\ \emph {et~al.}(2014)\citenamefont {Asb\'oth}, \citenamefont {Tarasinski},\ and\ \citenamefont {Delplace}}]{Asboth_2014}%
  \BibitemOpen
  \bibfield  {author} {\bibinfo {author} {\bibfnamefont {J.~K.}\ \bibnamefont {Asb\'oth}}, \bibinfo {author} {\bibfnamefont {B.}~\bibnamefont {Tarasinski}},\ and\ \bibinfo {author} {\bibfnamefont {P.}~\bibnamefont {Delplace}},\ }\bibfield  {title} {\bibinfo {title} {Chiral symmetry and bulk-boundary correspondence in periodically driven one-dimensional systems},\ }\href {https://doi.org/10.1103/PhysRevB.90.125143} {\bibfield  {journal} {\bibinfo  {journal} {Phys. Rev. B}\ }\textbf {\bibinfo {volume} {90}},\ \bibinfo {pages} {125143} (\bibinfo {year} {2014})}\BibitemShut {NoStop}%
\bibitem [{\citenamefont {Su}\ \emph {et~al.}(1979)\citenamefont {Su}, \citenamefont {Schrieffer},\ and\ \citenamefont {Heeger}}]{su_1979}%
  \BibitemOpen
  \bibfield  {author} {\bibinfo {author} {\bibfnamefont {W.~P.}\ \bibnamefont {Su}}, \bibinfo {author} {\bibfnamefont {J.~R.}\ \bibnamefont {Schrieffer}},\ and\ \bibinfo {author} {\bibfnamefont {A.~J.}\ \bibnamefont {Heeger}},\ }\bibfield  {title} {\bibinfo {title} {Solitons in polyacetylene},\ }\href {https://doi.org/10.1103/PhysRevLett.42.1698} {\bibfield  {journal} {\bibinfo  {journal} {Phys. Rev. Lett.}\ }\textbf {\bibinfo {volume} {42}},\ \bibinfo {pages} {1698} (\bibinfo {year} {1979})}\BibitemShut {NoStop}%
\bibitem [{\citenamefont {Dal~Lago}\ \emph {et~al.}(2015)\citenamefont {Dal~Lago}, \citenamefont {Atala},\ and\ \citenamefont {Foa~Torres}}]{Lago_2015}%
  \BibitemOpen
  \bibfield  {author} {\bibinfo {author} {\bibfnamefont {V.}~\bibnamefont {Dal~Lago}}, \bibinfo {author} {\bibfnamefont {M.}~\bibnamefont {Atala}},\ and\ \bibinfo {author} {\bibfnamefont {L.~E.~F.}\ \bibnamefont {Foa~Torres}},\ }\bibfield  {title} {\bibinfo {title} {{F}loquet topological transitions in a driven one-dimensional topological insulator},\ }\href {https://doi.org/10.1103/PhysRevA.92.023624} {\bibfield  {journal} {\bibinfo  {journal} {Phys. Rev. A}\ }\textbf {\bibinfo {volume} {92}},\ \bibinfo {pages} {023624} (\bibinfo {year} {2015})}\BibitemShut {NoStop}%
\bibitem [{\citenamefont {Ryu}\ \emph {et~al.}(2010)\citenamefont {Ryu}, \citenamefont {Schnyder}, \citenamefont {Furusaki},\ and\ \citenamefont {Ludwig}}]{Ryu_2010}%
  \BibitemOpen
  \bibfield  {author} {\bibinfo {author} {\bibfnamefont {S.}~\bibnamefont {Ryu}}, \bibinfo {author} {\bibfnamefont {A.~P.}\ \bibnamefont {Schnyder}}, \bibinfo {author} {\bibfnamefont {A.}~\bibnamefont {Furusaki}},\ and\ \bibinfo {author} {\bibfnamefont {A.~W.~W.}\ \bibnamefont {Ludwig}},\ }\bibfield  {title} {\bibinfo {title} {Topological insulators and superconductors: tenfold way and dimensional hierarchy},\ }\href {https://doi.org/10.1088/1367-2630/12/6/065010} {\bibfield  {journal} {\bibinfo  {journal} {New J. Phys.}\ }\textbf {\bibinfo {volume} {12}},\ \bibinfo {pages} {065010} (\bibinfo {year} {2010})}\BibitemShut {NoStop}%
\bibitem [{\citenamefont {Jiang}\ \emph {et~al.}(2011)\citenamefont {Jiang}, \citenamefont {Kitagawa}, \citenamefont {Alicea}, \citenamefont {Akhmerov}, \citenamefont {Pekker}, \citenamefont {Refael}, \citenamefont {Cirac}, \citenamefont {Demler}, \citenamefont {Lukin},\ and\ \citenamefont {Zoller}}]{Jiang_2011}%
  \BibitemOpen
  \bibfield  {author} {\bibinfo {author} {\bibfnamefont {L.}~\bibnamefont {Jiang}}, \bibinfo {author} {\bibfnamefont {T.}~\bibnamefont {Kitagawa}}, \bibinfo {author} {\bibfnamefont {J.}~\bibnamefont {Alicea}}, \bibinfo {author} {\bibfnamefont {A.~R.}\ \bibnamefont {Akhmerov}}, \bibinfo {author} {\bibfnamefont {D.}~\bibnamefont {Pekker}}, \bibinfo {author} {\bibfnamefont {G.}~\bibnamefont {Refael}}, \bibinfo {author} {\bibfnamefont {J.~I.}\ \bibnamefont {Cirac}}, \bibinfo {author} {\bibfnamefont {E.}~\bibnamefont {Demler}}, \bibinfo {author} {\bibfnamefont {M.~D.}\ \bibnamefont {Lukin}},\ and\ \bibinfo {author} {\bibfnamefont {P.}~\bibnamefont {Zoller}},\ }\bibfield  {title} {\bibinfo {title} {{M}ajorana fermions in equilibrium and in driven cold-atom quantum wires},\ }\href {https://doi.org/10.1103/PhysRevLett.106.220402} {\bibfield  {journal} {\bibinfo  {journal} {Phys. Rev. Lett.}\ }\textbf {\bibinfo {volume} {106}},\ \bibinfo {pages} {220402} (\bibinfo {year} {2011})}\BibitemShut {NoStop}%
\bibitem [{\citenamefont {Roy}\ and\ \citenamefont {Harper}(2017)}]{Roy_2017}%
  \BibitemOpen
  \bibfield  {author} {\bibinfo {author} {\bibfnamefont {R.}~\bibnamefont {Roy}}\ and\ \bibinfo {author} {\bibfnamefont {F.}~\bibnamefont {Harper}},\ }\bibfield  {title} {\bibinfo {title} {Periodic table for {F}loquet topological insulators},\ }\href {https://doi.org/10.1103/PhysRevB.96.155118} {\bibfield  {journal} {\bibinfo  {journal} {Phys. Rev. B}\ }\textbf {\bibinfo {volume} {96}},\ \bibinfo {pages} {155118} (\bibinfo {year} {2017})}\BibitemShut {NoStop}%
\bibitem [{\citenamefont {Resta}(1998)}]{Resta_1998}%
  \BibitemOpen
  \bibfield  {author} {\bibinfo {author} {\bibfnamefont {R.}~\bibnamefont {Resta}},\ }\bibfield  {title} {\bibinfo {title} {Quantum-mechanical position operator in extended systems},\ }\href {https://doi.org/10.1103/PhysRevLett.80.1800} {\bibfield  {journal} {\bibinfo  {journal} {Phys. Rev. Lett.}\ }\textbf {\bibinfo {volume} {80}},\ \bibinfo {pages} {1800} (\bibinfo {year} {1998})}\BibitemShut {NoStop}%
\bibitem [{\citenamefont {Bardyn}\ \emph {et~al.}(2013)\citenamefont {Bardyn}, \citenamefont {Baranov}, \citenamefont {Kraus}, \citenamefont {Rico}, \citenamefont {{\.I}mamo{\u{g}}lu}, \citenamefont {Zoller},\ and\ \citenamefont {Diehl}}]{Bardyn_2013}%
  \BibitemOpen
  \bibfield  {author} {\bibinfo {author} {\bibfnamefont {C.-E.}\ \bibnamefont {Bardyn}}, \bibinfo {author} {\bibfnamefont {M.~A.}\ \bibnamefont {Baranov}}, \bibinfo {author} {\bibfnamefont {C.~V.}\ \bibnamefont {Kraus}}, \bibinfo {author} {\bibfnamefont {E.}~\bibnamefont {Rico}}, \bibinfo {author} {\bibfnamefont {A.}~\bibnamefont {{\.I}mamo{\u{g}}lu}}, \bibinfo {author} {\bibfnamefont {P.}~\bibnamefont {Zoller}},\ and\ \bibinfo {author} {\bibfnamefont {S.}~\bibnamefont {Diehl}},\ }\bibfield  {title} {\bibinfo {title} {Topology by dissipation},\ }\href {https://doi.org/10.1088/1367-2630/15/8/085001} {\bibfield  {journal} {\bibinfo  {journal} {New J. Phys.}\ }\textbf {\bibinfo {volume} {15}},\ \bibinfo {pages} {085001} (\bibinfo {year} {2013})}\BibitemShut {NoStop}%
\bibitem [{\citenamefont {Bravyi}(2005)}]{Bravyi_2005}%
  \BibitemOpen
  \bibfield  {author} {\bibinfo {author} {\bibfnamefont {S.}~\bibnamefont {Bravyi}},\ }\bibfield  {title} {\bibinfo {title} {Lagrangian representation for fermionic linear optics},\ }\href@noop {} {\bibfield  {journal} {\bibinfo  {journal} {Quantum Inf. Comput.}\ }\textbf {\bibinfo {volume} {5}},\ \bibinfo {pages} {216} (\bibinfo {year} {2005})}\BibitemShut {NoStop}%
\bibitem [{\citenamefont {Kohler}\ \emph {et~al.}(1997)\citenamefont {Kohler}, \citenamefont {Dittrich},\ and\ \citenamefont {H\"anggi}}]{Kohler_1997}%
  \BibitemOpen
  \bibfield  {author} {\bibinfo {author} {\bibfnamefont {S.}~\bibnamefont {Kohler}}, \bibinfo {author} {\bibfnamefont {T.}~\bibnamefont {Dittrich}},\ and\ \bibinfo {author} {\bibfnamefont {P.}~\bibnamefont {H\"anggi}},\ }\bibfield  {title} {\bibinfo {title} {{F}loquet-{M}arkovian description of the parametrically driven, dissipative harmonic quantum oscillator},\ }\href {https://doi.org/10.1103/PhysRevE.55.300} {\bibfield  {journal} {\bibinfo  {journal} {Phys. Rev. E}\ }\textbf {\bibinfo {volume} {55}},\ \bibinfo {pages} {300} (\bibinfo {year} {1997})}\BibitemShut {NoStop}%
\bibitem [{\citenamefont {Wustmann}(2010)}]{wustmann2010statistical}%
  \BibitemOpen
  \bibfield  {author} {\bibinfo {author} {\bibfnamefont {W.}~\bibnamefont {Wustmann}},\ }\emph {\bibinfo {title} {Statistical mechanics of time-periodic quantum systems}},\ \href@noop {} {Ph.D. thesis},\ \bibinfo  {school} {Technische Universit{\"a}t Dresden} (\bibinfo {year} {2010})\BibitemShut {NoStop}%
\bibitem [{\citenamefont {{\v{Z}}unkovi{\v{c}}}\ and\ \citenamefont {Prosen}(2010)}]{Zunkovic_2010}%
  \BibitemOpen
  \bibfield  {author} {\bibinfo {author} {\bibfnamefont {B.}~\bibnamefont {{\v{Z}}unkovi{\v{c}}}}\ and\ \bibinfo {author} {\bibfnamefont {T.}~\bibnamefont {Prosen}},\ }\bibfield  {title} {\bibinfo {title} {Explicit solution of the {L}indblad equation for nearly isotropic boundary driven {XY} spin 1/2 chain},\ }\href {https://doi.org/10.1088/1742-5468/2010/08/P08016} {\bibfield  {journal} {\bibinfo  {journal} {J. Stat. Mech.: Theory Exp.}\ }\textbf {\bibinfo {volume} {2010}}\bibinfo  {number} { (08)},\ \bibinfo {pages} {P08016}}\BibitemShut {NoStop}%
\bibitem [{\citenamefont {Prosen}\ and\ \citenamefont {Ilievski}(2011)}]{Prosen_2011}%
  \BibitemOpen
\bibfield  {number} {  }\bibfield  {author} {\bibinfo {author} {\bibfnamefont {T.}~\bibnamefont {Prosen}}\ and\ \bibinfo {author} {\bibfnamefont {E.}~\bibnamefont {Ilievski}},\ }\bibfield  {title} {\bibinfo {title} {Nonequilibrium phase transition in a periodically driven {$XY$} spin chain},\ }\href {https://doi.org/10.1103/PhysRevLett.107.060403} {\bibfield  {journal} {\bibinfo  {journal} {Phys. Rev. Lett.}\ }\textbf {\bibinfo {volume} {107}},\ \bibinfo {pages} {060403} (\bibinfo {year} {2011})}\BibitemShut {NoStop}%
\bibitem [{\citenamefont {Prosen}\ and\ \citenamefont {{\v{Z}}unkovi{\v{c}}}(2010)}]{Prosen_2010}%
  \BibitemOpen
  \bibfield  {author} {\bibinfo {author} {\bibfnamefont {T.}~\bibnamefont {Prosen}}\ and\ \bibinfo {author} {\bibfnamefont {B.}~\bibnamefont {{\v{Z}}unkovi{\v{c}}}},\ }\bibfield  {title} {\bibinfo {title} {Exact solution of {M}arkovian master equations for quadratic {F}ermi systems: thermal baths, open {XY} spin chains and non-equilibrium phase transition},\ }\href {https://doi.org/10.1088/1367-2630/12/2/025016} {\bibfield  {journal} {\bibinfo  {journal} {New J. Phys.}\ }\textbf {\bibinfo {volume} {12}},\ \bibinfo {pages} {025016} (\bibinfo {year} {2010})}\BibitemShut {NoStop}%
\bibitem [{\citenamefont {Prosen}(2010)}]{Prosen_2010_spectral}%
  \BibitemOpen
  \bibfield  {author} {\bibinfo {author} {\bibfnamefont {T.}~\bibnamefont {Prosen}},\ }\bibfield  {title} {\bibinfo {title} {Spectral theorem for the {L}indblad equation for quadratic open fermionic systems},\ }\href {https://doi.org/10.1088/1742-5468/2010/07/P07020} {\bibfield  {journal} {\bibinfo  {journal} {J. Stat. Mech.: Theory Exp.}\ }\textbf {\bibinfo {volume} {2010}}\bibinfo  {number} { (07)},\ \bibinfo {pages} {P07020}}\BibitemShut {NoStop}%
\bibitem [{\citenamefont {Kolisnyk}\ \emph {et~al.}(2024)\citenamefont {Kolisnyk}, \citenamefont {Quei{\ss}er}, \citenamefont {Schaller},\ and\ \citenamefont {Sch{\"u}tzhold}}]{kolisnyk_floquet_2024}%
  \BibitemOpen
\bibfield  {number} {  }\bibfield  {author} {\bibinfo {author} {\bibfnamefont {D.}~\bibnamefont {Kolisnyk}}, \bibinfo {author} {\bibfnamefont {F.}~\bibnamefont {Quei{\ss}er}}, \bibinfo {author} {\bibfnamefont {G.}~\bibnamefont {Schaller}},\ and\ \bibinfo {author} {\bibfnamefont {R.}~\bibnamefont {Sch{\"u}tzhold}},\ }\bibfield  {title} {\bibinfo {title} {{F}loquet analysis of a superradiant many-qutrit refrigerator},\ }\href {https://doi.org/10.1103/PhysRevApplied.21.044050} {\bibfield  {journal} {\bibinfo  {journal} {Phys. Rev. Applied}\ }\textbf {\bibinfo {volume} {21}},\ \bibinfo {pages} {044050} (\bibinfo {year} {2024})}\BibitemShut {NoStop}%
\bibitem [{\citenamefont {Molignini}(2019)}]{molignini_2019}%
  \BibitemOpen
  \bibfield  {author} {\bibinfo {author} {\bibfnamefont {P.}~\bibnamefont {Molignini}},\ }\emph {\bibinfo {title} {From topological to dissipative many-body phases out of equilibrium}},\ \href@noop {} {Ph.D. thesis},\ \bibinfo  {school} {ETH Zurich} (\bibinfo {year} {2019})\BibitemShut {NoStop}%
\bibitem [{\citenamefont {Schnell}\ \emph {et~al.}(2020)\citenamefont {Schnell}, \citenamefont {Eckardt},\ and\ \citenamefont {Denisov}}]{Schnell_2020}%
  \BibitemOpen
  \bibfield  {author} {\bibinfo {author} {\bibfnamefont {A.}~\bibnamefont {Schnell}}, \bibinfo {author} {\bibfnamefont {A.}~\bibnamefont {Eckardt}},\ and\ \bibinfo {author} {\bibfnamefont {S.}~\bibnamefont {Denisov}},\ }\bibfield  {title} {\bibinfo {title} {Is there a {F}loquet {L}indbladian?},\ }\href {https://doi.org/10.1103/PhysRevB.101.100301} {\bibfield  {journal} {\bibinfo  {journal} {Phys. Rev. B}\ }\textbf {\bibinfo {volume} {101}},\ \bibinfo {pages} {100301} (\bibinfo {year} {2020})}\BibitemShut {NoStop}%
\bibitem [{\citenamefont {Schnell}\ \emph {et~al.}(2021)\citenamefont {Schnell}, \citenamefont {Denisov},\ and\ \citenamefont {Eckardt}}]{Schnell_2021}%
  \BibitemOpen
  \bibfield  {author} {\bibinfo {author} {\bibfnamefont {A.}~\bibnamefont {Schnell}}, \bibinfo {author} {\bibfnamefont {S.}~\bibnamefont {Denisov}},\ and\ \bibinfo {author} {\bibfnamefont {A.}~\bibnamefont {Eckardt}},\ }\bibfield  {title} {\bibinfo {title} {High-frequency expansions for time-periodic {L}indblad generators},\ }\href {https://doi.org/10.1103/PhysRevB.104.165414} {\bibfield  {journal} {\bibinfo  {journal} {Phys. Rev. B}\ }\textbf {\bibinfo {volume} {104}},\ \bibinfo {pages} {165414} (\bibinfo {year} {2021})}\BibitemShut {NoStop}%
\bibitem [{\citenamefont {Dinc}\ \emph {et~al.}(2025)\citenamefont {Dinc}, \citenamefont {Eckardt},\ and\ \citenamefont {Schnell}}]{Dinc_2025}%
  \BibitemOpen
  \bibfield  {author} {\bibinfo {author} {\bibfnamefont {G.~D.}\ \bibnamefont {Dinc}}, \bibinfo {author} {\bibfnamefont {A.}~\bibnamefont {Eckardt}},\ and\ \bibinfo {author} {\bibfnamefont {A.}~\bibnamefont {Schnell}},\ }\bibfield  {title} {\bibinfo {title} {Effective {F}loquet {L}indblad generators from spectral unwinding},\ }\href {https://doi.org/10.1103/x2qh-12gg} {\bibfield  {journal} {\bibinfo  {journal} {Phys. Rev. A}\ }\textbf {\bibinfo {volume} {111}},\ \bibinfo {pages} {062216} (\bibinfo {year} {2025})}\BibitemShut {NoStop}%
\bibitem [{\citenamefont {van Caspel}\ \emph {et~al.}(2019)\citenamefont {van Caspel}, \citenamefont {Tapias~Arze},\ and\ \citenamefont {P{\'e}rez~Castillo}}]{Caspel_2019}%
  \BibitemOpen
  \bibfield  {author} {\bibinfo {author} {\bibfnamefont {M.}~\bibnamefont {van Caspel}}, \bibinfo {author} {\bibfnamefont {S.~E.}\ \bibnamefont {Tapias~Arze}},\ and\ \bibinfo {author} {\bibfnamefont {I.}~\bibnamefont {P{\'e}rez~Castillo}},\ }\bibfield  {title} {\bibinfo {title} {Dynamical signatures of topological order in the driven-dissipative {K}itaev chain},\ }\href {https://doi.org/10.21468/SciPostPhys.6.2.026} {\bibfield  {journal} {\bibinfo  {journal} {SciPost Phys.}\ }\textbf {\bibinfo {volume} {6}},\ \bibinfo {pages} {026} (\bibinfo {year} {2019})}\BibitemShut {NoStop}%
\bibitem [{\citenamefont {Lieu}\ \emph {et~al.}(2020)\citenamefont {Lieu}, \citenamefont {McGinley},\ and\ \citenamefont {Cooper}}]{Lieu2020}%
  \BibitemOpen
  \bibfield  {author} {\bibinfo {author} {\bibfnamefont {S.}~\bibnamefont {Lieu}}, \bibinfo {author} {\bibfnamefont {M.}~\bibnamefont {McGinley}},\ and\ \bibinfo {author} {\bibfnamefont {N.~R.}\ \bibnamefont {Cooper}},\ }\bibfield  {title} {\bibinfo {title} {Tenfold way for quadratic {L}indbladians},\ }\href {https://doi.org/10.1103/PhysRevLett.124.040401} {\bibfield  {journal} {\bibinfo  {journal} {Phys. Rev. Lett.}\ }\textbf {\bibinfo {volume} {124}},\ \bibinfo {pages} {040401} (\bibinfo {year} {2020})}\BibitemShut {NoStop}%
\end{thebibliography}%

\end{document}